\documentclass[aps,prx,reprint,twocolumn,superscriptaddress,showpacs,showkeys,longbibliography]{revtex4-1}

\usepackage{amsmath,amssymb,amstext}
\usepackage[usenames,dvipsnames]{color}
\usepackage{graphicx}
\usepackage{bm,bbold,braket}
\usepackage{natbib}
\usepackage{txfonts, comment,stmaryrd}
\usepackage{dcolumn}

\usepackage[english]{babel}

\usepackage{wasysym}
\usepackage[colorlinks,bookmarks=false,citecolor=blue,linkcolor=red,urlcolor=blue]{hyperref}

\definecolor{mygreen}{rgb}{0,0.5,0}
\definecolor{myblue}{rgb}{0,0,0.75}
\definecolor{mymagenta}{cmyk}{0,1,0,0.12}

\definecolor{darkblue}{rgb}{0.0,0.0,0.7}

\definecolor{darkgreen}{rgb}{0.0,0.7,0.0}

\def\be{\begin{equation*}}
\def\ee{\end{equation*}}
\def\bea{\begin{eqnarray*}}
\def\eea{\end{eqnarray*}}

\def\vec{\mathbf}
\def\bs{\boldsymbol}
\def\mc{\mathcal}
\newcommand{\ra}[1]{\renewcommand{\arraystretch}{#1}} 
\newcolumntype{"}{@{\hskip\tabcolsep\vrule width 1pt\hskip\tabcolsep}}

\def\wt#1{\inner(#1)}
\def\inner(#1,#2,#3,#4,#5,#6){\ensuremath\left(\begin{array}{ccc} #1 & #2 & #3 \\ #4 & #5 & #6 \end{array}\right)}

\def\ws#1{\innerv(#1)}
\def\innerv(#1,#2,#3,#4,#5,#6){\ensuremath\left\{\begin{array}{ccc} #1 & #2 & #3 \\ #4 & #5 & #6 \end{array}\right\}}

\begin{document}

\title{Quantum Spin Ice and dimer models with Rydberg atoms}

\author{A.~W.~Glaetzle}
    \email{alexander.glaetzle@uibk.ac.at}    
    \affiliation{Institute for Quantum Optics and Quantum Information of the Austrian Academy of Sciences, A-6020 Innsbruck, Austria}
    \affiliation{Institute for Theoretical Physics, University of Innsbruck, A-6020 Innsbruck, Austria}
\author{M.~Dalmonte}
    \affiliation{Institute for Quantum Optics and Quantum Information of the Austrian Academy of Sciences, A-6020 Innsbruck, Austria}
        \affiliation{Institute for Theoretical Physics, University of Innsbruck, A-6020 Innsbruck, Austria}
\author{R.~Nath}
    \affiliation{Institute for Quantum Optics and Quantum Information of the Austrian Academy of Sciences, A-6020 Innsbruck, Austria}
     \affiliation{Institute for Theoretical Physics, University of Innsbruck, A-6020 Innsbruck, Austria}
        \affiliation{Indian Institute of Science Education and Research, Pune 411 008, India}
 \author{I.~Rousochatzakis}
    \affiliation{Max Planck Institute for the Physics of Complex Systems, D-01187 Dresden, Germany}
 \author{R.~Moessner}
    \affiliation{Max Planck Institute for the Physics of Complex Systems, D-01187 Dresden, Germany}
\author{P.~Zoller}
    \affiliation{Institute for Quantum Optics and Quantum Information of the Austrian Academy of Sciences, A-6020 Innsbruck, Austria}
    \affiliation{Institute for Theoretical Physics, University of Innsbruck, A-6020 Innsbruck, Austria}

\date{\today}

\begin{abstract}
Quantum spin ice represents a paradigmatic example on how the physics of frustrated magnets is related to gauge theories. In the present work we address the problem of approximately realizing quantum spin ice in two dimensions with cold atoms in optical lattices. The relevant interactions are obtained by weakly admixing van der Waals interactions between laser admixed Rydberg states to the atomic ground state atoms, exploiting the strong angular dependence of interactions between Rydberg $p$-states together with the possibility of designing step-like potentials. This allows us to implement Abelian gauge theories in a series of geometries, which could be demonstrated within state of the art atomic Rydberg experiments. We numerically analyze the family of resulting microscopic Hamiltonians and find that they exhibit both classical and quantum order by disorder, the latter yielding a quantum plaquette valence bond
solid. We also present strategies to implement Abelian gauge theories using both $s$- and $p$-Rydberg states in exotic geometries, e.g. on a 4-8 lattice.
\end{abstract}



\maketitle

\section{Introduction\label{sec:intro}}
\begin{figure}[tb]  
\centering 
\includegraphics[width= 0.9\columnwidth]{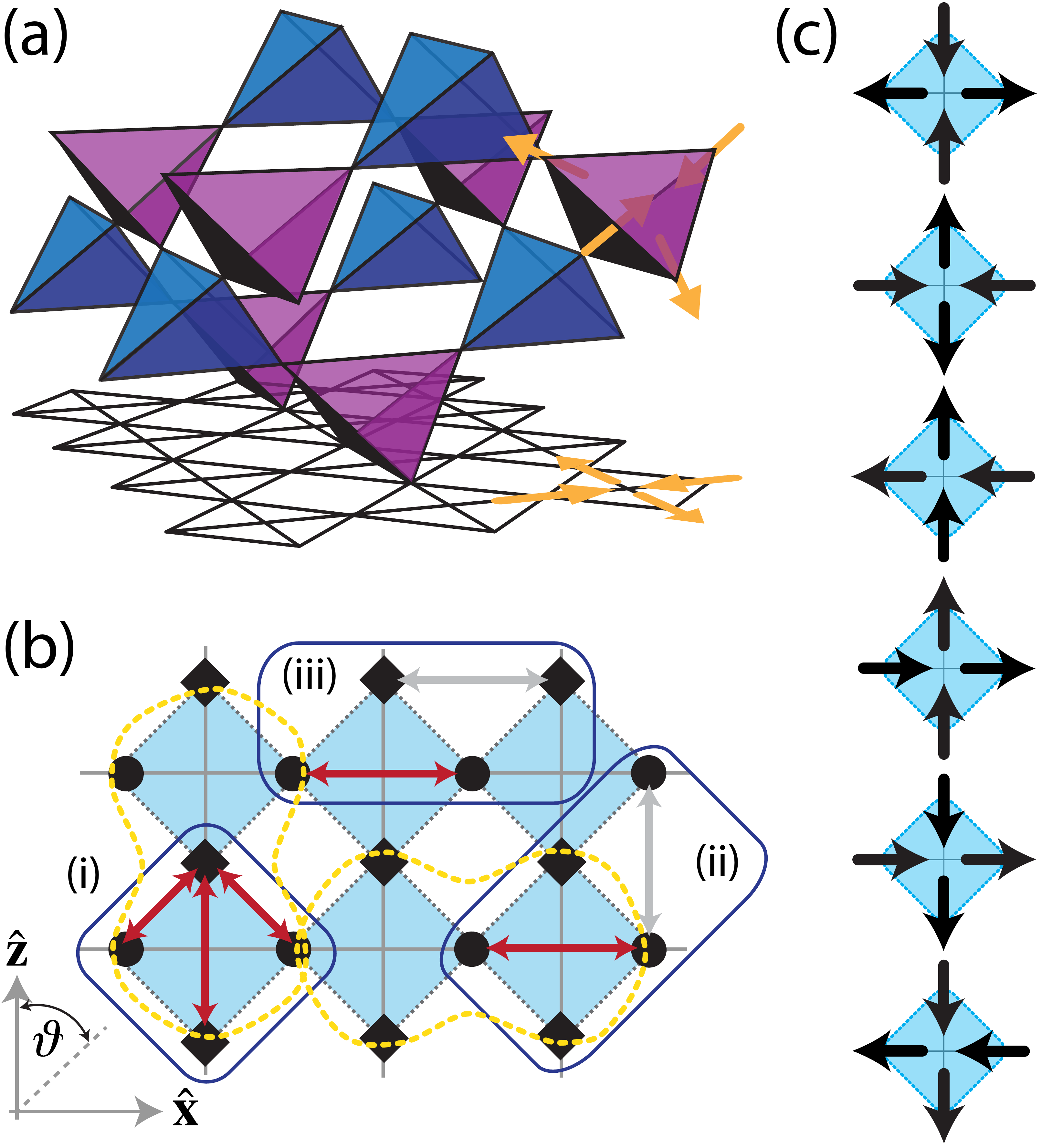}\\
\caption{\small{(a) In spin ice materials the magnetic moments  (yellow arrows) of rare-earth ions are located on the corners of a pyrochlore lattice, which is a network of corner-sharing tetrahedra. They behave as almost perfect Ising spins and point along the line from the corner to the centre of the tetrahedron, either inward or outward. Due to the different Ising-axes of the spins this results in an effectively antiferromagnetic interaction which is frustrated. (b)  Projecting the 3D pyrochlore lattice onto a 2D square lattice yields a checkerboard lattice where tetrahedrons are mapped onto crossed-plaquettes (light-blue). Interactions between two spins located on $\medbullet$ or $\blacksquare$ lattice sites have to be (i) step-like as a function of the distance, (ii) anisotropic and (iii) require a bipartite labelling of the lattice sites. (c) Degenerate ground state configurations of spins on a crossed-plaquette. They obey the {\it ice-rules}, which enforce two spins pointing inward and two spins pointing outward at each vertex.}}
\label{fig:pyro} 
\end{figure}

The ice model has been fundamental in furthering our understanding of collective phenomena in condensed matter and statistical physics: in 1935  Pauling provided an explanation of the `zero-point entropy' of water ice ~\cite{Pauling:1935hm} as measured by Giauque and Stout~\cite{Giauque:1933cd,*Giauque:1936gz}, while Lieb demonstrated with his exact solution of the ice model in two dimension ~\cite{Lieb:1967bv} that there exist phase transitions with critical exponents different from those of Onsager's solution of the Ising model.
The experimental discovery~\cite{R12} of a classical spin version of the ice model~\cite{R13} has in turn generated much interest in the magnetism community~\cite{R14a, Balents2010, lacroix2011introduction,R14c}.

More recently, quantum ice models~\cite{R1a,R1b,R1c,R1d,R1e} have attracted a great deal of attention in the context of phases exhibiting exotic types of orders, such as resonating valence bond liquids~\cite{R2a,R2b} or quantum Coulomb phases~\cite{R1c,R3b,R3c,R3d}. They form part of a broader family of models, which also includes quantum dimer models or other quantum vertex models~\cite{R4}, in which locally a hard constraint is imposed, such as the ice rules defined below. Such a constraint can then endow the configuration space with additional structure -- most prominently, an emergent gauge field which can be the basis of the appropriate effective description at low energies~\cite{Castelnovo:2012kk,eduardo1997field,R4}. This is an important phenomenon as it is perhaps the simplest way of obtaining gauge fields as effective degrees of freedom in condensed matter physics. More broadly, this is part of a long-running search for magnetic materials hosting quantum spin liquids~\cite{Balents2010}.

In the present work we address the problem of physically realizing models of quantum spin ice and quantum dimer models in two spatial dimensions with ultracold atoms in optical lattices. Our proposal builds on the recent experimental advances, and opportunities in engineering  many-body  interactions with laser excited Rydberg states~\cite{Heidemann:2008bg,Anderson:1998hn, Hofmann:2013gm, Maxwell:2013km,Schwarzkopf:2013fx,Schempp:2014jm,Tauschinsky:2010ep, Carter:2013vi,Kubler:2010dw,Dudin:2012hm, Schauss:2012hh,Li:2013gg, Ebert:2013tr,Valado:2013ez,Barredo:2014tb}. In particular, this will allow us to develop a Rydberg toolbox for the complex interactions required in 2D quantum ice models~\cite{lacroix2011introduction}. Our investigation also fits into the broader quest for the realization of synthetic gauge fields with cold atoms. While much effort is being devoted to the generation of static gauge fields~\cite{Dalibard2011}, e.g. on optical flux lattices, here we follow the strategy of generating a {\it dynamical} gauge field~\cite{R1c,R3b,R3c,R3d,Castelnovo:2012kk,eduardo1997field, Tewari:2006js,Banerjee:2013fl,Tagliacozzo:2013bv,Zohar:2013eo} emerging upon imposition of the ice rule.

While in condensed matter systems the interactions underlying ice and spin ice arise naturally in a 3D context [see Fig.~\ref{fig:pyro}(a)], the 2D quantum ice on a square lattice requires a certain degree of  fine tuning of the relevant interactions [see Fig.~\ref{fig:pyro}(b)]. In 3D spin ice materials, for example, the ions of magnetic rare-earth atoms reside on a pyrochlore lattice, representing a network of corner-sharing tetrahedra. Magnetic interactions in combination with crystal fields give rise to a low energy manifold of states on each tetrahedron consisting of six configurations, in which two spins point inward and two spins point outward [compare Fig.~\ref{fig:pyro}(c)]. In a similar way, in water ice each O$^{2-}$ atom in a tetrahedrally coordinated framework has two protons attached to it, giving rise to a manifold of energetically degenerate configurations. 2D models of ice and spin ice can be understood as projection of the pyrochlore on a square lattice (see Fig.~\ref{fig:pyro}), where again the low energy configurations of spins residing on the links obey the ``ice rule" two-in and two-out at each vertex. While these 2D ice models play a fundamental role in our theoretical understanding of frustrated materials, a physical realization requires a precise adjustment of the underlying interactions -- different local configurations which are symmetry distinct need to be at least approximately degenerate; the required fine-tuning however needs to be delicately directionally dependent, as in the pure ice model, the ratio of some interactions between different pairs of equidistant spins vanishes, see Fig.~\ref{fig:pyro}(b). Things are not all hopeless, however, as there exist a number of settings in which partial progress has been made to realizing such models. In two dimensions, artificial structures using nanomagnetic~\cite{Nisoli:2013uk} or colloidal arrays~\cite{Libal:2006ie} have been proposed, including strategies for tuning the interactions appropriately~\cite{R5}. The present proposal with Rydberg atoms is unique, however, as it combines both the possibilities of engineering the complex interactions using Rydberg interactions with the accessibility of the quantum regime in cold atom experiments. 

Alkali atoms prepared in their electronic ground state can be excited by laser light to Rydberg states, i.e. states of high principal quantum number $n$~\cite{gallagher2005rydberg,Saffman:2010ky,Comparat:2010cb,Low:2012ct}. These Rydberg atoms interact strongly via the van der Waals interaction exhibiting the remarkable scaling $V_{\rm VdW}\sim n^{11}$, and which exceed typical ground state interactions of cold atoms by several orders of magnitude. In an atomic ensemble the large level shifts associated with these interactions implies that only a single atom can be excited to the Rydberg state, while multiple excitations are suppressed within a blockade radius determined by the van der Waals interactions and laser parameters~\cite{Jaksch:2000eg,Lukin:2001bua}. This blockade mechanism results in  novel collective and strongly correlated many-particle phenomena such as the formation of superatoms, and Rydberg quantum crystals~\cite{gallagher2005rydberg,Saffman:2010ky,Comparat:2010cb,Low:2012ct}. In present experiments the emphasis is on {\it isotropic} van der Waals interactions~\cite{Heidemann:2008bg,Anderson:1998hn, Hofmann:2013gm, Maxwell:2013km,Schwarzkopf:2013fx,Schempp:2014jm,Tauschinsky:2010ep, Carter:2013vi,Kubler:2010dw,Dudin:2012hm, Schauss:2012hh,Li:2013gg, Ebert:2013tr,Valado:2013ez}, which, for example, can be obtained by exciting Rydberg $s$-states using a two-photon excitation scheme. In contrast, we will be interested below in excitations of Rydberg $p$-states, where the van der Waals interactions can be highly {\it anisotropic}, and we will discuss below in detail the controllability of this anisotropy, shape and range of these interactions via atomic and laser parameters for the case of Rb atoms. In the present context this will provide us with the tools to engineer the required complex interaction patterns for 2D quantum spin ice and dimer models. The setup we will discuss will consist of cold atoms in optical lattices, where the strong Rydberg interactions are weakly admixed to the atomic ground states~\cite{Santos:2000ec,Honer:2010je,Pupillo:2010bt,Henkel:2010il,Macri:2014jy}, thus effectively dressing ground state atoms to obtain the complex interactions in atomic Hubbard models required for the realization of 2D quantum spin ice and dimer models. 

We then proceed to numerically analyze the family of Hamiltonian realizable with this toolbox. We verify that it contains two phases exhibiting distinct types of order by disorder~\cite{R6,R7}. Most remarkably, quantum order by disorder -- due to the presence of quantum dynamics in the ice model -- realizes the plaquette valence bond solid as an unusual non-N\'eel phase of a frustrated magnet. This terminates when classical degeneracy lifting takes over.

The latter phase is conventional in that it is diagnosed by a conventional spin order parameter, which would manifest itself in a Bragg peak in the structure factor. By contrast, the valence bond solid would be diagnosed by higher order `string' correlators, and is thus fundamentally different from some other instances of quantum order by disorder, where finally the mechanism, but not quite so much the order parameter, is exotic~\cite{R7,R8}. Notably, precisely such order parameters have become accessible to experimental measurements recently~\cite{R9}, so that our proposal not only covers the setting for realizing quantum order by disorder, but also the means for detecting it.

In addition, at finite temperature, we find that the classically ordered state, even in the absence of quantum dynamics, melts into a classical version of a Coulomb phase, namely a Coulomb gas in which thermally activated plaquettes violating the ice rule play the role of positive and negative charges. As these interact via an entropic two-dimensional (logarithmic) Coulomb law, this phase is only marginally confined~\cite{R10a,R10b}.

While throughout the paper we will be mostly interested in quantum ice models, which require the development of advanced interaction-pattern design, we will also discuss a second strategy to implement constrained dynamics, and in particular quantum dimer models, with Rydberg atoms. It relies on combining simple interaction patterns, such as the ones generated by s-states, with complex lattice structures, which can be realized either via proper laser combination or by the recently developed optical lattice design with digital micromirror devices~\cite{Fukuhara:2013hq}. These models extend the class of dynamical gauge fields in AMO systems to non hyper-cubic geometries. 
Overall, the ability to synthetically design Abelian dynamical gauge fields with discrete variables also establishes interesting connections with high energy physics, where these theories are usually refereed as quantum link models~\cite{Banerjee:2013dy,Wiese:2013kk}. Within this context, the key developments in engineering pure gauge theories can be combined with other schemes, where dynamical matter is included, which have already been proposed in the context of cold atom gases~\cite{Wiese:2013kk}.

The the paper is structured as follows. In Sec.~\ref{sec:iceback}, we briefly provide the background on quantum ice models needed for digesting the remainder of the material. Sec.~\ref{sec:amo} outlines the implementation of a family of model Hamiltonians approximating the quantum ice model using atoms in optical lattices weakly admixed with a Rydberg $p$-state. Our model Hamiltonians are then analyzed for their phase diagram in Sec.~\ref{sec:mb}. Sec.~\ref{sec:beyond} presents strategies to implement simpler Abelian gauge theories using both $s$- and $p$-Rydberg states in exotic geometries, e.g. a 4-8 lattice. The paper closes with a summary and outlook in Sec.~\ref{sec:conc}.

\section{The quantum ice model\label{sec:iceback}} 

This section provides a brief overview over the statistical mechanics of the ice model, the emergence of a gauge field, and the challenges in realizing such a model experimentally. 

\subsection{The configuration space: ice rules and emergent gauge fields} 

\begin{figure}[tb]  
\centering 
\includegraphics[width= 0.99\columnwidth]{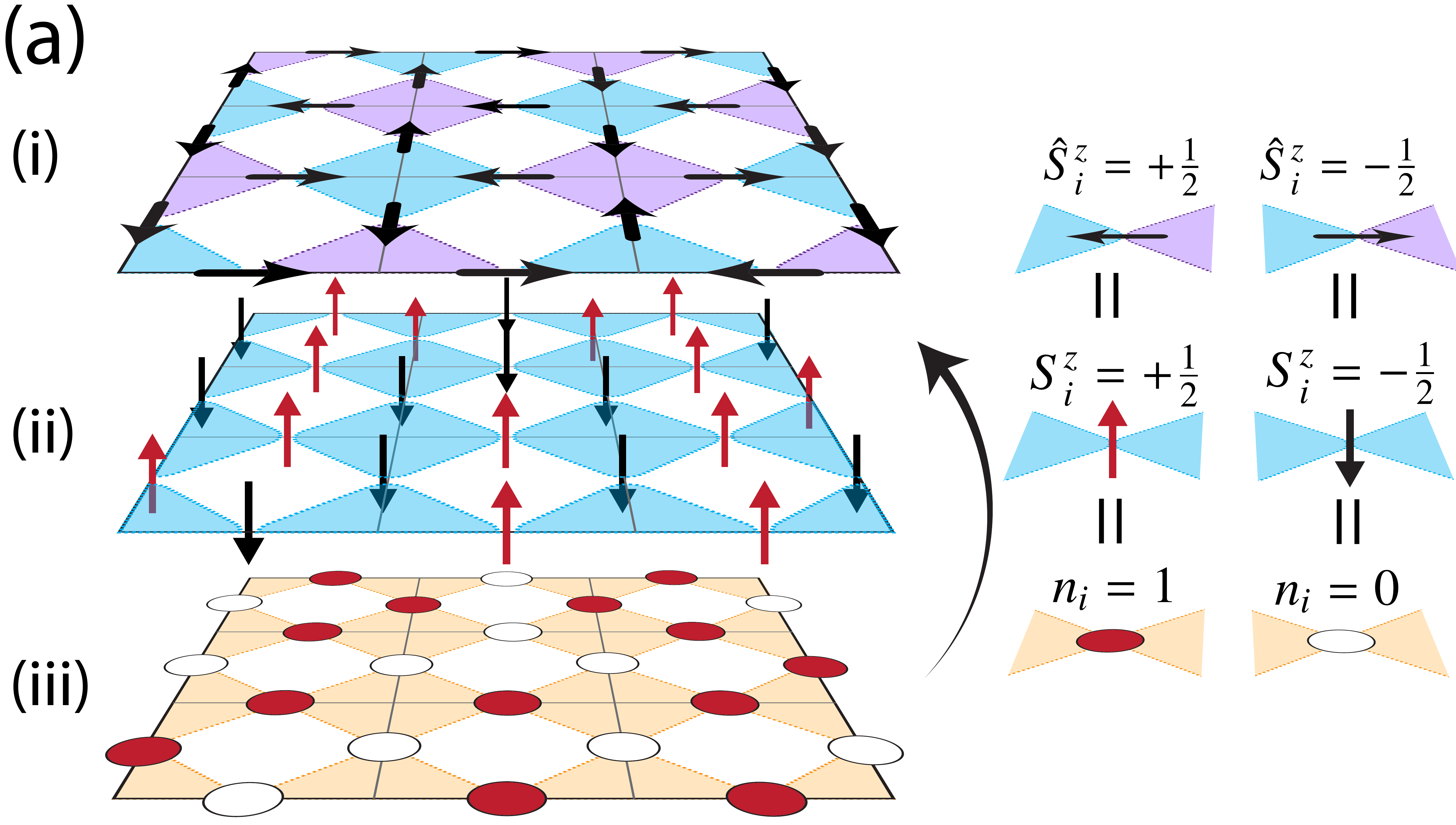}\\
\vspace{.3 cm}
\includegraphics[width= 0.98\columnwidth]{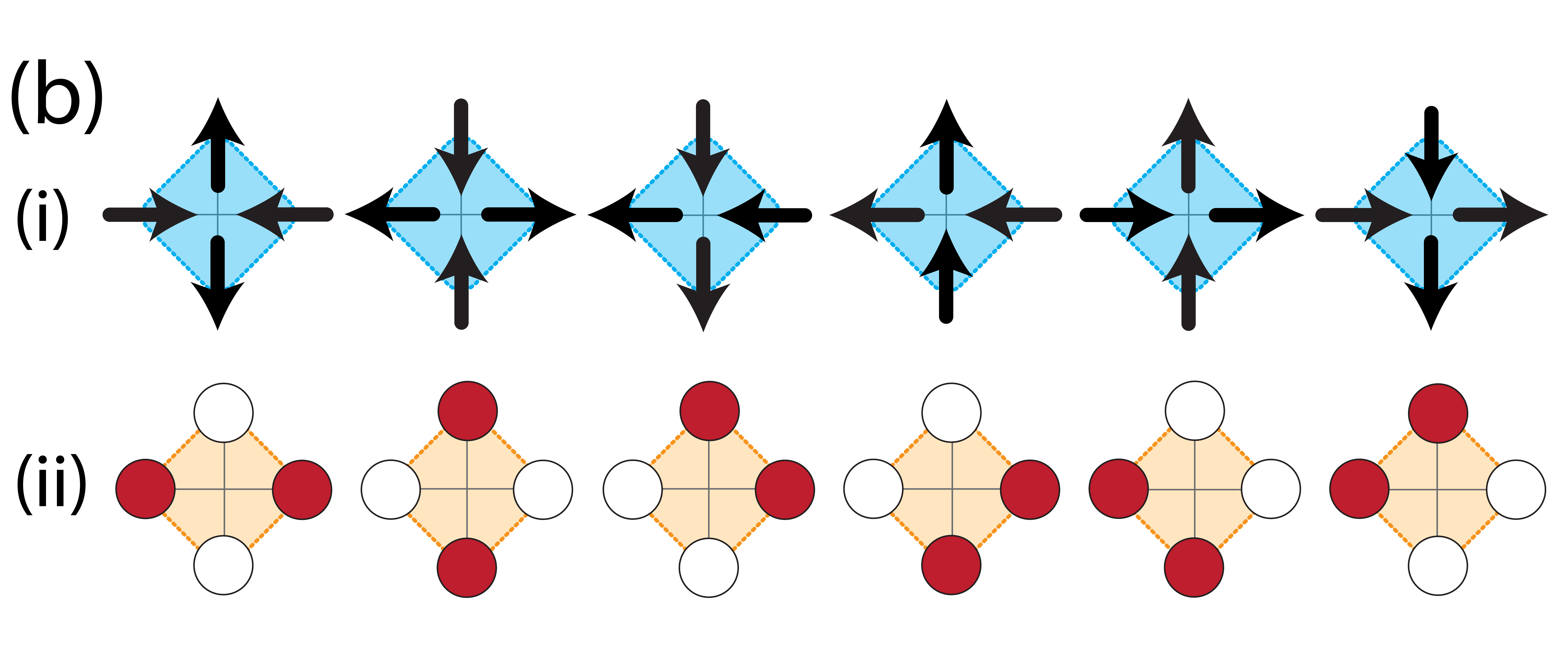}\\
\caption{\small{(a)  The spins of the 2D ice model on a checkerboard lattice can be interpreted as (i) ``fluxes'', $\{\hat S_i^z \}$, pointing either inward or outward of a specific crossed plaquette. This requires a bipartite labelling of the plaquettes (light blue and light magenta plaquettes) since an outward pointing flux vector corresponds to an inward pointing flux vector for the neighboring plaquette. (ii)  They can be interpreted as spins, $\{S_i^z \}$ aligned perpendicular to the plane, pointing either up (red arrows) or down (black arrows). In the right inset we identify a spin pointing up, $S_i^z=+\frac{1}{2}$, with a flux vector pointing from the magenta to the blue plaquette, $\hat S_i^z=+\frac{1}{2}$ and {\it vice versa}. 
(iii) Spins, $\{S_i^z \}$, can be mapped onto hard-core bosons, $n_i\in\{0,1\}$. Here, e.g. a particle (red circle), $n_i=1$, corresponds to a spin pointing upward, $S_i^z=+\frac{1}{2} $, while an empty lattice site (white circle), $n_i=0$, corresponds to a spin pointing downward, $S_i^z=-\frac{1}{2} $. (b) The six {\it ice-rule} states correspond to vertex configurations with two hard-core bosons and two empty lattice sites. }}
\label{fig:mapping} 
\end{figure}

The ice model on the square lattice, also known the six-vertex model or simply square ice~\cite{Lieb:1967bv}, has Ising degrees of freedom residing on the {\em links} of a square lattice. They can either be thought of as `fluxes', $\left\{\hat S^z_i\right\}$, pointing in either of the directions along the bond $i$, see layer (i) of Fig.~\ref{fig:mapping}(a); or equivalently can be mapped onto spins $\left\{S^z_i\right\}$ which point up or down depending on the direction of the flux [see layer (ii) of Fig.~\ref{fig:mapping}(a)].

Only configurations satisfying the ice rule are permitted, which stipulate that the spins on each vertex add up to zero -- there are $\left({4 \atop 2}\right)=6$\ ways of arranging this, see panel~(i) of Fig.~\ref{fig:mapping}(b). The number of configurations satisfying the ice rule grows exponentially with the size of the system -- for a lattice of $N$ spins, there are $\left(4/3\right)^\frac{3N}{4}$ ice states~\cite{Lieb:1967bv}.

The origin of the emergent gauge field is transparent in flux language, where it implies that the lattice divergence of the flux field vanishes: defining the x(y) component of a two-dimensional vector flux ${\bf b}$\ to be the flux along the corresponding links emanating from a vertex in the positive x(y) direction, one has
\begin{equation}
\nabla\cdot{\bf b}=0 \Rightarrow\ {\bf b}=\nabla\times {\bf a} .
\end{equation}

Note that a gauge field ${\bf a}$ has appeared naturally as a consequence of enforcing the ice rule, just as it does in magnetostatics, where Maxwell's law for the magnetic field $\nabla\cdot{\bf B}=0$ leads to the introduction of the familiar vector potential ${\bf A}$.

In the present example in two dimensions, where the flux is a two component vector ${\bf b}$, and the scalar constraint $\nabla\cdot{\bf b}=0$ fixes one degree of freedom, ${\bf a}$ only has one physical degree of freedom left -- it can be thought of as a scalar, usually referred to as a height: ${\bf a}=h{\bf z}$  `in the z-direction'~\cite{R10a,R10b}. Defects in this height field -- forbidden in the six vertex model but allowed when violating the ice rule comes only with a finite energy penalty -- are then known as charges or monopoles, which carry a gauge charge with respect to the emergent gauge field. 

Having enforced the ice rule, the natural degree of freedom is thence an {\em emergent} gauge field ${\bf a}$ -- it is in this way that gauge fields quite generically emerge in condensed matter physics, with a constraint arising either from the need to satisfy a dominant term in the Hamiltonian, or a microscopic relation on the local Hilbert space~\cite{R4}.

\subsection{Realization, and fine-tuning in $d=2$} 
\label{sec:2d}

The ice rule on a given vertex involves four spins, but it can be enforced via a pairwise interaction: if all four spins on a vertex interact antiferromagnetically {\em and equally} -- described by the Hamiltonian 
\begin{equation}
H_{\rm ice}=V\sum_{i,j\in +}S_i^zS_j^z,
\label{eq:spinice}
\end{equation}
where $+$ denotes a crossed plaquette in 2D or a tetrahedron in 3D -- the resulting ground states are those which obey the ice rules, see panel (i) of Fig.~\ref{fig:mapping}(b). In three dimensions, equality of the pairwise interactions can be symmetry-generated -- by placing the spins on the corners of a tetrahedron, any antiferromagnetic interaction depending only on the distance between the spins will yield the ice rule. By contrast, in two dimensions, a tetrahedron becomes a square with interactions also across the diagonal (Fig.~\ref{fig:pyro}), which are no longer symmetry equivalent to those along the edges~\cite{R11}.

In particular, interactions, $V_{ij}(\mathbf{r})$, between two spins $i$ and $j$ located on the bonds of a checkerboard lattice separated by a distance $\mathbf{r}$, have to fulfill three demeaning properties [see Fig.~\ref{fig:pyro}(b)] in order to map onto the Spin ice Hamiltonian of Eq.~\eqref{eq:spinice}

\begin{enumerate}

\item {\bf Anisotropy:} Interactions have to be {\it strongly anisotropic}. This is illustrated in panel (ii) of Fig.~\ref{fig:pyro}(b). Particles which belong to the same vertex interact strongly (red arrow), while particles which do not belong to the same vertex do not interact (gray arrow).  Thus, for $\medbullet$ particles in panel (ii) one needs an interaction which satisfied $V_{\medbullet\medbullet}(\vartheta=0)=0$ (gray arrow) and $V_{\medbullet\medbullet}(\vartheta=\pi/2)=\tilde V_0$ (red arrow), where the angle $\vartheta$ is defined in the inset.
 
 \item {\bf Step-like potentials:} All four particles which belong to the same vertex (enclosed by light blue squares) interact with the same strength $\tilde V_0$, independent of their distance, either $a$ or $a\sqrt{2}$, where $a$ is the lattice spacing. Obviously, an interaction of the form $1/|\mathbf{r}|^\alpha$ would not suffice. It is therefore necessary to have {\it step-like potentials} which fulfill $V_{ij}(|\mathbf{r}|<r_c)=\tilde V_0\neq 0$ and $V_{ij}(|\mathbf{r}|>r_c)=0$ for $\sqrt{2}a<r_c<2a$. 

\item {\bf Bipartite-lattice structure:} Furthermore, panel (iii) of Fig.~\ref{fig:pyro}(b), shows that the desired interaction properties cannot be satisfied by a homogeneous interaction pattern, but require a bipartite structure [squares and circles in Fig.~\ref{fig:pyro}(b)] where the angular dependence on the interaction depends on the lattice bipartition. For example, in the last paragraph we enforced that $V_{\medbullet\medbullet}(\vartheta=0)=0$ [see panel (ii)] but the opposite is true for $\blacksquare$ particles, see panel (iii). Here, $V_{\blacksquare\blacksquare}(\vartheta=\pi/2)=0$ but  $V_{\blacksquare\blacksquare}(\vartheta=0)=\tilde V_0$. On top of that, mixed interactions between $\medbullet$ and $\blacksquare$ particles on the 45 degree lines should obey $V_{\medbullet\blacksquare}(\vartheta=\pm\pi/4)=\tilde V_0$ in order to ensure that all six possible interactions at a specific vertex are the same, see panel (i).
\end{enumerate}

It is these three countervailing requirements that we manage to satisfy approximately  by using Rydberg dressed atoms to engineer an appropriate quantum Hamiltonian (Sec.~\ref{sec:amo}).

\subsection{Adding quantum dynamics, and quantum order by disorder}
While the properties of the two-dimensional ice model were broadly understood a long time ago, the question what a quantum version of it would look like was not posed until much later~\cite{R1a}. Unlike in, say, a transverse field Ising model, where the simplest quantum dynamics consists of reversing a single spin, the ice model does not permit such single-site configuration changes, as these would lead to a violation of the ice rule. 

\begin{figure}[tb]  
\centering 
\includegraphics[width= \columnwidth]{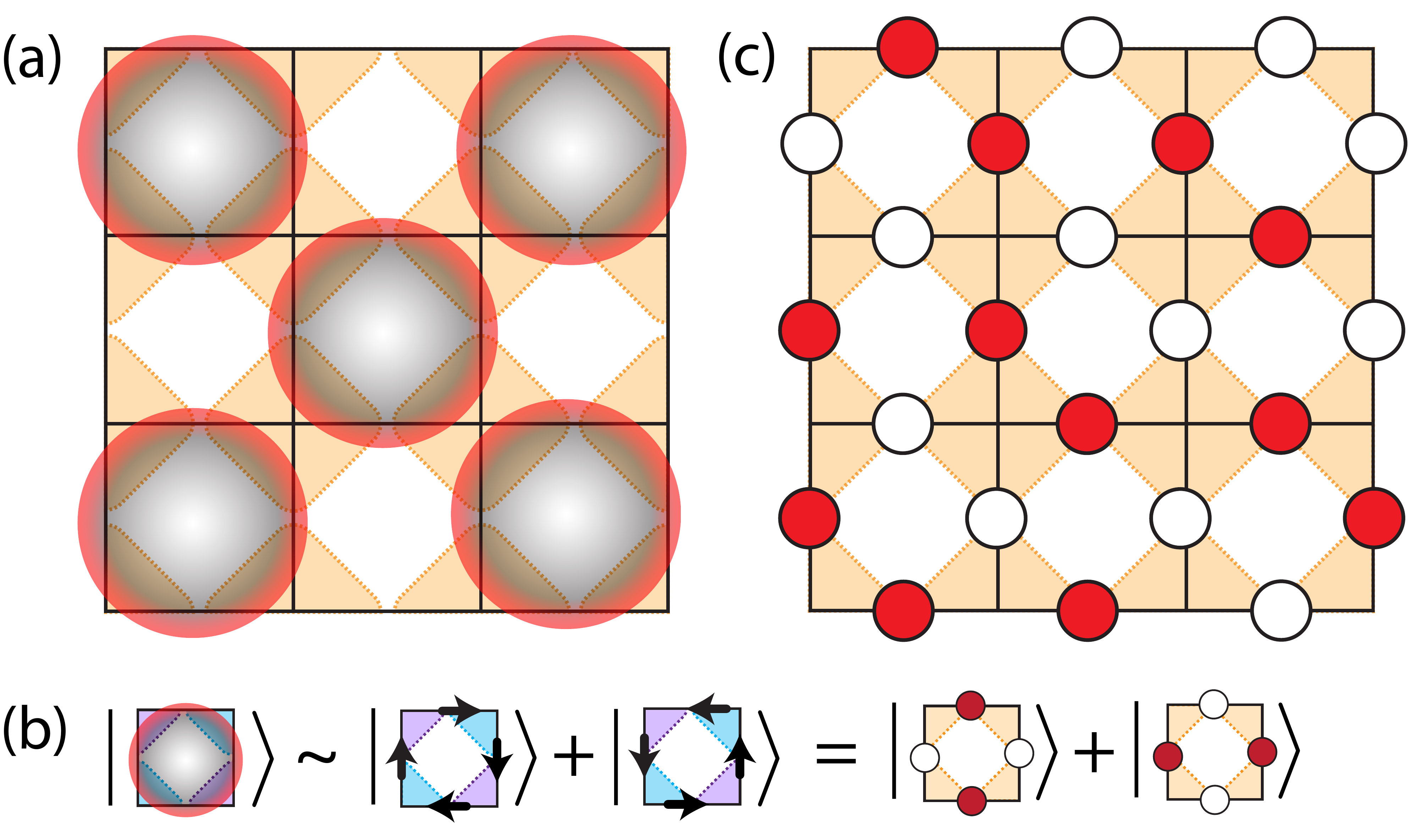}
\caption{\small{(a) Cartoon state of a plaquette RVB solid. An alternating pattern of plaquettes (shaded red circles) are resonating, i.e., they are in an eigenstate $|\varocircle\rangle$ [see panel (b)] of the plaquette Hamiltonian of Eq.~\eqref{eq:plaq},  $H_\Box|\varocircle\rangle=-t|\varocircle\rangle$. The GS in the thermodynamic limit is two-fold degenerate, reflecting the different coverings of the square lattice with alternating plaquettes. (c) cartoon of one of the degenerate ground states with $(-\pi/2, \pi/2 )$ order. Along the bottom-left / top-right diagonal, there is antiferromagnetic order. Along the other, the order has a double period, $\uparrow\uparrow\downarrow\downarrow$.}}
\label{fig:rvb} 
\end{figure}

The smallest cluster which may flip consists of a closed flux loop around a plaquette, denoted by $\Box$ (see Fig.~\ref{fig:rvb}), 
\begin{equation}
H_\Box=-t\sum_{i,j,k,l\,\in\,\Box}\left(S_i^+S_j^-S_k^+S_l^-+{\rm h.c.}\right)
\label{eq:plaq}
\end{equation}
and this will be the second ingredient that our work will implement (Sec.~\ref{sec:amo}). What this amounts to in the language of gauge theory is the addition of a field conjugate to the height/gauge field ${\bf a}=h{\bf z}$ -- or in more familiar parlance of electromagnetism, the appearance of an (emergent) magnetic field alongside an (emergent) electric one~\cite{eduardo1997field}.

In two dimensions, adapting a celebrated result by Polyakov (which does not apply straightforwardly as it is based on Lorentz invariance which does not hold {\it a priori} for our emergent field), it is known that the (emergent) electromagnetism is confining. As a consequence, the emergent excitations cannot spread freely over the system, being bounded by an effective string tension due to the gauge fields.
Concretely, one finds a phenomenon known as order by disorder~\cite{R7}. The quantum dynamics mixes the degenerate ice states into a superposition to form the quantum ground state. Even though the quantum dynamics induces fluctuations (`disorder'), the resulting ground state exhibits long-range order. This order takes the form known as a plaquette valence bond solid (Fig.~\ref{fig:rvb})~\cite{R15}, which breaks translational symmetry. Such valence bond solids occur frequently in the theory of quantum magnets, but they are not commonly realized in experiment. 

In Sec.~\ref{sec:mb}, we show that the model Hamiltonian we provide a recipe for does exhibit this kind of order-by-disorder plaquette phase, and we discuss how to detect this kind of exotic order. In addition, we find that for weak quantum dynamics, a different, classical type of symmetry-breaking occurs (see in Fig.~\ref{fig:rvb}(c)). This happens because different ice states are only approximately degenerate for our engineered Hamiltonian, and the residual energy differences are sufficient to select a particular ordered configuration.

\subsection{Relation between quantum ice, Bose-Hubbard models and dimer models}

As a starting point for our implementation, we will consider a hard-core extended Bose-Hubbard Hamiltonian on a 2D checkerboard lattice:
\begin{equation}\label{eq:BH}
H=-J_h\sum_{\langle i,j\rangle}\left(b_i^\dag b_j+{\rm h.c.}\right)+\sum_{i,j}\tilde V_{ij} n_i n_j.
\end{equation}
Here, $b_i^\dag$ ($b_i$) is an operator that creates (annihilates) a hard-core boson on site $i$ which obey an on-site contained $b_i^2=b_i^{\dag 2}=0$.  The rate $J_h$ is the nearest neighbor (NN) hopping amplitude and $V$ describes a repulsion between all atoms sitting close to the same vertex. The operator  $n_i=b_i^\dag b_i$ counts the number of bosons at site $i$ and can be either zero or one, $n_i\in\{0,1\}$. The summation runs over nearest neighbors only. The hard-core boson model can be mapped to a spin-1/2 model using the transformation ~\cite{sachdev2001quantum} $b_i^\dag\rightarrow S_i^+$, $b_i\rightarrow S_i^-$, $n_i\rightarrow S_i^z+1/2$, $J_h\rightarrow J_\perp$ and $V_{ij}\rightarrow J_{ij}$, 
which yields
\begin{equation}\label{eq:XXZ}
H=-J_\perp\sum_{\langle ij\rangle}\left(S_i^+ S_j^-+{\rm h.c.}\right)+\sum_{\langle ij\rangle}J_{ij} \left(S_i^z+\frac{1}{2}\right) \left(S_j^z+\frac{1}{2}\right).
\end{equation}
Expanding the last term gives the two-body interaction proportional to $J_{ij} S_i^z S_j^z$ and an additional magnetic field term proportional to $J_{ij} S_i^z$ which is constant after fixing an initial number of particles. This will fix the gauge sector in the gauge theory description~\cite{R4}. 

In order to implement the constrained model of Eq.~\eqref{eq:spinice} we demand (i) {\it anisotropic} and (ii) {\it step-like} interactions between (iii) {\it two species} of particles, as discussed in Sec~\ref{sec:2d}, that is $\tilde V_{ij}$ has to fulfill
\begin{equation}
\sum_{ ij}\tilde V_{ij} n_i n_j=\tilde V_0\sum_{+}\sum_{i,j\in +} n_i n_j,
\label{eq:constraint}
\end{equation}
with $\tilde V_0$ a constant interaction between all particles belonging to the same vertex denoted by $+$. Under this assumptions, in the limit $\tilde{V}_0 \gg J_h$ the Bose-Hubbard Hamiltonian of Eq.~\eqref{eq:BH} maps onto the spin ice Hamiltonian of Eq.~\eqref{eq:spinice}. The specific form of $V_{ij}$ ensures that all six interactions between particles which belong to the same vertex are equal and interactions between particles which do not belong to the same vertex vanish. In the case of total half-filling of the initial bosons, $N = L /2$, one has $
\sum_i S^z_i =0 $: this fixes the effective dynamics on the aforementioned ice manifold of interest. In the case different initial fillings are considered, one has access to different quantum dynamics: a notable case is the $N = L/4 $ case, which defines a constrained dynamics on a manifold where a single boson sits close to each vertex~\cite{R1a}. The effective description is then the same of hard-code quantum dimer models on a square lattice.


\section{Quantum ice with Rydberg-dressed atoms: exploiting p-states\label{sec:amo}}

We now turn to the realization of the extended 2D Bose Hubbard Hamiltonian of Eq.~\eqref{eq:BH} with cold  atoms in optical lattices. The key challenge is the implementation of the interactions $\tilde V_{ij}$ with constraints represented in Eq.~(\ref{eq:constraint}). We will show below that this can be achieved via the very anisotropic Rydberg interactions involving laser excited $p$-states of Rubidium atoms.

\subsection{Single-particle Hamiltonian on a  bi-partite lattice}\label{sec:single}
\begin{figure*}[tb]  
\centering 
\includegraphics[width= 0.98\textwidth]{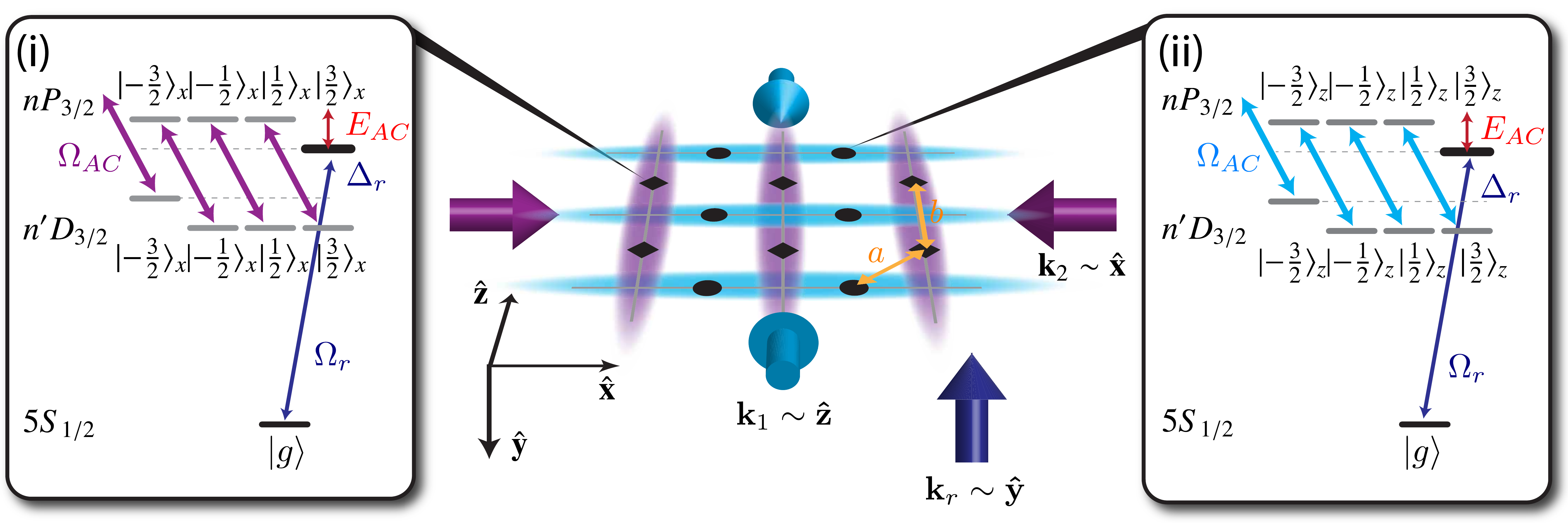}
\caption{\small{We consider $^{87}$Rb atoms loaded in a square optical lattice with lattice spacing $a$ and lattice sites labeled alternating as $\medbullet$ and $\blacksquare$. Additional AC Stark lasers (magenta and light blue arrows) with wave vectors $\mathbf{k}_1=\pm2\pi/\lambda_{AC}\,\mathbf{\hat z}$ and $\mathbf{k}_2=\pm2\pi/\lambda_{AC}\,\mathbf{\hat x}$, respectively, form two pairs of standing waves each with periodicity $b=\lambda_{AC}/2$, which are rotated by 45 degrees with respect to the initial lattice. In order to create local quantization axes along $\mathbf{\hat z}$ or $\mathbf{\hat x}$ for $\medbullet$ or $\blacksquare$ lattice sites, respectively, we require that that atoms located a $\medbullet$ lattice site only feel the intensity maxima of the light blue laser with $\mathbf{k}_1\sim\mathbf{z}$, while $\blacksquare$ lattice site only feel the intensity maxima of the magenta laser with $\mathbf{k}_2\sim \mathbf{\hat x}$.
This can be achieved by adjusting the initial trapping lattice by tilting the corresponding trapping lasers with an angle $\alpha$ such that $a=\lambda/[2\sin(\alpha/2)] \ge \lambda/2$~\cite{Viteau:2011jz} in order to fulfill $b=\sqrt{2}a$. The two AC Stark lasers have a polarization $\sigma_+$ and resonantly couple the $n{}^2P_{3/2}$ manifold to a lower lying $n'D_{3/2}$ manifold thereby inducing an AC Stark shift on each Zeeman $m$-level in the $n{}^2P_{3/2}$ manifold except for the maximum stretched $|n{}^2P_{3/2},3/2\rangle_{z,x}$ states. This locally isolates the $|n{}^2P_{3/2},3/2\rangle_{z}$ and $|n{}^2P_{3/2},3/2\rangle_{x}$ state at lattice sites $\medbullet$ and $\blacksquare$, respectively, in energy by at least $E_{AC}$ (see left and right panels). A global Rydberg laser (dark blue arrow) with detuning $\Delta_r\ll E_{AC}$ propagating along the $y$-direction then selectively admixes the states $|n{}^2P_{3/2},3/2\rangle_z$ and $|n{}^2P_{3/2},3/2\rangle_x$ at lattice sites $\medbullet$ and $\blacksquare$, respectively, to the ground state $|g\rangle$}.}
\label{fig:tensor} 
\end{figure*}

In our setup we consider Rubidium $^{87}$Rb atoms prepared in an internal ground state, which we choose as \mbox{$|g\rangle\equiv|F=2$, $m_F=2\rangle$}. The atoms are trapped in a 2D square optical lattice in the $xz$-plane created by two pairs of counter propagating laser beams of wavelength $\lambda$ and wave vector $k=2\pi/\lambda$, and strongly confined in the $y$-direction by an additional laser. Note that by tilting the laser beams by an angle $\alpha$ we can adjust the lattice spacing in the $xz$-plane to any value $a=\lambda/[2\sin(\alpha/2)] \ge \lambda/2$~\cite{Viteau:2011jz} (see below).  Quantum tunneling allows the atoms to hop between different lattice sites, thus realizing the kinetic energy term with hopping amplitude $J_{h}$ of the single band Hubbard model \eqref{eq:BH}. Furthermore, we work in the hardcore boson limit, i.e.  $U\gg J_h$, in which multiple occupancy in a single site is energetically prohibited.

As already discussed in the context of Fig.~\ref{fig:pyro}(b) we want to distinguish between $\medbullet$ and $\blacksquare$  sites in the 2D lattice. This bipartite labeling of the optical lattice is essential for realizing the complex interaction pattern $\tilde V_{\medbullet\medbullet}$, $\tilde V_{\medbullet\blacksquare}$ and $\tilde V_{\blacksquare \blacksquare}$ discussed in Sec.~\ref{sec:2d},  which underlies the second term of the extended Bose Hubbard Hamiltonian \eqref{eq:BH}. In our scheme,  we assume that atoms on lattice sites  $\medbullet$ are excited by laser light to the Rydberg ${}^{2}$P${}_{3/2}$-state \mbox{$|r_\medbullet\rangle=|n{}^2P_{3/2},m=3/2\rangle_z$}, whereas atoms at sites $\blacksquare$ are excited to \mbox{$|r_\blacksquare\rangle=|n{}^2P_{3/2},m=3/2\rangle_x$}. Here the subscripts $x$ and $z$ indicate the {\em different local quantization axes} for the $\medbullet$ and $\blacksquare$  sites. Note that the Rydberg states of interest are the {\em stretched states} of the fine structure manifold, i.e. states with maximum $m=3/2$ value for the given angular momentum $j=3/2$. We will show in the next section that the van der Waals interactions between these polarized Rydberg $p$-states realize naturally the complex interaction pattern required by Eq.~\eqref{eq:constraint} for  quantum spin ice. By weakly admixing these Rydberg states to the atomic ground state with a laser [see Sec.~\ref{sec:softcore}], the ground state atoms will inherit these interaction patterns, thus realizing the interaction term in the extended Bose Hubbard Model of Eq.~\eqref{eq:BH}, including the constraints enforced by the interactions satisfying Eq.~\eqref{eq:constraint}. 

It is essential in our scheme that we energetically isolate the stretched states $|n{}^2P_{3/2},m=3/2\rangle_{x,z}$ from the other $m$-states in the given fine structure manifold. This is necessary to protect these states from mixing with other Zeeman $m$-levels. Such unwanted couplings can be induced by van der Waals interactions (see Sec.~\ref{sec:pstates} below), or via the light polarization of the Rydberg laser. This energetic protection requires an (effective)  {\em local magnetic field}, which for the $\medbullet$ and $\blacksquare$  sites points in the $x$ and $z$-direction, respectively. Strong local fields with spatial resolution on the scale given by the lattice spacing can be obtained via AC Stark shifts, combining the $m$-dependence of atomic AC tensor polarizabilities with spatially varying polarization gradients. Fig.~\ref{fig:tensor} outlines a scheme, where we superimpose two pairs of counter propagating laser beams of wavelength $\lambda_{AC}$ (light blue and magenta arrows). They create a standing wave pattern (light blue and magenta gradients), such that $\medbullet$ lattice sites only see the intensity maxima of the standing wave propagating along the $z$-direction (light blue laser), while $\blacksquare$ lattice sites see the intensity maxima of the standing wave propagating along the $x$-direction (magenta laser).\\
\\
\begin{figure*}[tb]  
\centering 
\includegraphics[width= 0.99\textwidth]{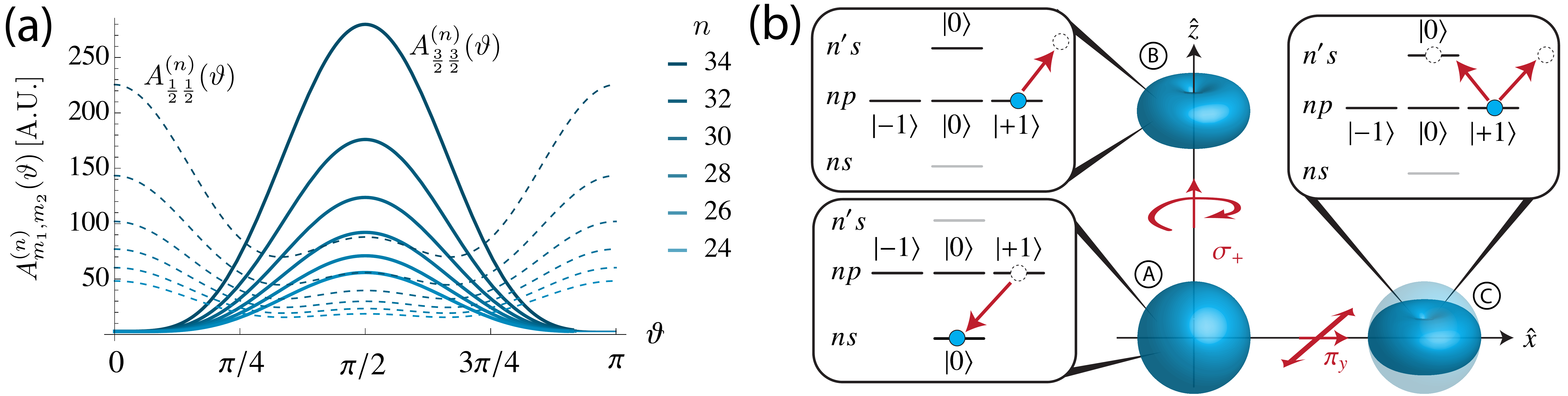}
\caption{\small{(a) Angular part, $A^{(n)}_{m_1,m_2}(\vartheta)$, of the van der Waals interaction, $V^{(n)}_{m_1,m_2}(r,\vartheta)=\langle m_1,m_2|\hat V_{\rm vdW}| m_1,m_2\rangle=(n-\delta_{n\ell j})^{11} A^{(n)}_{m_1,m_2}(\vartheta)/r^6$, between a pair of $^{87}$Rb atoms in the $|3/2,3/2\rangle_z\equiv|n{}^2P_{3/2},3/2\rangle_z\otimes|n{}^2P_{3/2},3/2\rangle_z$ state (solid lines) and in the  $|1/2,1/2\rangle_z\equiv|n{}^2P_{3/2},1/2\rangle_z\otimes|n{}^2P_{3/2},1/2\rangle_z$ state (dashed lines). We plot the angular part of the rescaled interaction energy, $A^{(n)}_{m_1,m_2}$ as a function of the angle $\vartheta$ for various values of the principal quantum number $n$ in atomic units. Here, $A^{(n)}_{3/2,3/2}$ (solid lines) corresponds to the angular part of the interaction, $V_{\medbullet\medbullet}$, between two atoms excited to the $|r_\medbullet\rangle$ Rydberg state and shows a characteristic $\sim\sin^4\vartheta$ shape due to the dominant $S_{1/2}$-channel. Residual interactions at $\vartheta=0$ are very small and arise from channels coupling to virtual $D$-states.  (see Tab.~\ref{tab:p32}). (b) The angular characteristic of the interaction between two atoms both in the ``stretched'' Rydberg state, $|r_\medbullet\rangle=|n{}^2P_{3/2},3/2\rangle_z=|np,1\rangle_z|\frac12\frac12\rangle_z$, can be qualitatively understood from its angular part $|np,1\rangle_z$ and the dominating $S$-channel: atom A prepared in a $|np,\textstyle{1}\rangle_z$ state can make a virtual transition to a lower-lying $|ns,0\rangle_z$ state (red arrow in the lower left panel) while emitting a photon. If this photon propagates along the $z$-direction it has polarization $\sigma_+$ and cannot be absorbed by atom B. Therefore, atom A and atom B will not interact, i.e. $V_{\medbullet\medbullet}(\vartheta=0)=0$. If this photon propagates alone the $x$-direction it is linear polarized with a polarization vector along the $y$-direction. In the frame of atom C this photon will drive both $\sigma_+$- and $\sigma_-$-transitions and thus can be absorbed. Hence, atom A can interact with atom C, i.e.  $V_{\medbullet\medbullet}(\vartheta=\pi/2)\neq 0$. }}
\label{fig:pstateint} 
\end{figure*}

The AC Stark lasers have polarization $\sigma_+$ and resonantly couple the $nP_{3/2}$ manifold to a lower lying $n'D_{3/2}$ manifold (see magenta and blue arrows in the left and right panels of Fig.~\ref{fig:tensor}, respectively). This induces an AC-Stark shift on each Zeeman $m$-level in the $n{}^2P_{3/2}$ manifold. The Rabi frequency is proportional to \mbox{$\Omega_{AC}\sim\sqrt{(m-\frac{3}{2})(m+\frac{5}{2})}\langle nP_{3/2}||r||n'D_{3/2}\rangle\mathcal{E}$} with $\mathcal{E}$ the electric field strength of the AC Stark lasers. In this configuration the stretched states $|n{}^2P_{3/2},m=3/2\rangle_{z,x}$ of interest are not affected by the AC Stark lasers. The minimum shift (as a function of $m\neq 3/2$) is denoted $E_{\rm AC}$, which has to obey $E_{\rm AC}\gg V_{\rm off}$ and $E_{\rm AC}\gg \Delta_r$ in order to suppress mixing between different $m$-states due to van der Waals interactions and the excitation laser. Here, $V_{\rm off}$ is the largest off-diagonal van der Waals matrix element in the $n{}^2P_{3/2}$ manifold (see App.~\ref{app:vdw}).

The AC Stark lasers will create an additional trapping potential, $V_{AC}(\mathbf{r}_i)|g\rangle\langle g|_i$, for ground state atoms with minima not commensurate with the initial trapping lattice. In order to not distort the desired lattice structure this additional potential must not be larger than the initial lattice potential, see App.~\ref{app:aclattice}.

It is then possible to dress the ground state atoms with either the $|r_\medbullet\rangle=|n{}^2P_{3/2},m=3/2\rangle_z$ or the $|r_\blacksquare\rangle=|n{}^2P_{3/2},m=3/2\rangle_x$ Rydberg state by a  single, global laser with Rabi frequency $\Omega_r$ and detuning $\Delta_r$ propagating in the direction perpendicular to the plane, i.e. $\mathbf{k}_r \sim \mathbf{y}$ (dark blue arrow in Fig.~\ref{fig:tensor}). In the local $x$- and $z$-basis this laser will couple to all four $|m\rangle_{z,x}$ levels with different weights (see App.~\ref{app:laser}). Since the states $|m\neq 3/2\rangle_{z,x}$ are energetically separated by at least $E_{\rm AC}$ from the $|m=3/2\rangle$ state a laser with detuning $\Delta_r\ll E_{\rm AC}$ and wave vector $\mathbf{k}\sim \mathbf{y}$ will selectively admix the states $|3/2\rangle_z$ and $|3/2\rangle_x$ at lattice sites $\medbullet$ and $\blacksquare$, respectively, to the ground state $|g\rangle$ with an effective Rabi frequency $\Omega_r'=\Omega_r/(2\sqrt{2})$. The single particle Hamiltonian describing the laser dressing in a frame rotating with the laser frequency for an atom $i$ then becomes
\begin{equation}
\label{eq:Hsingle}
H_i=-\Delta_r|r_{\alpha_i}\rangle\langle r_{\alpha_i}|_i+\frac{1}{2}\Omega_r'\left(|g\rangle\langle r_{\alpha_i}|_i+|r_{\alpha_i}\rangle\langle g|_i\right),
\end{equation}
where $\alpha_i\in\{\medbullet,\blacksquare\}$ depends on the lattice site of the $i$-th atom. In the weakly-dressing regime, $\Delta_r\gg\Omega_r$, the new dressed ground states are $|\medbullet\rangle_i\equiv|g\rangle_i+\Omega_r/(2\Delta_r)|r_\medbullet\rangle_i$ or $|\blacksquare\rangle_i\equiv|g\rangle_i+\Omega_r/(2\Delta_r)|r_\blacksquare\rangle_i$ if $\alpha_i=\medbullet$ or $\blacksquare$, respectively. Thus, each ground state atom gets a small admixture of one of the Rydberg states, depending on the sublattice. 
 Due to the weak admixture of the Rydberg states, the dressed ground states get a comparatively small decay rate $\tilde \Gamma=(\Omega_r/2\Delta_r)^2\Gamma$, where $\Gamma$ is the decay rate of the bare Rydberg state, which has to be much smaller than the relevant system energy scales discussed below.  

\subsection{Interactions between $p$-states}\label{sec:pstates}

Below we will consider the van der Waals interactions,  $V_{\medbullet\medbullet}$, $V_{\medbullet\blacksquare}$ and $V_{\blacksquare \blacksquare}$, between pairs of atoms prepared in the bare Rydberg states  \mbox{$|r_\medbullet\rangle=|n{}^2P_{3/2},m=3/2\rangle_z$} and \mbox{$|r_\blacksquare\rangle=|n{}^2P_{3/2},m=3/2\rangle_x$}. For Rubidium atoms excited to Rydberg $p$-states, these van der Waals forces are strongly anisotropic~\cite{Walker:2008bm,Walker:2005kj,Reinhard:2007hm,Singer:2005ct}. Fig.~\ref{fig:pstateint}(a) shows the angular part of the van der Waals interaction, $V_{\medbullet\medbullet}$, for different $n$-values, which is in very good approximation proportional to 
\begin{equation}
V_{\medbullet\medbullet}(r,\vartheta)\sim \frac{(e a_0)^4n^{11}}{r^6}\sin^4\vartheta,
\end{equation}
while the actual strength depends on the principal quantum number $n$ and scales as $n^{11}$ away from the F\"orster resonance at $n=38$. Similarly, one finds for the interaction between \mbox{$|r_\blacksquare\rangle=|n{}^2P_{3/2},m=3/2\rangle_x$} Rydberg states
\begin{equation}
V_{\blacksquare\blacksquare}(r,\vartheta)\sim\frac{(e a_0)^4n^{11}}{r^6}\cos^4\vartheta,
\end{equation}
which can be obtained by rotating the coordinate system by $\pi/2$.
Mixed interactions such as
\begin{equation}
V_{\medbullet\blacksquare}(r,\vartheta)\sim \frac{(e a_0)^4n^{11}}{r^6} \left(3 \sin 2 \vartheta +2\right)^2,
\end{equation}
are shown in Fig.~\ref{fig:int45} (App.~\ref{app:mixed}) and have two asymmetric maxima at \mbox{$\vartheta=\pm \pi/4$}. The Rydberg states  \mbox{$|r_\medbullet\rangle$} and \mbox{$|r_\blacksquare\rangle$} therefore realize the desired angular interaction properties as discussed in Sec.~\ref{sec:2d}. Together with the possibility of creating soft-core potentials (see the following subsection), the anisotropy of these interactions leads naturally to the desired interaction pattern illustrated in Fig.~\ref{fig:pyro}(b) and demanded by Eq.~\eqref{eq:constraint}. These interactions underly our realization of the Bose Hubbard Hamiltonian \eqref{eq:BH}. 

\begin{table*}[tb]
\begin{center}
\begin{tabular}{lcccccc}
\toprule
Channel &  \multicolumn{5}{c}{$C_6^{(\nu)}$ [A.U.]}  & $\langle\frac32\frac32|\mathcal{D}_\nu(\vartheta)|\frac32\frac32\rangle$ \\
$\nu$ & $n=26$ & $n=28$ & $n=30$ & $n=32$ & $n=34$    \\
\colrule
$S_{1/2}+S_{1/2}$ & $ 1.58\times 10^{17} $ & $ 5.07\times 10^{17} $ & $ 1.60\times 10^{18} $& $ 4.88\times 10^{18} $& $ 1.61\times 10^{19} $ & $\sin^4 \vartheta/4$\\	
$S_{1/2}+D_{3/2}$ & $ 6.38\times 10^{15} $ & $ 1.60\times 10^{16} $ & $ 3.72\times 10^{16} $& $ 8.15\times 10^{16} $& $ 1.70\times 10^{17} $ &$(2+\cos 2\vartheta)\sin^2\vartheta/50$\\	
$S_{1/2}+D_{5/2}$ & $ 6.46\times 10^{15} $ & $ 1.62\times 10^{16} $ & $ 3.76\times 10^{16} $& $ 8.26\times 10^{16} $& $ 1.72\times 10^{17} $ &$(209+84 \cos 2\vartheta+27\cos 4\vartheta)/2400$\\			
$D_{3/2}+D_{3/2}$ & $-1.17\times 10^{15} $ & $-2.71\times 10^{15} $ & $-5.85\times 10^{15} $& $-1.20\times 10^{16} $& $-2.34\times 10^{16} $ &$(5+2\cos 2\vartheta+\cos 4\vartheta)/1250$\\
$D_{3/2}+D_{5/2}$ & $-1.06\times 10^{15} $ & $-2.43\times 10^{15} $ & $-5.20\times 10^{15} $& $-1.06\times 10^{16} $& $-2.05\times 10^{16} $ &$(358+186\cos 2\vartheta-27\cos 4\vartheta)/15000$\\
$D_{5/2}+D_{5/2}$ & $-9.50\times 10^{14} $ & $-2.15\times 10^{15} $ & $-4.56\times 10^{15} $& $-9.16\times 10^{15} $& $-1.76\times 10^{16} $ &$3 (1745-876 \cos 2 \vartheta +27 \cos 4 \vartheta )/20000 $\\		
\botrule
\end{tabular}
\caption{Two atoms both in the $|r_\medbullet\rangle=|n{}^2P_{3/2},3/2\rangle_z=|np,1\rangle_z|\frac12\frac12\rangle_z$ state can couple to six channels. Each channel $\nu$ has a characteristic angular dependency $\left(\mathcal{D}_\nu\right)_{\frac32\frac32}$ which contributes with weight $C_6^{(\nu)}$. The total interaction can be obtained by summing over all channels, i.e. $V_{\medbullet\medbullet}(r,\vartheta)=\sum_\nu C_6^{(\nu)}\langle\frac32\frac32|\mathcal{D}_\nu(\vartheta)|\frac32\frac32\rangle/r^6$. It turns out that two atoms in the $|r_\medbullet\rangle$ Rydberg state dominantly couple to the $S_{1/2}+S_{1/2}$ channel with a characteristic angular dependence $\sim\sin^4\vartheta$.}
\label{tab:p32}
\end{center}
\end{table*}

We now detail the physical mechanism which generates these anisotropic interactions, and describe how to derive the aforementioned results. Van der Waals interactions between two atoms $i$ and $j$ prepared in a given Rydberg state arise from the exchange of virtual photons: atom $i$ in a Rydberg state $|r_i\rangle$ can for example virtually undergo a dipole allowed transition to a lower-lying electronic state $|\alpha\rangle$ while emitting a photon. If this virtual photon reaches atom $j$ during its lifetime, it can excite the second atom to an electronic state $|\beta\rangle$. This then leads to correlated oscillations of instantaneously induced dipoles in both atoms which give rise to the non-retarded van der Waals force~\cite{milonni1994quantum}. For the familiar case of $s$-states these interactions are isotropic, $V_{\rm VdW}(r)=C_{6}/r^{6}$ with the van der Waals coefficient $C_{6}$ scaling as  $C_{6}\sim n^{11}$~\cite{Walker:2008bm,Walker:2005kj,Reinhard:2007hm,Singer:2005ct}. Here, $n$ is the principal quantum number and $r$ the distance between atoms. These van der Waals interactions between Rydberg states exceed ground state interactions by several orders of magnitude and have been observed and explored in recent experiments ~\cite{Heidemann:2008bg,Anderson:1998hn, Hofmann:2013gm, Maxwell:2013km,Schwarzkopf:2013fx,Ebert:2013tr,Valado:2013ez,Schempp:2014jm,Barredo:2014tb,Tauschinsky:2010ep, Carter:2013vi,Kubler:2010dw,Dudin:2012hm, Schauss:2012hh,Li:2013gg}.

In the case of Rydberg $p$-states, the angular distribution of these emission and absorption processes of virtual photons in combination with the angular momentum structure of the atomic orbitals leads to nontrivial anisotropic van der Waals interactions. We now focus on the van der Waals interaction $V_{\medbullet\medbullet}$, between both atoms in the \mbox{$|r_\medbullet\rangle=|n{}^2P_{3/2},m=3/2\rangle_z$} Rydberg state with quantization axis along the $z$-direction. Mixed interactions, $V_{\medbullet\blacksquare}$, and interactions between both atoms in the \mbox{$|r_\blacksquare\rangle=|n{}^2P_{3/2},m=3/2\rangle_x$} Rydberg state, $V_{\blacksquare \blacksquare}$, will be derived in App.~\ref{app:vdw}. The latter can simply be obtained by rotating the $xz$-plane by 90 degrees, i.e. $V_{\blacksquare \blacksquare}(r,\vartheta)=V_{\medbullet \medbullet}(r,\vartheta-\pi/2)$, while in order to calculate mixed interactions one has to calculate off-diagonal matrix elements in the $n{}^2P_{3/2}$ manifold.

In general, van der Waals interactions arise as a second order process from dipole-dipole interaction,  $\hat V_{\rm dd}^{(ij)}(\mathbf{r})=\left(\mathbf{d}^{(i)}\cdot \mathbf{d}^{(j)}-3(\mathbf{d}^{(i)}\cdot\mathbf{n})(\mathbf{d}^{(j)}\cdot\mathbf{n})\right)/r^3$, where $V_{\rm dd}$ couples the initial Rydberg states $|r_i,r_j\rangle$ to virtual intermediate states $|\alpha,\beta\rangle$ and back. Here, $\mathbf{d}^{(i)}$ is the dipole operator of the  $i$-th atom and $\mathbf{r}=r \mathbf{n}=(r,\vartheta,\varphi)$ is the relative distance between atom $i$ and atom $j$ with $\mathbf{n}$ a unit vector and $(r,\vartheta,\varphi)$ the spherical coordinates.  It is convenient to rewrite the latter expression in a spherical basis ~\cite{schwinger1998classical}
\begin{equation}
\hat V_{\rm dd}^{(ij)}(\mathbf{r})=-\sqrt{\frac{24 \pi}{5}}  \frac{1}{r^{3}} \sum_{\mu,\nu}C_{\mu,\nu;\mu+\nu}^{1,1;2}
 Y_{2}^{\mu+\nu}(\vartheta,\varphi)^* d^{(i)}_{\mu}d^{(j)}_{\nu},
 \label{eq:dipdip}
\end{equation}
with $d^{(i)}_\mu$ the spherical components ($\mu,\nu\in \{-1,0,1\}$) of $\mathbf{d}^{(i)}$, $C_{m_1,m_2;M}^{j_1,j_2;J}
$ the Clebsch-Gordan coefficients, and $Y_l^m$ the spherical harmonics. 

Due to the dipole selection rules, states in the $n{}^2P_{3/2}$ manifold can only couple to states in a $n'S_{1/2}$, $n'D_{5/2}$ or $n'D_{3/2}$ manifold. It turns out that for $^{87}$Rb the dominating channel is $P_{3/2}+P_{3/2}\longrightarrow S_{1/2}+S_{1/2}$, which can be explicitly seen from Table~\ref{tab:p32} for various $n$ levels. In order to simplify the following discussion we will first focus on this channel and neglect all other channels including $D_{3/2}$ and $D_{5/2}$ states which lead to small imperfections discussed in App.~\ref{app:vdw}. 

For a single atom, the $|n{}^2P_{3/2},\frac{3}{2}\rangle$ state is a {\it stretched-state} which reads in the uncoupled basis $|n{}^2P_{3/2},\frac{3}{2}\rangle\equiv|np,1\rangle\otimes|\frac{1}{2},\frac{1}{2}\rangle$. Thus, it can be factorized into an angular and a spin degree of freedom. Since the dipole-dipole interaction, $\hat V_{\rm dd}(\mathbf{r})$, does not couple spin degrees of freedom, the angular dependence of the van der Waals interaction is determined solely by the angular part of the wave function, which is $|np,1\rangle$.

\begin{figure*}[htb]  
\centering 
\includegraphics[width= \textwidth]{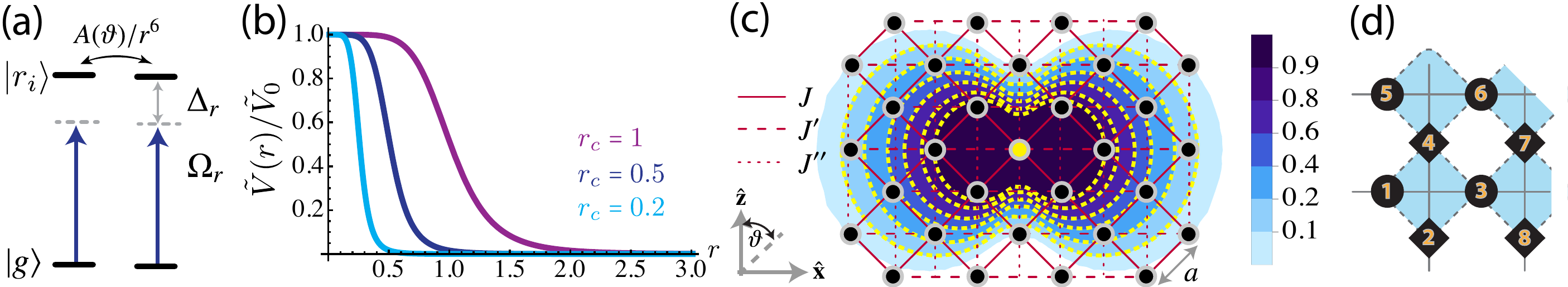}
\caption{\small{(a) Qualitative sketch of the energy levels (black lines) and lasers (thick solid dark-blue arrows) required for the Rydberg dressing scheme. The ground state $|g\rangle$ of each atom is off-resonantly coupled to a Rydberg state $|r_i\rangle$ with a c.w. laser of Rabi frequency $\Omega_r$ and detuning $\Delta_r$ (see also Fig.~\ref{fig:tensor}). Pairwise interactions between the energetically well-isolated Rydberg states can be anisotropic, i.e. $V_{ij}(\mathbf{r})=A(\vartheta)/r^6$. (b) Energy eigenvalues $\tilde V(\mathbf{r})$ of Eq.~\eqref{eq:dressedpot1} (dressed Born-Oppenheimer potential surfaces) of Rydberg-dressed ground state atoms for different values of the Condon radius $r_c$ defined in Eq.~\eqref{eq:rc}. The potential has a step-like shape and saturates for small distances at $\tilde V_0$, while the onset of the steep slope is given by $r_c$. (c) Contour plot of the dressed ground state energy $\tilde V_{ij}(\mathbf{r})/\tilde V_0$ between the  atom in the middle (yellow circle) and the surrounding atoms (black circles) all in the $|r_\medbullet\rangle$ Rydberg state. In this case $A(\vartheta)\sim\sin^4\vartheta$ which gives rise to a figure-eight shaped interaction plateau (yellow dashed lines). Residual interactions along $\vartheta=0$ come from virtual transitions to $D$-states (see Sec.~\ref{sec:pstates}). Interactions along the $z$- (red dotted line), $x$- (red dashed line) and along the 45 degree lines (red solid lines) can be adjusted by varying the principal quantum number $n$ and the detuning $\Delta_r$. (d) Labeling of the lattice sites for the example of Sec.~\ref{sec:numbers}.  }}
\label{fig:softcore} 
\end{figure*}

Figure~\ref{fig:pstateint}(b)  illustrates the interaction between two atom initially prepared in a $|np,1\rangle$ state as a function of $\vartheta$ for $\vartheta=0$ (atom A and B) and $\vartheta=\pi/2$ (atom A and C). We first consider atom A in the lower left corner of Fig.~\ref{fig:pstateint}(b). Initially prepared in a $|np,\textstyle{1}\rangle$ state it can make a virtual transition to a lower-lying $|ns,0\rangle$ state [red arrow in the lower left panel of Fig.~\ref{fig:pstateint}(b)] while emitting a photon. The corresponding angular distribution of the spontaneously emitted photon has the same characteristic as  light emitted by a classical dipole tracing out a circular trajectory in the $x$-$y$ plane~\cite{schwinger1998classical}. In general, it is elliptically polarized with cylindrical symmetry, but in particular there are two specific directions: 

(i) light emitted along the $z$ direction ($\vartheta=0$) is circularly polarized, rotating in the same way as the dipole. Thus, a photon emitted in the $z$ direction has polarization $\sigma_+$ and carries one unit of angular momentum such that the total angular momentum of the combined system atom-photon is conserved. A second atom, labeled as atom B in Fig.~\ref{fig:pstateint}(b), located on the $z$ axis ($\vartheta=0$) cannot absorb this photon [see red arrow in the upper left panel of Fig.~\ref{fig:pstateint}(b)], since only a $|n's,0\rangle$ state is available. The same result can be derived from Eq.~\eqref{eq:dipdip}, which for $\vartheta=0$ simplifies to
\begin{equation}
\hat V_{\rm dd}^{(ij)}(\vartheta=0)=-\frac{2}{r^3}\sum_\mu \frac{d^{(i)}_{\mu}d^{(j)}_{-\mu}}{(1-\mu)!(1+\mu)!},
\end{equation}
and couples only states with initial magnetic quantum number $m_1$, $m_2$ to states with $m_1\pm 1$, $m_2\mp 1$, such that the total angular momentum $M=m_1+m_2$ is conserved. Therefore, the dipole-dipole matrix element vanishes, $\langle np1,np1 |\hat V_{\rm dd}^{(AB)}(\vartheta=0)| ns0,n's0\rangle=0$, and hence atoms $A$ and $B$ do not interact. 

(ii) light emitted into the $x$-$y$ plane ($\vartheta=\pi/2$) is linearly polarized, with a polarization vector lying in the $x$-$y$ plane and perpendicular to the emission direction. A third atom, labeled as atom C  in Fig.~\ref{fig:pstateint}(b), located on the $x$ axis is able to absorb this linearly polarized photon emitted by atom A, which in the frame of atom C corresponds to a superposition of  $\sigma_+$ and $\sigma_-$ polarized light [see red arrows in the right panel of Fig.~\ref{fig:pstateint}(b)]. For $\vartheta=\pi/2$, Eq.~\eqref{eq:dipdip} contains a sum over $\cos[\frac{\pi}{2}(\mu+\nu)]$, and the only non-vanishing combinations for $\mu=-1$ are $\nu=\pm 1$. Thus, the dipole matrix element $\langle np1,np1 |\hat V_{\rm dd}^{(AC)}(\vartheta=\pi/2)| ns0,n's0\rangle$ is non zero and atom $A$ and $C$ will interact. 

In general, for the dominant channel $P_{3/2}+P_{3/2}\longrightarrow S_{1/2}+S_{1/2}$ only the term $d^{(1)}_{-1}d^{(2)}_{-1}$ with $\mu=\nu=-1$ in Eq.~\eqref{eq:dipdip} can contribute to the dipole-dipole matrix element and thus the van der Waals interaction between both atoms in a $n{}^2P_{3/2},m=3/2$ states becomes $V^{(n)}_{ \frac{3}{2},\frac{3}{2}}(r,\vartheta,\varphi)\sim (Y_{2}^{-2})^2\sim \sin^4\vartheta$ for this channel. Residual interactions at $\vartheta=0$ and $\pi$ come from couplings to $D_{3/2,5/2}$ channels which are small, see Tab.~\ref{tab:p32}. Thus, using Rubidium $n{}^2P_{3/2}$ states with $m_j=3/2$ allows an almost perfect realization of an anisotropic interaction with vanishing interaction along one axis and large interaction along a perpendicular axis. Note that interactions between two atoms in a $|n{}^2P_{3/2},m=3/2\rangle$ are negative (attractive) for $n>38$ and positive (repulsive) for $n<38$, where a F\"orster resonance at $38P_{3/2}+38P_{3/2}\longrightarrow 38S_{1/2}+39S_{1/2}$ changes the sign of the interaction~\cite{Singer:2005ct,Reinhard:2007hm,Walker:2008bm,Walker:2005kj}. Figure~\ref{fig:pstateint}(a) shows the result of the the full calculation of the van der Waals interactions between $n{}^2P_{3/2},3/2$ states including all channels and summing over $n'$ and $n''$ levels between $ n \pm 10$. The full calculation agrees well with the simplified picture discussed above and illustrated in Fig.~\ref{fig:pstateint}(b) since the dominating channel is the one coupling to $S_{1/2}$ states.

\subsection{Soft-core potentials}\label{sec:softcore}

In the previous section we showed how to engineer the anisotropic part of the interactions required by Eq.~\eqref{eq:constraint}. We now discuss how to create soft-core potentials by weakly admixing the Rydberg state $p$-states to the atomic ground state. Following the $s$-state case~\cite{Henkel:2010il}, this leads to an effective interaction between dressed ground state atoms, $\tilde V_{ij}$, with a soft-core shape and an anisotropic plateau radius. This guarantees that interactions between atoms sitting on a square lattice at different distances $a$ and $\sqrt{2}a$ experience the same interaction potential~\cite{Dauphin:2012jo}, as required by Eq.~\eqref{eq:constraint} and illustrated in Fig.~\ref{fig:pyro}(b).

The single atom configuration we have in mind was introduced in Sec.~\ref{sec:single} and is governed by the Hamiltonian of Eq.~\eqref{eq:Hsingle}. Pairwise interactions [see panel (a) of Fig.~\ref{fig:softcore}] between $N$ atoms both excited to the Rydberg states $|r_i\rangle|r_j\rangle$ are described by
\begin{equation}
H_{ij}(\mathbf{r}_{ij})=V_{ij}(\mathbf{r})|r_i\rangle\langle r_i|\otimes  |r_j\rangle\langle r_j|,
\end{equation}
where $V_{ij}(\mathbf{r}_{ij})=A_{ij}(\vartheta_{ij},\varphi_{ij})/r_{ij}^6$ is the van-der-Waals interaction potential between the Rydberg states of atom $i$ and atom $j$ discussed in the previous section and $(r_{ij},\vartheta_{ij},\varphi_{ij})$ are the spherical coordinates of the relative vector. In the dressing limit, $\Omega_r\ll\Delta_r$,  atoms initially in their electronic ground states $|g\rangle_1\ldots|g\rangle_N$ are off-resonantly coupled to the Rydberg states $|r\rangle_1\ldots|r\rangle_N$. As a consequence, the new {\it dressed} ground states inherit a tunable fraction of the Rydberg interaction. The effective interaction potential between $N$ atoms in their dressed ground states, $|\tilde g\rangle_1\ldots|\tilde g\rangle_N$, can be obtained by diagonalizing the Hamiltonian 
\begin{equation}
\label{eq:Htot}
H=\sum_{i=1}^N H_i+\sum_{i\neq j}H_{ij}(\mathbf{r}_{ij})
\end{equation}
for a fixed relative position and zero kinetic energy. The total Hamiltonian $H$ has block structure
\begin{equation}
\begin{split}
H&=\left(\begin{array}{ccccc}
\mathbf{H}_0 & \mathbf{\Omega}_1 & 0 & 0 &\\
\mathbf{\Omega}_1^\dag & \mathbf{H}_1 & \mathbf{\Omega}_2 & 0 &  \\
0 & \mathbf{\Omega}_2^\dag & \mathbf{H}_2 & \mathbf{\Omega}_3 & \\
0 & 0 & \mathbf{\Omega}_3^\dag & \mathbf{H}_3 &\\
 &  & &  & \ddots
 \end{array}\right)
 \end{split}
\end{equation}
where $\mathbf{H}_n$ governs the dynamics in the subspace with $n$-Rydberg excitations present (see App.~\ref{app:softcore}), while the $\mathbf{\Omega}_n$ matrices describe the coupling between adjacent sectors $n$ and $n-1$ due to the laser. Only subspaces $\mathbf{H}_{n\geq 2}$ contain the interaction potentials $V_{ij}$ since we assume that ground and Rydberg states do not significantly interact. Here, $\mathbf{H}_0$ described the dynamics within the ground state manifold. 

In the following we will use Brillouin-Wigner perturbation theory (see App.~\ref{app:softcore}) based on the small parameter $\Omega_r/(2\Delta_r)\ll 1$ in order to derive the effective potential between dressed ground state atoms. For red detunings, $\Delta_r<0$, and repulsive Rydberg interactions $V_{ij}>0$, one finds up to fourth order in $\Omega_r/(2\Delta_r)$  for the position dependent energy shift of the dressed ground states a sum of binary interactions of the form
\begin{equation}
\begin{split}
\tilde V_{ij}=2\Delta_r\left(\frac{\Omega_r}{2\Delta_r}\right)^4\frac{r_c^6(\vartheta_{ij},\varphi_{ij})}{r_c^6(\vartheta_{ij},\varphi_{ij})+r^6_{ij}},
\label{eq:dressedpot1}
\end{split}
\end{equation}
with
\begin{equation}
\begin{split}
r_c(\vartheta_{ij},\varphi_{ij})=\left(\frac{A_{ij}(\vartheta_{ij},\varphi_{ij})}{2|\Delta_r|}\right)^{1/6}
\label{eq:rc}
\end{split}
\end{equation}
being the {\it anisotropic} Condon radius. In the case of anisotropic Rydberg interactions, the Condon radius depends on the angular pattern of the van der Waals interaction $A(\vartheta,\varphi)$ which can be tuned by choosing a particular Rydberg state. Additionally, the Condon radius can be scaled by changing the detuning $\Delta_r$ of the dressing laser. Figure~\ref{fig:softcore}(b) shows typical examples of the dressed ground state potential, $\tilde V_{ij}$, for different Condon radii, $r_c$. For large distances, $r\gg r_c$, the dressed ground state potential is proportional to the Rydberg interaction, $\tilde V_{ij}=\Omega_r^4/(2\Delta_r)^4V_{ij}\sim 1/r^{6}_{ij}$, reduced by a factor $[\Omega_r/(2\Delta_r)]^4$ arising from the small probability to excite the atomic ground state to the subspace of two atoms in the Rydberg state, governed by $\mathbf{H}_2$. However, for small distances, $r< r_c$, when two atoms are within the Condon radius, the excitation to the Rydberg states becomes ineffective due to the large total detuning $|\Delta_r|+V_{ij}$ (Dipole blockade), and the effective ground state interaction, $\tilde V_{ij}\approx\tilde V_0 [1-(r/r_c)^6]$ for $r< r_c$, saturates at a constant value 
\begin{equation}
\tilde V_0=\Omega_r^4/(2\Delta_r)^3, 
\label{eq:V0}
\end{equation}
which is independent of the strength or form of the Rydberg-Rydberg interactions~\cite{Santos:2000ec,Honer:2010je,Pupillo:2010bt,Henkel:2010il,Macri:2014jy}.
The resulting specific step-like form of $\tilde V_{ij}$ is shown in Fig.~\ref{fig:softcore}(b). The presence of a plateau at short distances, $r<r_c$ and a rapid decrease of the potential at $r\sim r_c$, where $\tilde V_{ij}\sim 1/r_{ij}^6$, allows to engineer approximately equal interactions between atoms within $r_c$ independent of their specific distance. At the same time, long-range interactions, such as next-nearest-neighbor (NNN), are substantially suppressed. 

\begin{figure}[bt]  
\centering 
\includegraphics[width= 0.91\columnwidth]{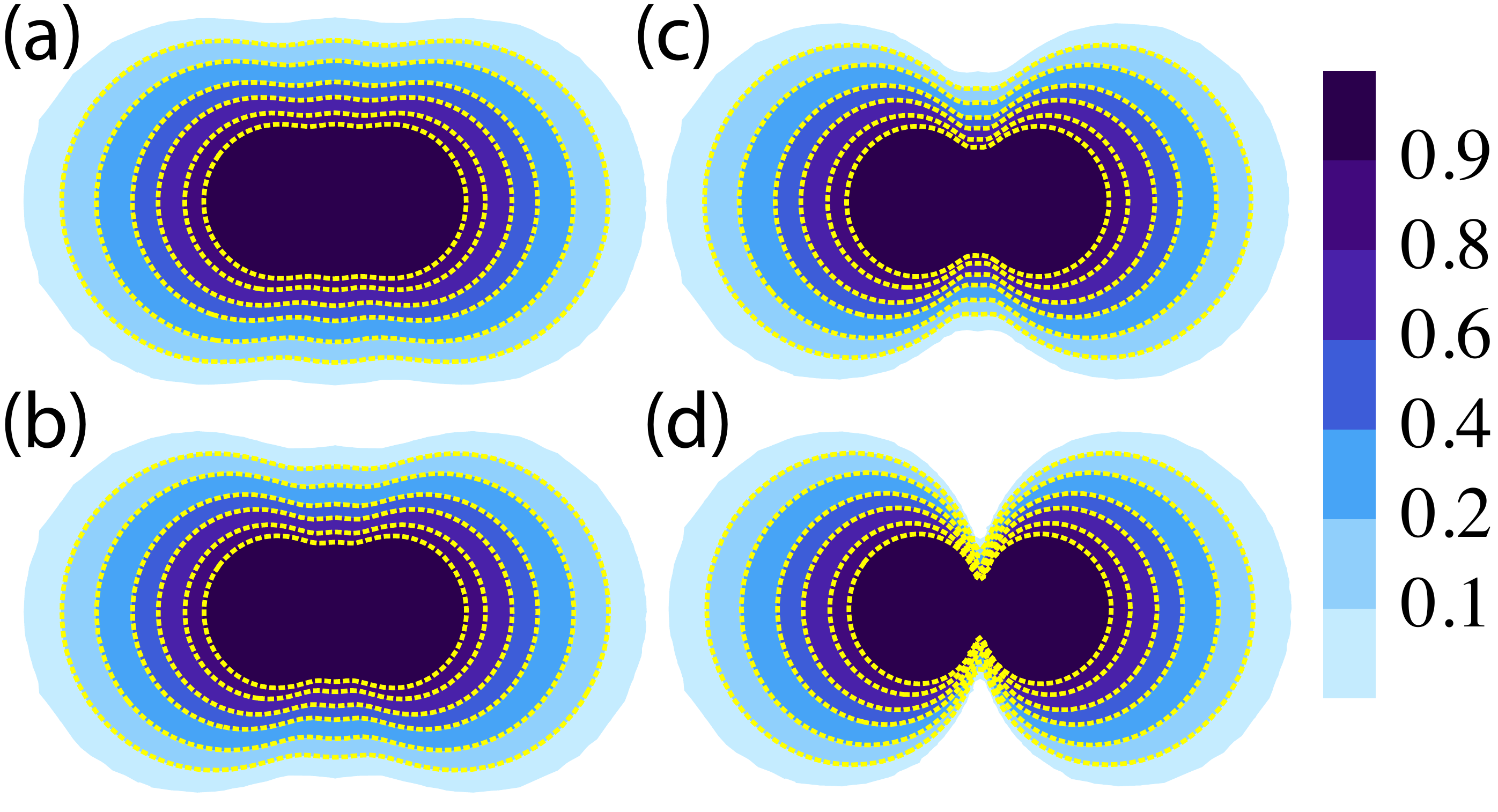}
\caption{\small{Soft core potentials of Eq.~\eqref{eq:dressedpot1} between $n{}^2P_{3/2}$, $m_j=3/2$ Rydberg states with (a) $n=25$, (b) $n=30$, (c) $n=36$ and (d) $n=38$. Residual interactions along the vertical direction comes from virtual transitions to $D$-channels and depends on the principal quantum number $n$, see Tab.~\ref{tab:p32}.}}
\label{fig:plat} 
\end{figure}

In the following we want to combine the effective step-like ground state potentials of Eq.~\eqref{eq:dressedpot1} with anisotropic interactions discussed in the previous section. Panel (c) of Fig.~\ref{fig:softcore} shows a  2D contour plot of $\tilde V_{ij}/\tilde V_0$ in the $x$-$z$ plane for $^{87}$Rb ground state atoms dressed with the Rydberg state $|r_\medbullet\rangle=|32 {}^2P_{3/2},m_j=3/2\rangle_z$. The dashed yellow equipotential lines have the form of a figure-eight which follows from the $V_{ij}\sim \sin^4\vartheta$ dependency of the Rydberg interaction, discussed in Sec.~\ref{sec:pstates}. Residual interactions along the $z$-axis $(\vartheta=0)$ comes from channels coupling to $D$-states (see Tab.~\ref{tab:p32}) and depend on the principal quantum number $n$ as shown in Fig.~\ref{fig:plat}. 

Once defined on the top of an underlying lattice, the combination of anisotropic interactions between Rydberg $p$-states and step-like potentials via dressing techniques allows to design interaction potentials as the ones shown in Fig.~\ref{fig:softcore}(c). Here, atoms (black circles) regularly arranged on a square lattice with lattice spacing $a$ interact with $J$, $J'$ and $J''$ along the $\pm$ 45-degree lines, the $x$-axis and the $z$-axis, respectively. It is possible to tune the interaction strength $J$, $J'$ and $J''$ over a large range by e.g. changing the detuning, $\Delta_r$, or the principal quantum number, $n$, of the Rydberg state. In particular one can realize an interaction pattern where atoms sitting at  different distances $a$ and $\sqrt{2}a$ interact with equal strength, that is $J\approx J'$, while $J''\ll J$, thus realizing a frustrated $J-J'$ model. Note that the interaction symmetry in this case is triangular on top of an square lattice.

\subsection{Explicit numbers and discussion of imperfections}
\label{sec:numbers}
As an explicit example we consider the $29\,{}^2P_{3/2}$ Rydberg manifold of $^{87}$Rb. We resonantly couple the $29\,{}^2P_{3/2}$ manifold to the lower-lying $7\,{}^2D_{3/2}$  manifold, as illustrated in Fig.~\ref{fig:tensor}  ($n=29$ and $n'=7$), with a laser of wavelength $\lambda_{AC}=3.296\,\mu$m. This results in a lattice spacing $a=\lambda_{AC}/(2\sqrt{2})=1.16\,\mu$m which can be adjusted by tilting the trapping lasers by an angle $\alpha=39$ degrees (see Sec.~\ref{sec:single}).

For the van der Waals interactions between the Rydberg states $|r_\medbullet\rangle=|29\,{}^2P_{3/2},3/2\rangle_z$ and $|r_\blacksquare\rangle=|29\,{}^2P_{3/2},3/2\rangle_x$ we find 
\begin{equation}
\begin{split}
&V_{\medbullet\medbullet}(r,\vartheta)=2\pi\times\frac{25.4-31.9\cos2\vartheta+8.2 \cos4\vartheta}{(r/\mu {\rm m})^6}{\rm MHz},\\
&V_{\blacksquare\blacksquare}(r,\vartheta)=2\pi\times\frac{25.4+31.9\cos2\vartheta+8.2 \cos4\vartheta}{(r/\mu {\rm m})^6}{\rm MHz},\\
&V_{\medbullet\blacksquare}(r,\vartheta)=2\pi\times\frac{16.8-8.2\cos4\vartheta+20.3 \sin2\vartheta}{(r/\mu {\rm m})^6}{\rm MHz},
\end{split}
\end{equation}
including all channels of Tab.~\ref{tab:p32} and summing over $\pm10$ $n$-values (see App.~\ref{app:vdw} and~\ref{app:mixed}). They are plotted in Fig.~\ref{fig:pstateint}(a) and Fig.~\ref{fig:int45}. The largest off-diagonal matrix element coupling different Zeeman $m$-levels is $V_{\rm off}(a)=\langle\frac32\frac32|\hat V_{\rm vdW}(a,\pi/2)|\frac12\frac12\rangle=2\pi\times 11.2$ MHz. Using an AC Stark laser with power \mbox{$P=1.0$ mW} focused on a area \mbox{$A=50\,\mu{\rm m}^2$} yields a Rabi frequency \mbox{$\Omega_{AC}=2 d_{7D-29P}\mathcal{E}_{AC}/\hbar=2\pi\times 205.5$ MHz}, where \mbox{$d_{7D-29P}=\langle7D_{3/2},3/2|d|29P_{3/2},1/2\rangle=0.065\,ea_0$} is the smallest transition dipole moment and $\mathcal{E}_{AC}=\sqrt{2 P/c\epsilon_0 A}$ is the electric field strength. The AC Stark lasers will create an additional ground state potential with depth $V_{AC}=2\pi\times 27.8$~kHz thus the initial trapping potential, $V_{\rm trap}$, must be larger than $V_{AC}$, see App.~\ref{app:aclattice}.

Adjusting the detuning $\Delta_r$ of the Rydberg laser allows to tune the length scale and the imperfections in Eq.~\eqref{eq:dressedpot1}. These are (i) small long-range interactions between nearest-neighbor lattice sites and (ii) deviations form the constraint model of Eq.~\eqref{eq:constraint}. Here, for example we use $\Delta_r=2\pi\times 400$~kHz which yields the following interaction pattern between particles labeled in Fig.~\ref{fig:softcore}(d): $\tilde V_{14}/\tilde V_0=\tilde V_{23}/\tilde V_0=0.96$, $\tilde V_{13}/\tilde V_0=\tilde V_{24}/\tilde V_0=0.80$, $\tilde V_{12}/\tilde V_0=\tilde V_{}/\tilde V_0=0.70$ around a vertex and small imperfect interactions between different vertices, e.g. $\tilde V_{15}/\tilde V_0=\tilde V_{28}/\tilde V_0=0.09$ and next-nearest-neighbor interactions e.g. $\tilde V_{16}/\tilde V_0=\tilde V_{27}/\tilde V_0=0.12$ or $\tilde V_{18}/\tilde V_0=\tilde V_{25}/\tilde V_0=0.01$.

By varying the Rabi frequency of the Rydberg laser
$\Omega_r=2 \pi \times (80,120,160)$~kHz one obtains 
$\epsilon=\Omega_r/2\Delta_r=(0.10,0,15,0.20)$ which gives rise to an effective ground state interaction 
$\tilde V_0=\Omega_r^4/8\Delta_r^3=2\pi\times (80,410,1290)$~Hz. This is much larger than the effective decay rate from the dressed ground state $\tilde \Gamma=\epsilon^2\Gamma=2 \pi\times (33,75,133)$~Hz, and larger than a corresponding tunneling rate between the minima. Here, $\Gamma=2\pi\times 3.3$~kHz is the decay rate form the Rydberg states.

There is an ample choice in the parameter regimes available as a function of the $n$-level. Away from the F\"orster resonance at $n=38$, it is possible to engineer infra-red lattices which allow for comparable timescales between the interactions induced by the dressing, and the tunneling matrix elements of the atoms on the original square lattice. Going higher in $n$, closer to the F\"orster resonance, allows faster timescales and slower decays: however, in this case the infra-red laser has a strong influence on the underlying lattice, excluding the possibility of using conventional single particle tunneling to induce quantum fluctuations. On the other hand, one can profit here from the richness of the Rydberg manifolds involved, realizing the hopping matrix element as a spin-exchange coupling between different atoms sitting at different potential minima~\footnote{A.~W.~Glaetzle, {\it et. al., in preparation}}.
In both cases above, the interaction pattern will depend on the specific targeted $n$, as discussed in Sec.~\ref{sec:pstates}. As the qualitative (and in many respects quantitative, as indicated in Table~\ref{tab:p32} and Fig.~\ref{fig:plat}) shape of the interactions will be very similar in the interval of interest $n = 25-37$, we will focus in the following on a single case sample to underpin the stability of the many-body effects we are interested in.

\section{Numerical results \label{sec:mb}}
In this section, we consider the properties of the approximate realization of the quantum spin ice model Hamiltonian proposed above. We demonstrate that, as a function of the strength of the quantum dynamics, the ground state has two regimes reflecting two distinct forms of ordering (Sec.~\ref{sec:2phases}). One, stabilized via a quantum order by disorder mechanism, generates the above mentioned plaquette phase for sufficiently strong quantum dynamics. As it is weakened, there is a transition into a phase with classical ordering, which is stabilized by the long-range parts of the dipolar couplings and which breaks translational symmetry in a different way. In addition, we show that even without quantum dynamics, there is an interesting thermal phase transition to an approximate realization of a (classical) Coulomb phase, with only a very small density of defects (plaquettes violating the ice rule) of around 5\% (Sec.~\ref{sec:TT}). We discuss signatures of these items in various quantities, in particular proposing a simple correlation function in which the quantum plaquette order will be visible, and which should be accessible in cold atom experiments via {\it in-situ} parity measurements~\cite{Bakr:2009bx,Sherson:2010hg,Schauss:2012hh}.

\begin{figure*}[!t]
\includegraphics[width=0.9\linewidth]{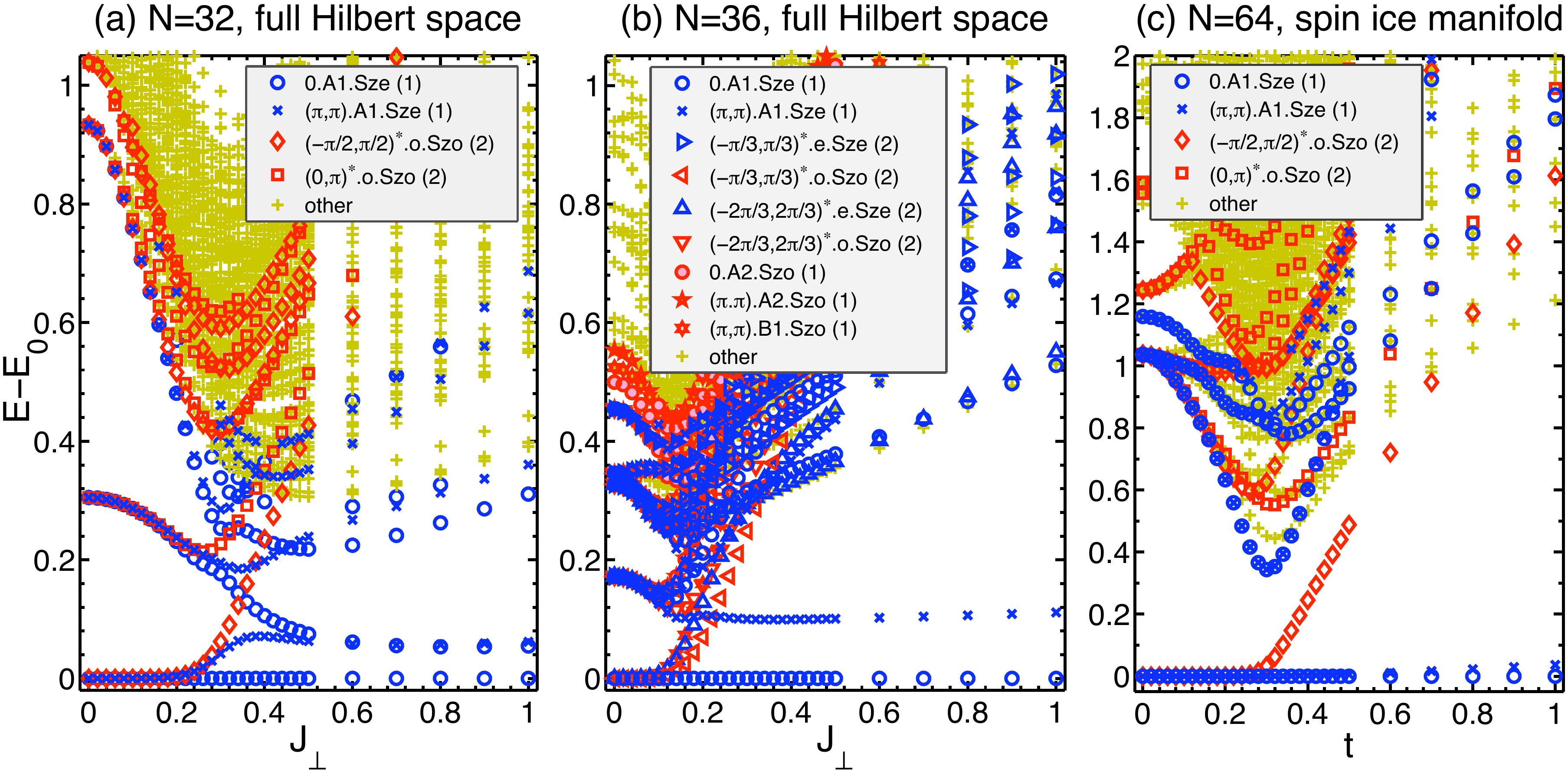}
\caption{(a-b) Low energy spectra of $H$ versus $J_\perp$ for N=32 (a) and 36 (b), in the total magnetization sector $S_z\!=\!0$ and for cutoff $J_c\!=\!0.001$. (c) Low energy spectra of the constrained, spin-ice model $H_2$ versus $t$ for 64 sites. The eigenstates are labeled by linear momentum $\vec{k}$, the irreducible representations of the point group of $\vec{k}$ (the point group of the model is $C_{2v}$), and parity under spin inversion ("Sze'' and "Szo'' stand for even and odd parity, respectively).}
\label{fig:en}
\end{figure*}

\subsection{General definitions and conventions}\label{sec:dfns}
We begin with some general definitions and technical details of our numerical study. We consider both the unconstrained spin-1/2 model $H$ from by Eq.~\eqref{eq:XXZ}, as well as the projected model $H_2$ inside the spin ice manifold: 
\bea
H&=&\sum_{i<j} J_{ij} S_{i}^zS_{i}^z + J_\perp \sum_{\langle ij\rangle} \Big( S_{i}^+S^-_{j}+S_{i}^-S^+_{j} \Big) \\
H_2&=&\sum_{i<j} J_{ij} S_{i}^zS_{j}^z - t \sum_{_i^j\square_l^k} \left( S_i^+ S_j^- S_k^+ S_l^- + h.c. \right)~.
\eea
Here $\langle ij\rangle$ denote nearest-neighbor (NN) sites on the 2D checkerboard lattice and $(ijkl)$ label the four sites around empty square plaquettes. 
In our Exact Diagonalizations (ED) we have considered finite-size clusters with periodic boundary conditions and N=16, 32, 36, 64 and 72 sites, see details in App.~\ref{app:Clusters}. To treat these clusters with ED, we exploit translational symmetry, point group operations (the model has $C_{2v}$ symmetry), as well as spin inversion ($S_z\!\to\!-S_z$) inside the total magnetization sector $S_z\!=\!0$. 
Consequently, the eigenstates are labeled by linear momentum $\vec{k}$, the irreducible representations of the point group of $\vec{k}$, and the parity under spin inversion.

We note that, whereas the quantum phase is quite robust, the classical phase is considerably less so, reflecting the many nearly-degenerate classical ice states. We illustrate this {in App.~\ref{app:ClassMin}} by imposing a variable cut-off on the long-range aspect of the dipolar couplings $J_{ij}$: by neglecting terms weaker than a cutoff $J_c$, we find a set of  states with different classical orders, which settle down into the correct ground state without truncation for $J_c$ no larger than $0.001$.

\subsection{The two zero-temperature phases:  Low-energy spectroscopy and ground state diagnostics}\label{sec:2phases}
Figure~\ref{fig:en} shows the low energy spectra of $H$ as a function of $J_\perp$ for N=32 (a) and N=36 (b), and that of $H_2$ as a function of $t$ for N=64 (c). All spectra correspond to the total magnetization sector $S_z=0$ and a cutoff value of $J_c\!=\!0.001$. In all spectra, there is a manifold of low-lying states that is well separated from higher-energy excitations. Provided they become degenerate in the thermodynamic limit, these states are the finite-size fingerprints of the spontaneously symmetry broken phases~\cite{Anderson,Bernu1992,*Bernu1994,Claire, Gregoire2007,ioannis2012}:  their multiplicities and symmetry content reveal the nature of the ground state. The structure of the low-lying energy states show consistently two qualitatively different phases. One, which is adiabatically connected to the classical limit $J_\perp\!=\!0$, and the other which is stabilized for large enough $J_\perp$ or $t$.

\begin{figure*}[!t]
\includegraphics[width=0.98\linewidth]{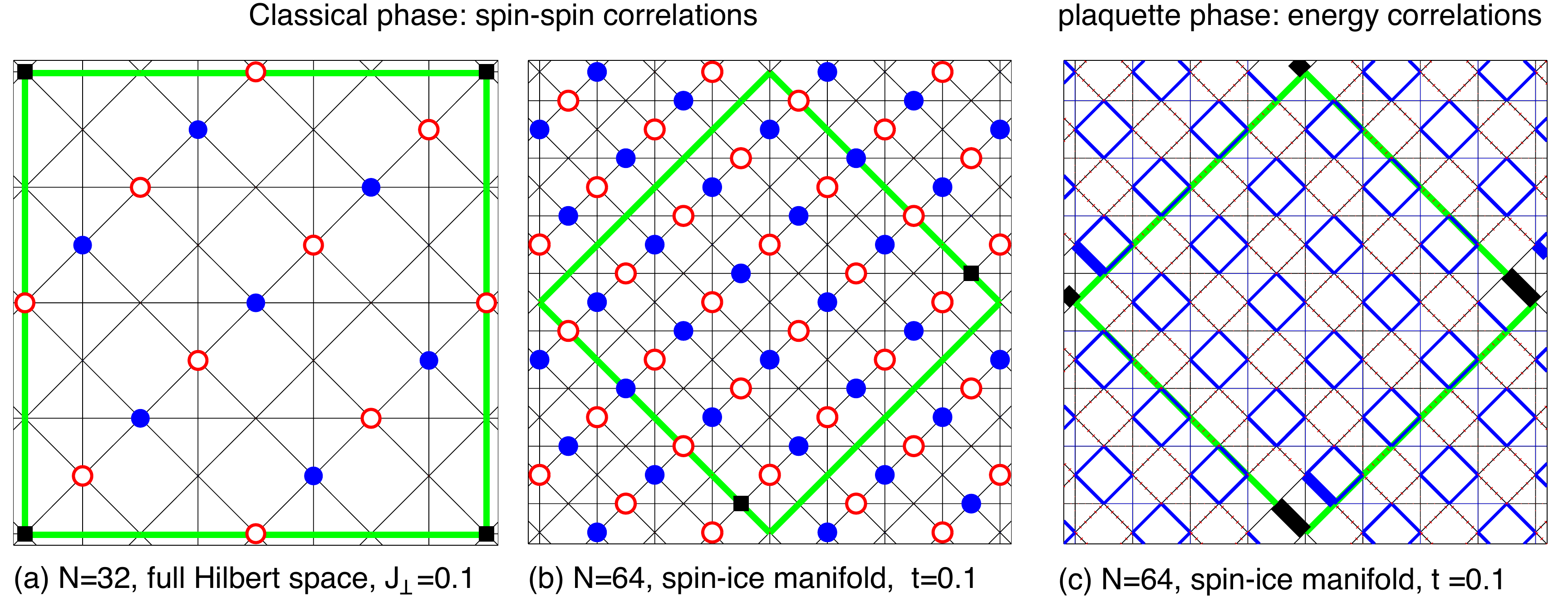}
\caption{Ground state diagnostics of the classical (a-b) and the QM plaquette (c) phases: Spin-spin correlation profiles (of the type $\langle S_i^z S_j^z\rangle$, where $i$ is the reference site, indicated by filled black square) in the ground state of (a) $H$ for N=32 and $J_\perp\!=\!0.1$, and (b) $H_2$ for N=64 and $t=0.1$. Filled blue (open red) circles corresponds to positive (negative) amplitude. (c) Connected energy correlation profiles (of the type $\langle S_i^zS_j^zS_k^zS_l^z\rangle\!-\!\langle S_i^zS_j^z \rangle \langle S_k^zS_l^z\rangle$, where the reference bond $(ij)$ is indicated by the thick black segment) in the ground state of $H_2$ for N=64 and $t=1$. Solid blue (dashed red) bonds indicate positive (negative) amplitudes, while the width of each bond scales with the magnitude. All data correspond to the symmetry sector ``0.A1.Sze'' and are taken for cutoff $J_c\!=\!0.001$.  }
\label{fig:GSDiagnostic}
\end{figure*}

We begin with the classical phase, focusing on the N=32 (a) and N=64 (c) results first. Here we find four low-lying states which become exactly degenerate as $J_\perp\!\to\!0$. We find translational symmetry breaking with ordering wavevector $\vec{Q}\!=\!(-\frac{\pi}{2},\frac{\pi}{2})$, as illustrated in Fig.~\ref{fig:GSDiagnostic}(d). The nature of this phase is revealed by the spin-spin correlation profiles of Figs.~\ref{fig:GSDiagnostic}(a-b), with alternating up-down spins along one of the two diagonal directions of the lattice. The vanishing of correlations on every second diagonal line arises due to the existence
of two states compatible with the non vanishing correlations on the other diagonals. For a finite cluster, these appear with equal weight and thus average out, while in the thermodynamic limit, symmetry breaking selects either one of the two spontaneously. 
Finally, the N=36 cluster cannot accommodate the $\vec{Q}\!=\!(-\frac{\pi}{2},\frac{\pi}{2})$ phase (see App.~\ref{app:ClassMin}), which is why the low-lying sector of Fig.~\ref{fig:en}(b) has a different structure (and, in fact, higher ground state energy per site, see Table~\ref{tab:ClassMin}). 

Turning to the quantum phase, the N=32 and 64-site spectra give the onset of this phase around $J_\perp\!\simeq\!0.23$ and $t\!\simeq\!0.28$, respectively. Beyond this point, the spin structure factor (not shown) is completely structureless, indicative of the absence of magnetic (classical) ordering. Since the imperfections in the present spin-ice model are expected to become irrelevant for large enough $J_\perp$, this phase must be the plaquette phase of the pure spin-ice model~\cite{R1b,R15} and the pure Heisenberg model~\cite{Fouet2003}. The standard diagnostic for this phase is the dimer-dimer (or energy-energy) correlations, and indeed the correlation profiles of Fig.~\ref{fig:GSDiagnostic}(c) show a strong $\vec{Q}\!=\!(\pi,\pi)$ response within one sublattice of empty plaquettes. 
This is consistent with the structure of the low-lying spectra which show two low-lying states with momenta $\vec{k}=0$ and $(\pi,\pi)$ which come almost on top of each other for N=64, see Fig.~\ref{fig:en}(c). Note that for N=32, there is a third low-lying state (with $\vec{k}\!=\!0$) which is however not related to the physics at the thermodynamic limit but it is specific to the special topology of this cluster~\footnote{Such ``extra'' low-lying states are also present in the pure Heisenberg model and are related to extra symmetries of the 32-site cluster~\cite{Fouet2003}}.

\begin{figure}[!b]
\includegraphics[width=0.99\linewidth]{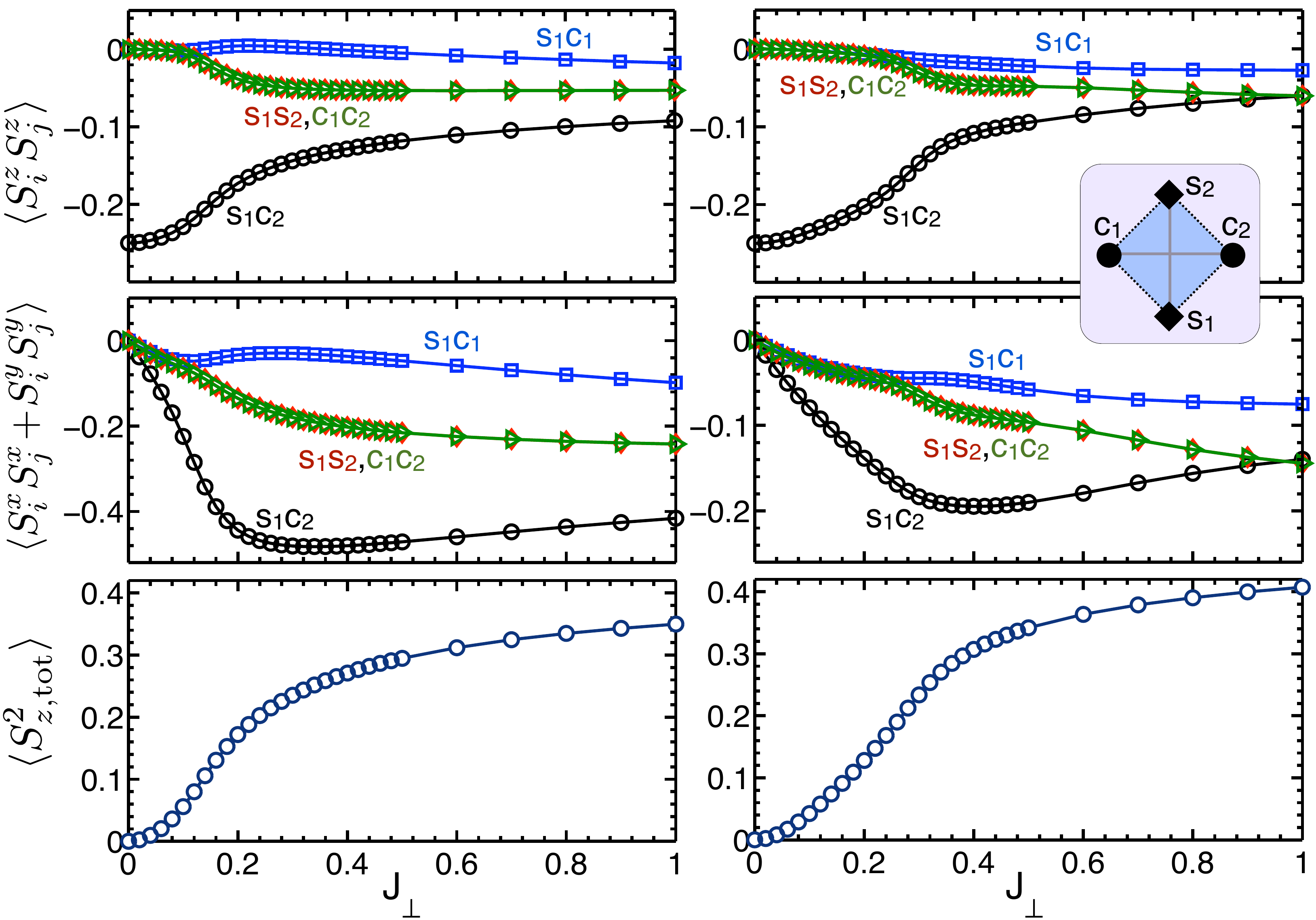}
\caption{Various expectation values in the ground state of $H$ for N=16 (left column) and 32 (right column), for cutoff $J_c\!=\!0.001$. The first two panels in each column show the NN spin-spin correlations for all symmetry inequivalent bonds (inset) in the longitudinal and transverse (xy) channel. The bottom panels show the square of the total magnetization per crossed plaquette, which is a measure of the weight from states outside the spin ice manifold.} 
\label{fig:NNCorrs}
\end{figure}

Further information about the two phases is given in Fig.~\ref{fig:NNCorrs}, which shows the GS expectation values of the longitudinal and transverse NN spin-spin correlations for all symmetry-inequivalent bonds, as well as the square magnetization of crossed plaquettes. The former describe how the energy is distributed over the bonds and over the different directions in spin space, while the latter is a measure of the admixture from states outside the spin ice manifold. First, the NN correlations show that the spins fluctuate mostly along the z-axis for small $J_\perp$, as expected. More importantly, most of the energy comes from antiferromagnetic bonds along one of the two diagonal directions (bonds labeled `$s_1c_2$' in the inset of Fig.~\ref{fig:NNCorrs}), which is a clear signature of the presence of strongly asymmetric spin-spin correlations in this regime. This asymmetry, which is inherited by the point group symmetry (C$_{2v}$) of the model, is more directly revealed in the spin structure factor discussed above. 
Second, the qualitative change in the behavior of the NN correlations around $J_\perp\!\sim\!0.2$, reflects the presence of the phase transition in this region.
Finally, the square of the total magnetization per crossed plaquette reveals that the spin-ice manifold remains well protected up to relatively high $J_\perp$. 

\begin{figure*}[!t]
\includegraphics[width=0.3\linewidth]{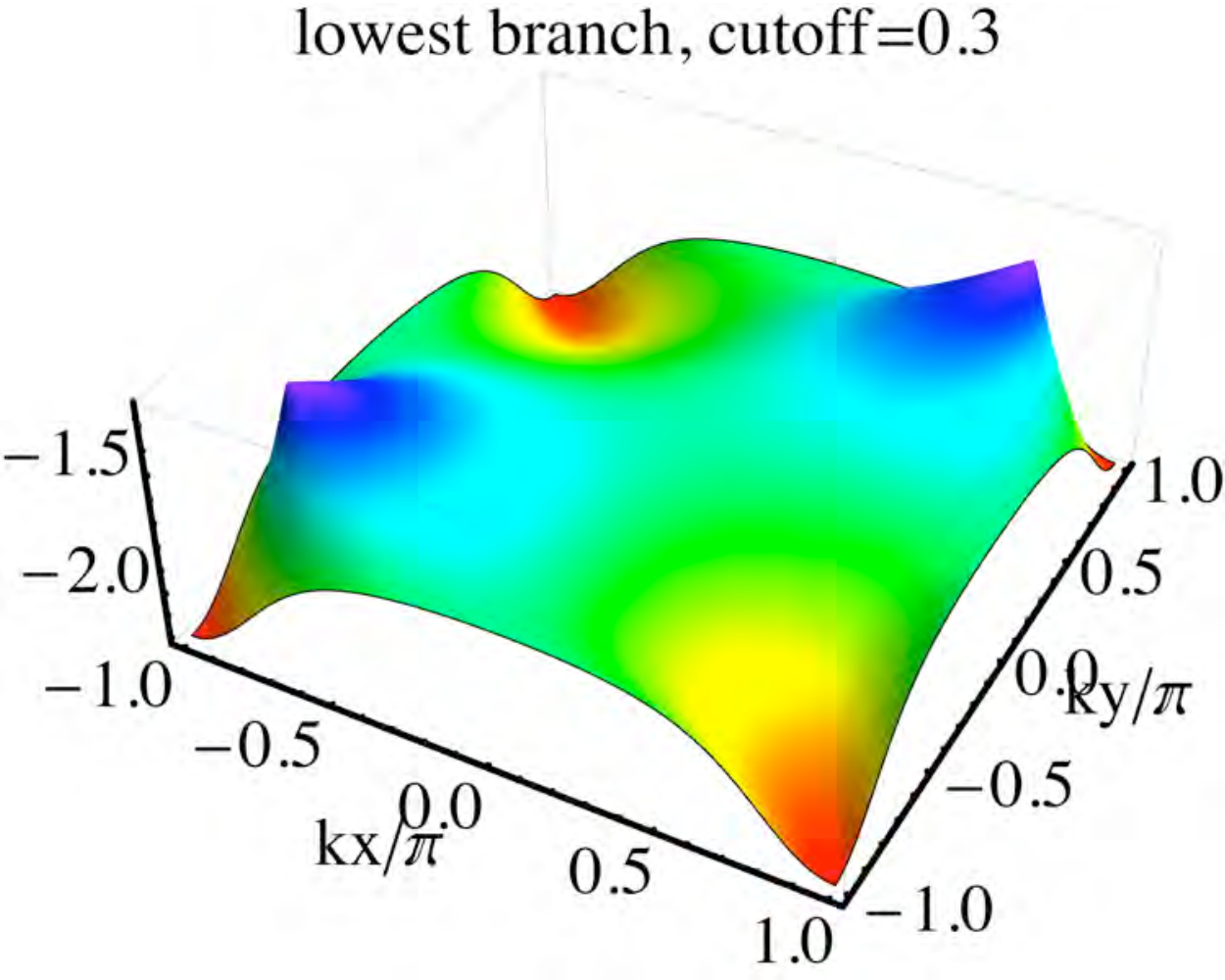}
\includegraphics[width=0.3\linewidth]{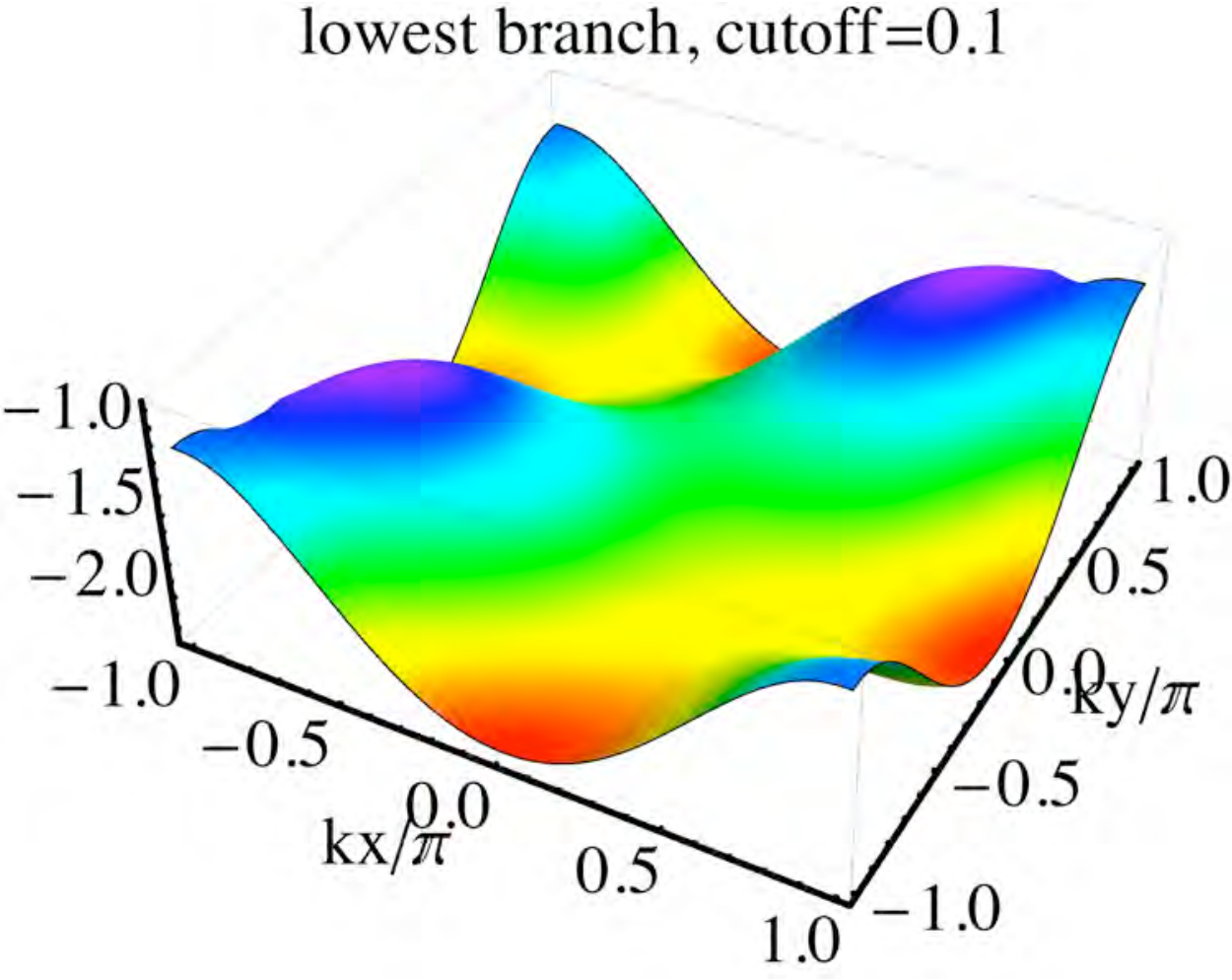}\\
\includegraphics[width=0.3\linewidth]{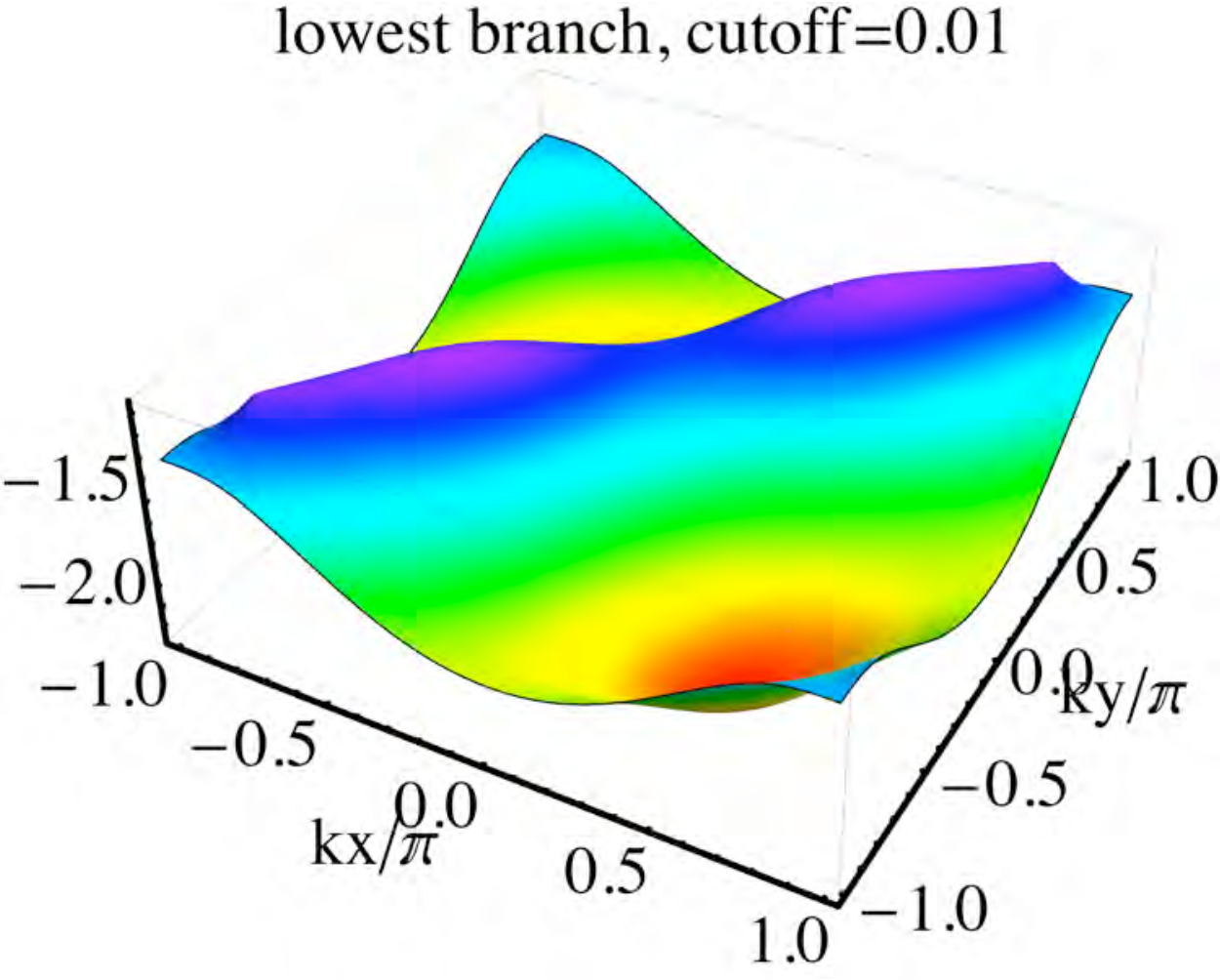}
\includegraphics[width=0.3\linewidth]{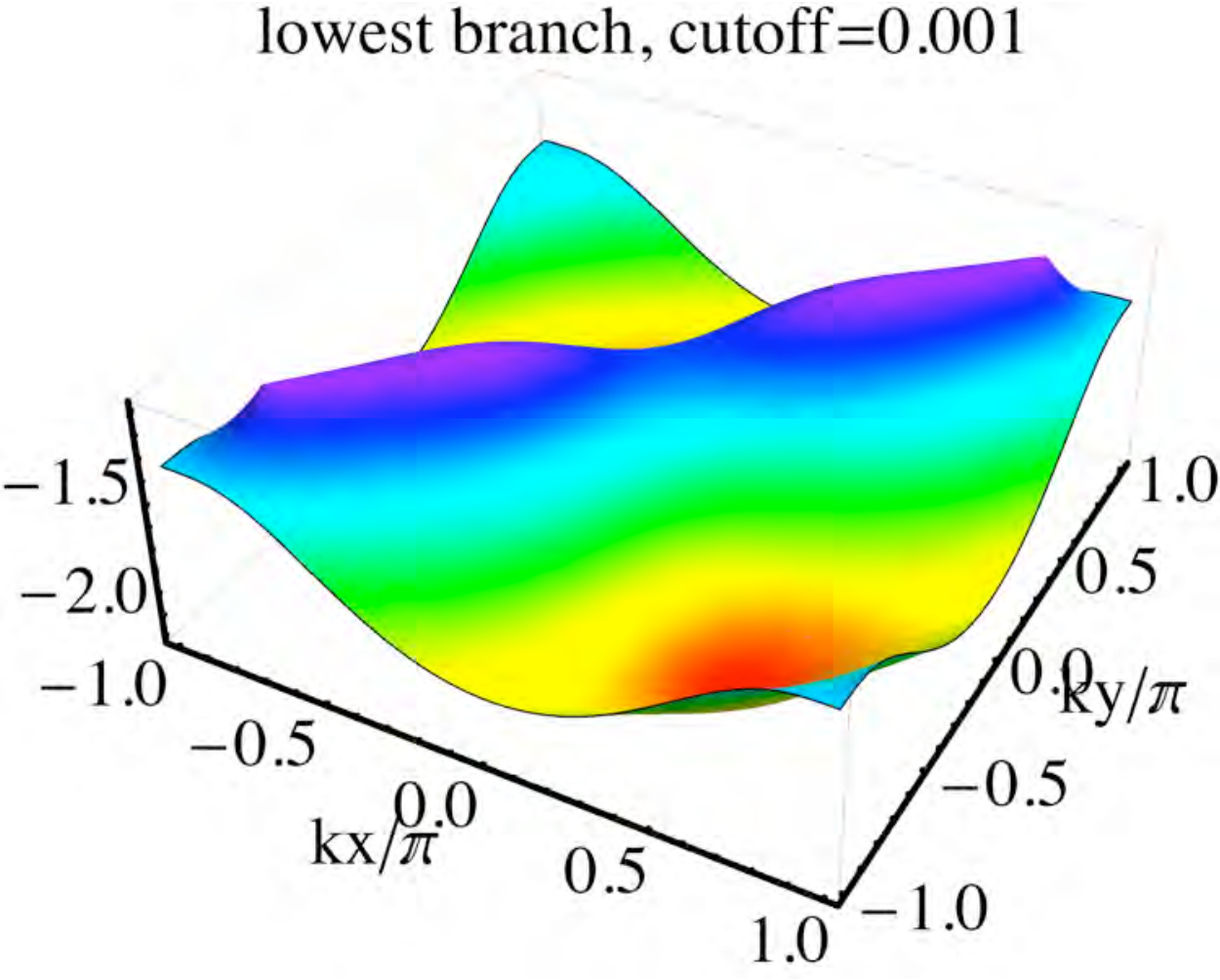}
\caption{Dispersion of the lowest eigenvalue $\lambda_1(\vec{k})$ of the dipolar interaction matrix $\bs{\Lambda}(\vec{k})$ (in the thermodynamic limit) for four different values of cutoffs. 
\label{fig:dispersions}}
\end{figure*}

\subsection{Further insights in the classical limit $J_\perp=0$}
\subsubsection{Momentum space minimization}\label{Sec:ClassMin}
The nature of the classical phase and the role of the dipolar couplings can be understood in more detail by a closer examination of the limit $J_\perp\!=\!0$ using a classical minimization treatment in momentum space~\cite{LT, Bertaut, Litvin, Kaplan}. The checkerboard lattice has a square Bravais lattice with two sites per unit cell. In the following, sites are labeled as $i\!\to\!(\vec{R},\alpha)$, where $\vec{R}$ gives the position of the unit cell, and $\alpha\!=\!1$-$2$. For $J_\perp\!=\!0$, we can replace $S_i^z\to\frac{1}{2}\sigma_i$, where $\sigma_i\!=\!\pm1$. The total energy then reads $E\!=\!\frac{1}{4}E'$, where
\be
E'=\frac{1}{2}\sum_{\vec{R}\vec{R}',\alpha\alpha'}  J_{\vec{R}\alpha,\vec{R}'\alpha'} \sigma_{\vec{R},\alpha} \sigma_{\vec{R}',\alpha'}~. 
\ee
Using $\sigma_{\vec{R},\alpha}\!=\!\frac{1}{\sqrt{N_{uc}}}\sum_{\vec{k}} e^{i\vec{k}\cdot\vec{R}} \sigma_{\vec{k},\alpha}$, where $N_{uc}\!=\!N/2$ is the number of unit cells, and $J_{\vec{R}\alpha,\vec{R}'\alpha'}\!=\!J_{\vec{R}-\vec{R}',\alpha\alpha'}$ (from translational invariance), yields
\be
E'=\frac{1}{2}\sum_{\vec{k}}\sum_{\alpha\alpha'} \sigma_{\vec{k},\alpha} \Lambda_{\alpha\alpha'}(\vec{k}) \sigma_{-\vec{k},\alpha'},\nonumber\\
\ee
where the $2\times2$ interaction matrix $\bs{\Lambda}(\vec{k})$ is given by
\be
\Lambda_{\alpha\alpha'}(\vec{k}) \equiv \sum_{\vec{r}} J_{\vec{r},\alpha\alpha'} e^{-i\vec{k}\cdot\vec{r}}~.
\ee
Let us denote by $\lambda_{1,2}(\vec{k})$ and $\vec{v}_{1,2}(\vec{k})$ the eigenvalues and the corresponding (normalized) eigenvectors of $\bs{\Lambda}(\vec{k})$, with $\lambda_1(\vec{k})\leq \lambda_2(\vec{k})$. Minimizing $\lambda_1(\vec{k})$ over the entire BZ of the model provides a lower bound for the energy~\cite{LT, Bertaut, Litvin, Kaplan}. The corresponding eigenvector is a faithful ground state provided it satisfies the spin length constraint at all sites.

The minimization can be done both for the infinite lattice and for the finite lattices studied by ED by simply scanning through the allowed momenta of each cluster. The latter are discussed in App.~\ref{app:ClassMin} and are useful for clarifying the various finite-size effects in our ED data. Here we focus on the infinite lattice case. Figure~\ref{fig:dispersions} shows the momentum dependence of the low-energy branch $\lambda_1(\vec{k})$ for cutoff values $J_c=$0.3, 0.1, 0.01 and 0.001. For $J_c=0.3$, which amounts to keeping only the dominant, NN couplings (i.e. three couplings per site), the minimum sits at $\vec{Q}\!=\!(\pi,\pi)$ and corresponds to the well-known N\'eel phase with AFM correlations along both the horizontal and the vertical directions of the lattice. This phase is stabilized by the imbalance in the NN imperfections, which favors the first two vertex configurations in Fig.~2(b). 
However, further-neighbor interactions destabilize the N\'eel phase and lead to a different minimum. For $J_c=0.1$, which amounts to keeping seven interactions per site, the minimum of $\lambda_1(\vec{k})$ now sits at the two M points of the BZ, $\vec{Q}\!=\!(\pi,0)$ and $(0,\pi)$, which correspond to a stripy AFM alignment of the spins in the horizontal or the vertical direction of the lattice. 

Lowering $J_c$ further shifts the minimum to two incommensurate (IC) positions, $\pm\vec{Q}_{IC}$, which are extremely close to the commensurate $\pm(-\pi/2,\pi/2)$ points. For example, for $J_c\!=\!0.01$ (12 interactions/site), $J_c\!=\!0.001$ (31 interactions/site) and $J_c\!=\!10^{-6}$ (299 couplings/site), the minima sit respectively at $\vec{Q}_{IC}\!=\!0.473(-\pi,\pi)$, $0.457(-\pi,\pi)$ and $0.462(-\pi,\pi)$. At the same time, the corresponding eigenvector $\vec{v}_1(\vec{Q}_{IC})$ cannot be used to construct a state satisfying the spin length constraint at all sites of the system simultaneously. This means that  the present method cannot deliver the true ground state of the system and that $\lambda_1(\vec{Q}_{IC})$ serves only as a lower energy bound. 

\begin{figure*}[!t]
\includegraphics[width=0.8\linewidth]{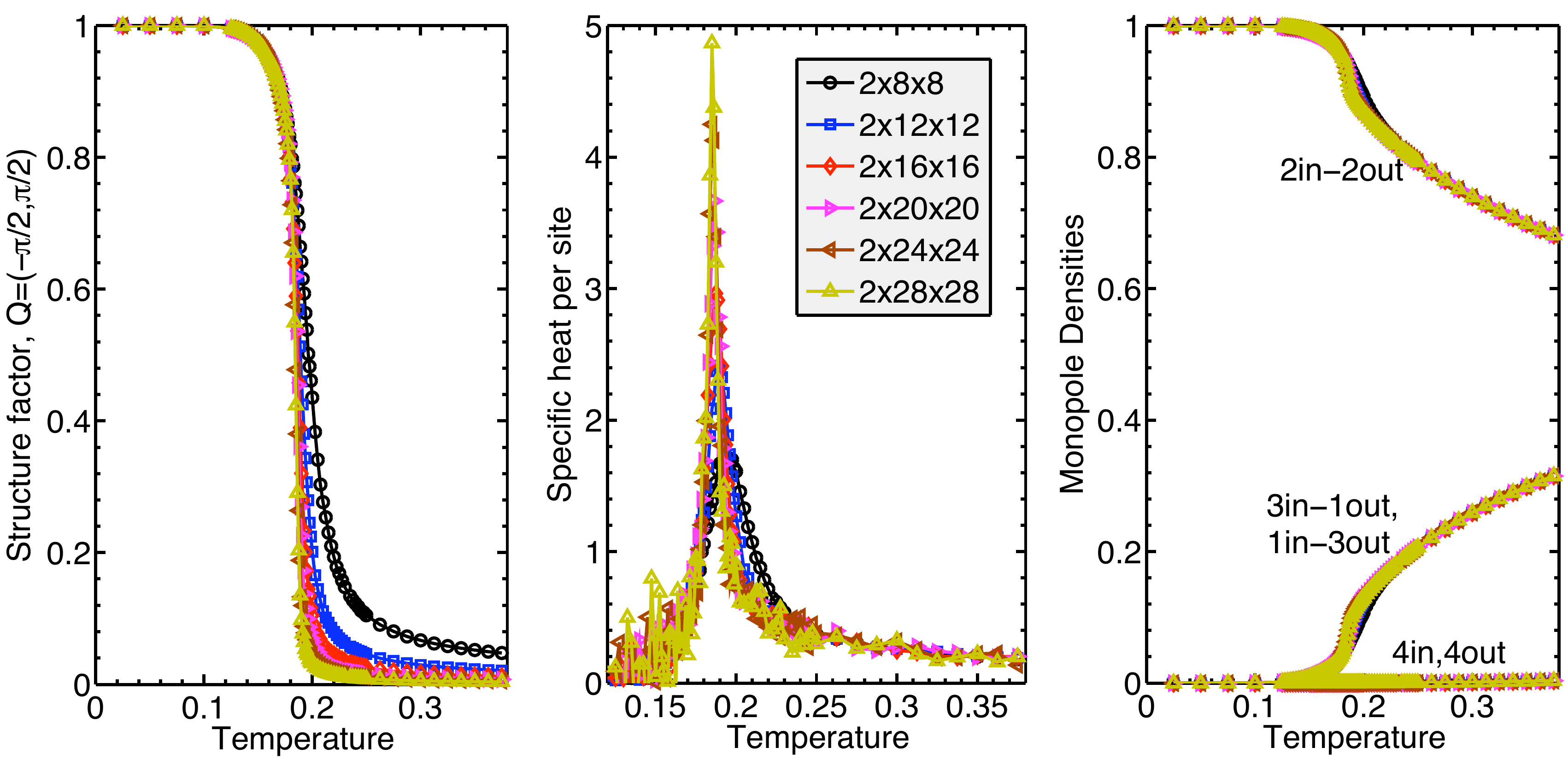}
\caption{Temperature dependence of the structure factor $\mc{S}(\vec{Q})$ (left), the specific heat per site (middle) and the monopole densities for systems up to N=2x28x28 sites, for $J_\perp=0$ and cutoff $J_c\!=\!0.001$.\label{fig:SofQ}}
\end{figure*}

Physically, the system may accommodate the tendency for incommensurate correlations by forming long-wavelength modulations of the local $(-\pi/2,\pi/2)$ order parameter, in analogy e.g. to the anisotropic Ising model with competing interactions (the so-called ANNNI model)~\cite{Elliott1961,Fisher:1980be,Bak1982,Selke1988}. We should remark however that the energy landscape around the IC minimum is very flat and its distance from $(-\frac{\pi}{2},\frac{\pi}{2})$ is very small, so in principle such discommensurations (if any) should appear at much longer distances than the ones considered in our finite-lattice calculations, and indeed the length scales over which cold atom realizations are uniform (in  both density and interaction patterns) on account of the parabolic confining potential. To confirm this point we have performed classical Monte Carlo (CMC) simulations on $N\!=\!2\!\times\!L\!\times\!L$-site clusters with periodic boundary conditions, see details in App.~\ref{app:CMC}. 
All results up to $L=36$ give consistently the $(-\frac{\pi}{2},\frac{\pi}{2})$ state without any sign of domain-wall discommensurations, implying that at least for these distances the system locks-in to the closest commensurate $\vec{Q}=(-\pi/2,\pi/2)$ phase.

\subsubsection{Thermal phase transition into a classical Coulomb phase}\label{sec:TT}
Given the finite energy gap above the commensurate $\vec{Q}\!=\!(-\frac{\pi}{2},\frac{\pi}{2})$ state at $J_\perp\!=\!0$ (see Fig.~\ref{fig:en}), one expects that this phase survives against thermal fluctuations up to a finite temperature $T_C$.
To confirm this picture,  and to find the numerical value of $T_C$, we have performed classical MC simulations at finite temperatures. The first two panels of Fig.~\ref{fig:SofQ} show the T-dependence of the structure factor $\mc{S}(\vec{Q})$ at $\vec{Q}\!=\!(-\pi/2,\pi/2)$, and the specific heat per site for systems up to N=2x28x28 sites. The results demonstrate clearly the thermal phase transition, with $T_C\simeq 0.185$. The third panel shows the T-dependence of the three different types of crossed plaquette configurations: the ice-rule 2in-2out states, and the defected 3up-1down (or 3down-1up) states and 4-in (or 4-out) states. The defects are almost entirely of the 3up-1down type, but their density remains very small up to the transition temperature (about 5\%). So the classical phase gives way to a Coulomb gas~\cite{R10a,R10b}, an approximate realization of a classical Coulomb phase, with only a very small density of defects.
This phase is marginally confined on account of the logarithmic nature of the interactions between the defects; given the
non-vanishing defect density above $T_C$, their correlations are expected to exhibit a screened (Debye) form~\cite{R16}.

\section{Quantum dimer models with Rydberg atoms: beyond quantum ice\label{sec:beyond}}

\subsection{Simple lattices, complicated interactions}\label{sec:routes}

As we have shown in the spin ice example, weakly Rydberg-dressed atoms in optical lattices provide a perfect platform to investigate quantum magnetism in AMO settings. From the one hand, the starting building block being XXZ models allows to use the $zz$ component to impose constraints, and to employ the exchange terms to generate the dynamics in perturbation theory. This is generally the best way to get effective many-body interactions from simple two- or one-body ones: the desired terms emerge at the lowest non-vanishing order in the expansion. From the other hand, the fact that interactions decay like van der Waals beyond a set radius makes imperfections intrinsically local in any dimension, circumventing additional dimensional effects which can emerge with power law interactions with a slower decay. 

The possibility of engineering the spin ice dynamics discussed above resulted from the combination of two features: (i) a simple lattice structure, a square lattice, on which the spins are arranged on, and (ii) a complicated interaction pattern resulting from the condition of equal $zz$ interaction strength around each vertex - requiring both anisotropic and plateau-like interactions. In AMO settings, a second strategy can be pursued, where simple interaction patterns are combined with exotic lattice geometries to induce emergent gauge invariance. This way, the complication of realizing fine-tuned pattern is transferred to a complicated lattice geometry, which might be realized provided the correspondent light-pattern is realizable. Below, we briefly summarize the key features of this second strategy, and discuss a specific application thereof. 

The first ingredient is {\it simple} interactions, which are in general repulsive and have an isotropic plateau-like structure. Within the context of Rydberg atoms, they are usually present in two cases. The first one are ground states atoms coupled via a two-photon transition to an $s$-state, whose interactions are isotropic in the full 3D space up to corrections of few percent~\cite{gallagher2005rydberg,Saffman:2010ky,Comparat:2010cb,Low:2012ct,Walker:2008bm,Walker:2005kj,Reinhard:2007hm,Singer:2005ct}. The corresponding frozen regimes have already been accessed in a series of experiments ~\cite{Heidemann:2008bg,Anderson:1998hn, Hofmann:2013gm, Maxwell:2013km,Schwarzkopf:2013fx,Ebert:2013tr,Valado:2013ez,Schempp:2014jm,Barredo:2014tb,Tauschinsky:2010ep, Carter:2013vi,Kubler:2010dw,Carr:2013cn,Dudin:2012hm, Schauss:2012hh,Li:2013gg}. A second way of realizing isotropic interaction is to consider atoms trapped on a 2D plane, and dressed to a p-state polarized along the direction orthogonal to the plane itself. This way, differently with respect to the case discussed above, the in-plane interaction is basically isotropic - virtual exchange of photons within the Rydberg manifold is always allowed. 

The second key ingredient is a set of {\it complicated} lattices where the aforementioned interactions, once considered as a {\it sharp} plateau ones, are sufficient to define a classical limit where there is a set of degenerate classical ground states, increasing extensively with system size. Given the shape of the interactions, one can identify the possible lattices as follows. First, we define as $b$ the largest distance between sites belonging to the same simplices~\cite{lacroix2011introduction} (or {\it gauge cell}), that is, the unit cell where the Gauss law of interest is defined (in the spin ice case, these are the squares with crosses). Secondly, we define as $c$ the smallest distance between sites which do not belong to the same gauge cell. It is then clear that, in the case where $b < c$, a plateau interaction of range $b < r_{bc} < c$ can generates the desired constraint in each gauge cell. In case this is not true (like, e.g., in the square ice case), additional features are needed such as angular dependence.  

\begin{figure}[b]  
\centering 
\includegraphics[width= 0.47\columnwidth]{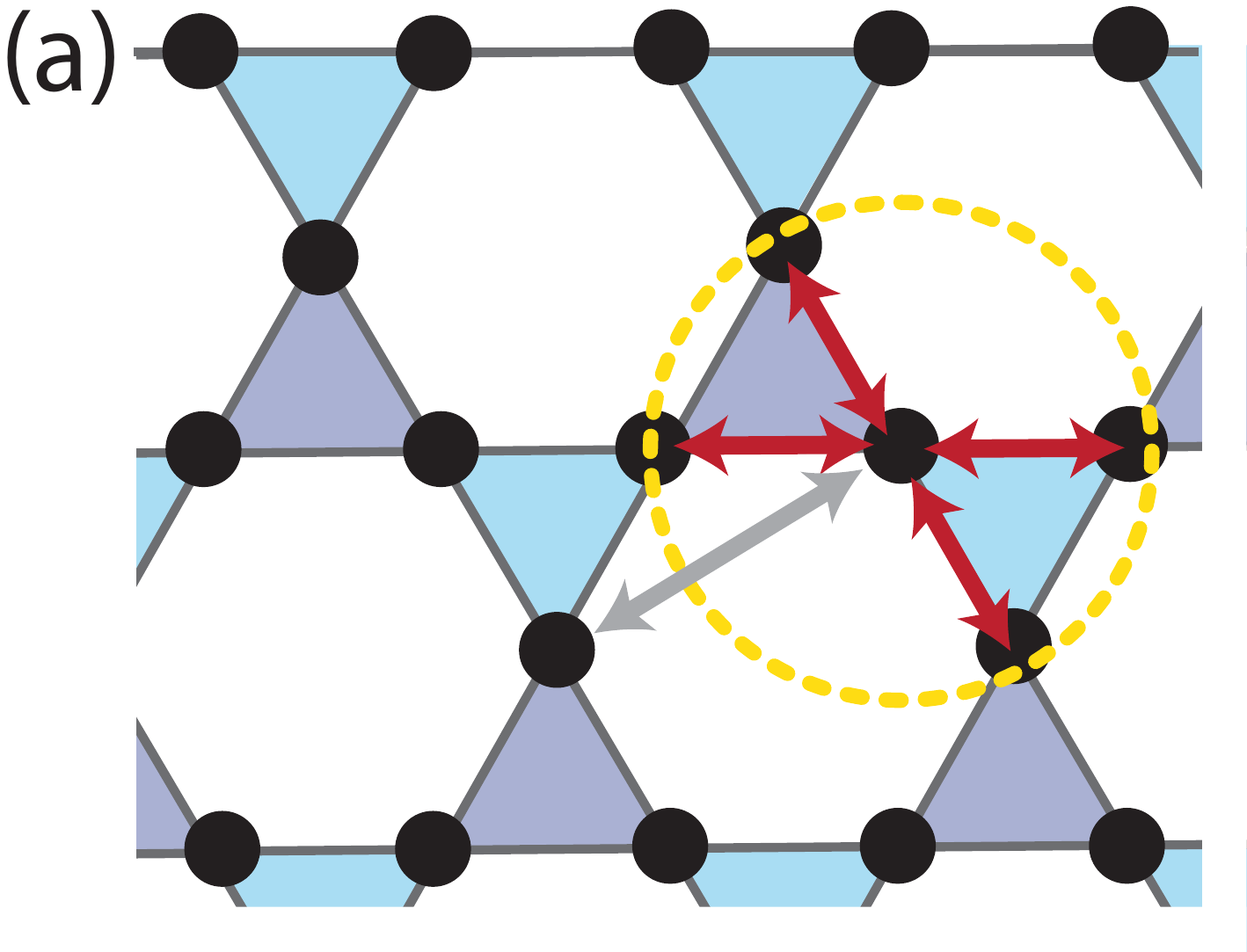}
\includegraphics[width= 0.47\columnwidth]{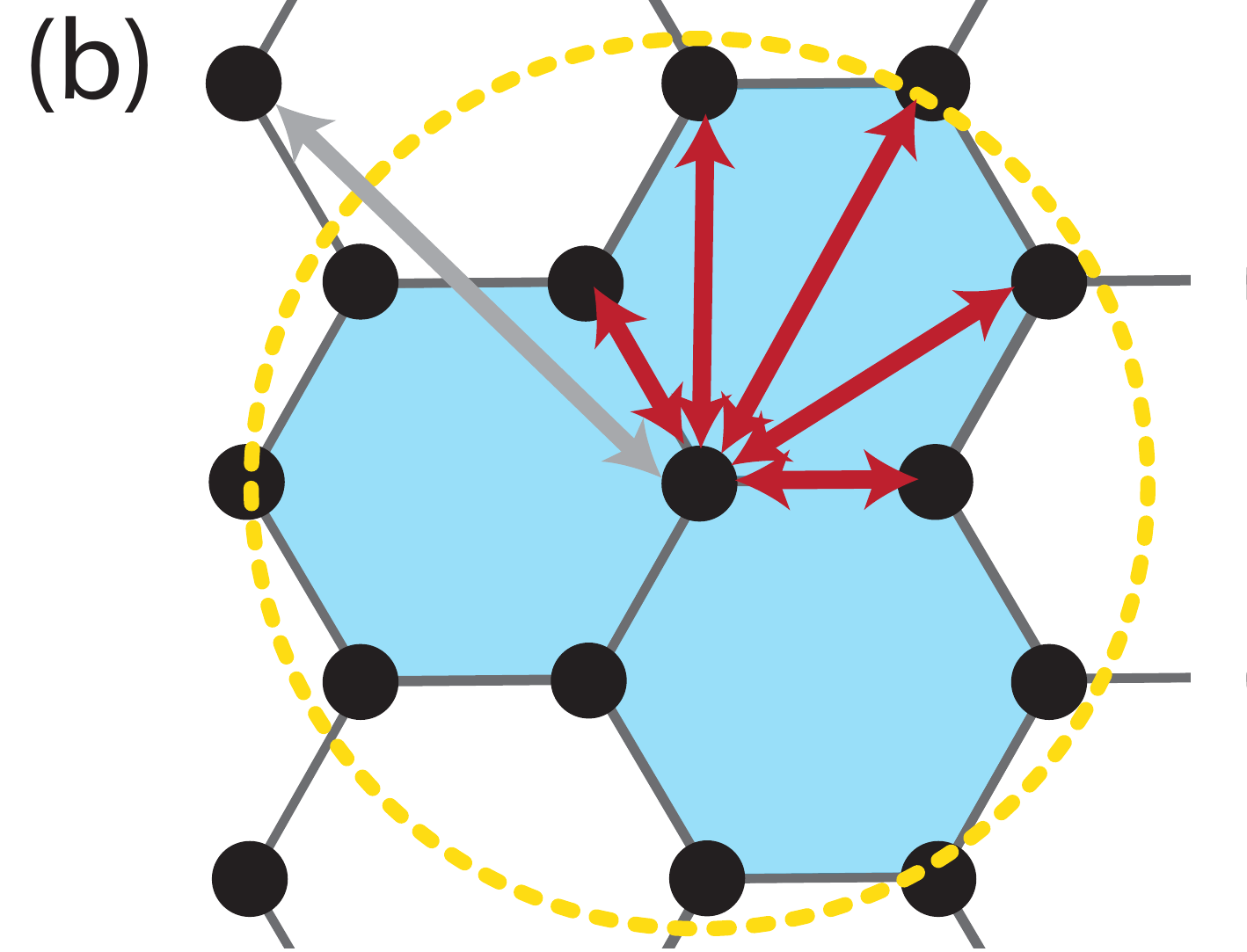}
\caption{\small{Lattices possessing the properties discussed in Sec.~\ref{sec:routes}: Panel (a) shows a kagome lattice with triangles as gauge cells (shaded area) and panel (b) a honeycomb lattice with hexagons as gauge cells. In both cases, the maximal intraplaquette euclidean distance (yellow dashed circle) is smaller than the minimal interplaquette distance (gray arrow). The radius of the needed plateau-like interaction is described by the yellow circles. } }
\label{fig:SecV_lattice} 
\end{figure}

Some examples of the lattices which satisfy the previous property are illustrated in Fig.~\ref{fig:SecV_lattice}, together with the corresponding gauge cells. The list includes several 2D lattices which have already been realized in AMO settings, such as the Kagome lattice with triangular gauge cells (but not with hexagonal gauge cells), the Ruby lattice and the Honeycomb lattice. In general, various lattices can be constructed satisfying the previous properties taking triangular cells as gauge cells. In the 3D case, a simple example is the pyrochlore lattice, already discussed in Ref.~\cite{Tewari:2006js} in the context of polar gases. Interestingly, in this former case, dipolar interactions behave in a very similar manner to simple plateau-like ones due to their symmetry content~\cite{Isakov2005}. 

Once taken at the proper filling factor for the underlying Bose-Hubbard Hamiltonian, all of those lattices generate in perturbation theory quantum dimer or quantum loop models~\cite{lacroix2011introduction}. These are naturally described by emergent gauge theories: however, the gauge symmetry itself is not always straightforwardly determined given the gauge symmetry of the microscopic constituents. While in our cases the latter is always of $U(1)$ type (the number of bosons in each gauge cell is preserved along the dynamics), the underlying gauge symmetries can be discrete. This is the case, e.g., of the Kagome Bose-Hubbard model with the hexagonal cell as gauge cell, where the proper low-energy theory is a $\mathbb{Z}_2$ gauge theory which can undergo deconfinement, and thus stabilize a spin-liquid phase.  

The procedure to derive the proper dimer model dynamics given a lattice and an Ising constraint is outlined in Ref.~\cite{lacroix2011introduction}. Below, we illustrate a simple example of how complicated lattices can meet simple interactions to let a quantum dimer model emerge by focusing on the concrete example of the 4-8 lattice.

\begin{figure*}[tb]  
\centering 
\includegraphics[width= \textwidth]{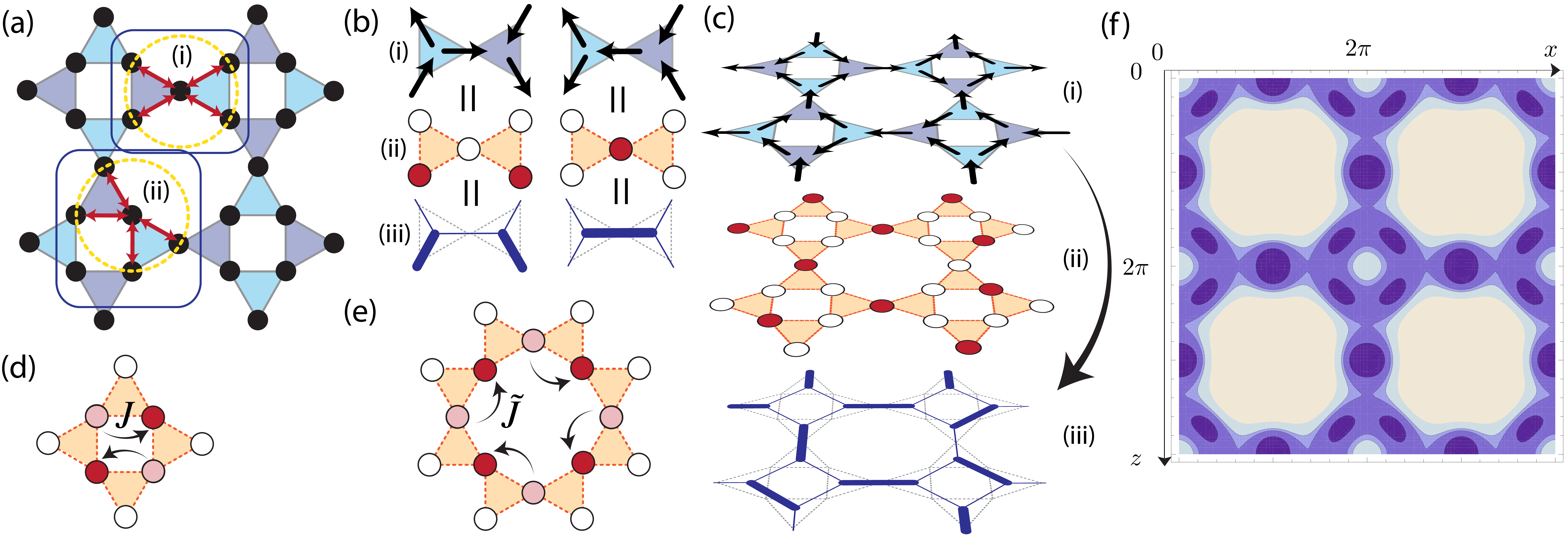}
\caption{\small{Configuration space of the squagome lattice and gauge invariant dynamics: (a) Atoms are disposed on a squagome pattern (filled-black circles) and interact with NN atoms (red arrows) which all are the same distance apart (yellow dashed circle). Triangular gauge cell are indicated as shaded areas in light blue and violet in an alternating pattern. Panel (b) describes possible gauge invariant configurations [from Eq.~\eqref{eq:G48}], where a configuration with (i) two flux vectors pointing outwards (inwards) on a light blue (violet) gauge cell map onto a configuration (ii)  where a single site on each triangle is occupied (red circle) and two lattice sites are unoccupied (white circles). Here, an arrow pointing from a violet triangle to a light blue triangle corresponds to an occupied lattice site and {\it vice versa} to an empty lattice site. (iii) This can be further mapped to a quantum dimer model on the 4-8 lattice. The new sites are defined at the centre of each triangle: the bond variable between them is either empty (thin blue line) or a dimer (thick blue line), depending on the original occupation of the site shared by the triangles. Panel (c) illustrates a full gauge invariant configuration. Quantum fluctuations induce non-trivial dynamics around both the square (d) and octagonal plaquettes (e), described by Eqs.~\eqref{eq:Hsquare} and~\eqref{eq:Hoctagon}. Panel (f) shows the optical lattice pattern of Eq.~\eqref{eq:48lasers} as described in the text. Darker areas correspond to deeper potentials.} }
\label{fig:4_8_slides} 
\end{figure*}

\subsection{Emergent quantum dimer dynamics on a 4-8 lattice from an XXZ model}

The 4-8 lattice (also known as CAVO lattice)~\cite{48lattice} 
represents a useful example to illustrate how the combination of a complicated lattice with simple Ising interactions can lead to intriguing quantum dynamics. The lattice structure for the underlying bosons we start from is the squagome lattice~\cite{squagome1,*squagome2,ioannissquagome}, illustrated in Fig.~\ref{fig:4_8_slides}: once the triangles are identified as the gauge cells, it is easy to see that $b = a$, $c=\sqrt{2}a$, so that a plateau interaction of range $ 1<r/a< \sqrt{2}$ can indeed enforce constraints on the gauge cells. Since each site is shared by 3 gauge cells, a filling fraction of $n=1/3$ atoms per site, combined with the plateau interactions, will generate a degenerate manifold $\mathcal{H}_{4-8}$ of classical ground states where for each triangle a single site is occupied [see Fig.\ref{fig:4_8_slides}(b)].

When formulated in spin language with $S^z_j = n_j-1/2$, the Hamiltonian
\begin{equation}
H^{Is}_{4-8} = J_z \sum_{\triangle}\sum_{\{i, j\}\in\triangle}S^z_j  S^z_i 
\end{equation}
has (trivially) a set of $U(1)$-like conserved charges at each triangle, that is:
\begin{equation}\label{eq:G48}
G_\triangle = \sum_{j\in\triangle}S^z_j + 1/2, \quad G_\triangle |\psi\rangle = 0 \qquad \forall |\psi\rangle \in \mathcal{H}_{4-8}.
\end{equation}
Once an additional, small term inducing quantum fluctuations is introduced:
\begin{equation}
H^{Ex}_{4-8} = J \sum_{\square}\sum_{\{i, j\}\in\square}S^+_j  S^-_i + \tilde{J} \sum_{\octagon}\sum_{\{i, j\}\in\octagon}S^+_j  S^-_i 
\end{equation}
tunneling between the different classical degenerate minima becomes possible within perturbation theory, while still preserving the set of conserved charges in Eq.~\eqref{eq:G48}. Notice that we used two different matrix elements for particle tunneling around the squares ($J$) and around the octagons ($\tilde{J}$). Two kind of moves are allowed. At second order, two particles sitting along a diagonal of a square plaquette can resonantly flip to sit on the other diagonal:
\begin{equation}
H^\square \simeq -\frac{J^2}{J_z} (S^+_1S^-_2S^+_3S^-_4 + \textrm{h.c.} )
\label{eq:Hsquare}
\end{equation}
where we have numbered the sites of the square plaquette in clockwise order. The next non-vanishing contribution takes place at fourth order, where particles sitting at the edges of each octagonal plaquette can re-arrange via an extended ring-exchange:
\begin{equation}
H^{\octagon} \simeq -3\frac{\tilde{J}^4}{J_z^3} (S^+_1S^-_2S^+_3S^-_4 S^+_5S^-_6S^+_7S^-_8 + \textrm{h.c.} ).
\label{eq:Hoctagon}
\end{equation}
where we have numbered the sites of the octagonal plaquette in clockwise order. The two terms are illustrated in Fig.~\ref{fig:4_8_slides}(d)-(e). We now reformulate the problem in terms of dimer models, which allows to set up a proper description in terms of effective degrees of freedom. In order to do that, we follow the procedure exemplified in Ref.~\cite{lacroix2011introduction,map1,map2a,map3}, and illustrated in Fig.~\ref{fig:4_8_slides}(b)-(c): we define a new lattice, the so called simplex lattice, whose vertices are the middle points of each gauge cell, and whose bonds connect vertices of gauge cells which share a single site: each bond sits on a vertex of the original lattice. Then, we introduce dimer variables on the bonds as follows: {\it i)} if a bond sits on a site which is occupied by a boson, we draw a dimer, {\it ii)} if not, we leave the bond empty. This way, the Gauss law of Eq.~\eqref{eq:G48} is easily reformulated as a conservation law of a single dimer at each vertex. 

The lattice on the top of which the quantum dimer model is defined is then a 4-8 lattice: as it is bipartite, the corresponding low-energy theory is a $U(1)$ gauge theory, which can then display different confined phases as a function of the two kinetic energy terms for the dimers $H^{\octagon} $ and $H^{\square} $. This setup might constitute then a perfect setting for the investigation of the competition between different RVB solid orders and the transitions between them. The corresponding periodic structure can be either realized using digital-micromirror-devices (DMD)~\cite{Fukuhara:2013hq}, or by using an optical potential of the form 
\begin{equation}
\begin{split}
V(x,y)=4 V_1(x+y,x-y)+ V_2(x,y),
\label{eq:48lasers}
\end{split}
\end{equation}
where
\begin{equation}
\begin{split}
V_1(x,y)=\cos(\pi x)^2+\cos(\pi y)^2-2\cos(0.55)\cos(\pi x)\cos(\pi y)
\end{split}
\end{equation}
is a 2D lattice created by two 1D standing waves with phase difference $\phi=0.55$ and anti-parallel polarizations $\mathbf{e}_1\cdot \mathbf{e}_2=-1$. The second 2D lattice is created by lasers with three times the frequency and orthogonal polarization,
\begin{equation}
\begin{split}
V_2(x,y)=\cos(3 \pi x)^2+\cos(3\pi y)^2.
\end{split}
\end{equation}
Both lattices are rotated by 45 degrees, respectively. 
The full lattice structure is illustrated in Fig.~\ref{fig:4_8_slides}(f), and realizes the squagome lattice potential of interest.

\section{Conclusions and Outlook\label{sec:conc}}

In summary, we have shown how dynamical gauge fields emerging from frustration can be ideally realized in cold atom systems by employing optical lattices combined with Rydberg interactions, allowing to probe gauge theory phenomena in  a variety of models. In particular, we analyzed in detail the case of quantum square ice, a paradigmatic example of frustrated statistical mechanics, both at the few- and at the many-body level. 

From the atomic physics side, the key element of our implementation is represented by the tunable interaction pattern generated by Rydberg $p$-states. Here, prominent atomic physics features can be exploited in order to generate (repulsive) anisotropic interactions that allow to enforce the complex gauge constraints of square ice models. The possibility of generating such anisotropic interaction patterns enriches the cold atom Hubbard toolbox of another potential feature, which can find different applications in many-body physics even beyond engineering complicated and fine tuned lattice constraints. A straightforward extension could be the realization of $U(1)$ discrete gauge theories on a cubic lattice, where the Polyakov argument is then circumvented and thus a stable deconfined phase, a gapless spin liquid, can be stabilized. Moreover, the non-trivial interactions within the Rydberg manifold can also be used to construct directly the spin model of interest, thus ensuring larger energy scales with respect to real tunneling dynamics.

From the many-body side, we have provided numerical evidence that typical imperfections generated by the Rydberg interactions still allow the observation of a non-trivial state of matter, a plaquette valence bond crystal. Moreover, we have shown how a cold atom suited detection technique can be identified, by performing parity measurements along the plaquettes, which directly identifies the spontaneous symmetry breaking of a discrete lattice symmetry. 
An additional point is that, even in the absence of quantum dynamics, the engineered interactions 
stabilise a magnetically ordered state with a large unit cell at low temperature, which gives way to a classical Coulomb gas, a marginally confining two-dimensional Coulomb phase with a small but nonzero density of charges in the form of thermally activated plaquettes violating the ice rule.

The advanced interaction engineering available within the Rydberg toolbox paves the way toward the realization of different constrained dynamics even beyond quantum ice. In the last section, we discussed some 2D examples of quantum dimer and quantum loop models (quantum link models) that can be realized by combining isotropic plateau-like interactions with exotic lattices. This different route, which replaces the complex interaction pattern of square ice with a complex optical lattice setting, represents an interesting, complementary approach for the microscopic realization of gauge theories, which can benefit from the recent developments of in situ imaging and digital-mirror-device optical lattice techniques. 

Different directions can be pursued further following the lines discussed here. A first, interesting extension would be to understand whether different kinds of anisotropic interactions can play a significant role in engineered Ising constraints in cold atom systems. In particular, anisotropic interactions between Rydberg $d$-states of $^{87}$Rb atoms have been recently demonstrated in Ref.~\cite{Barredo:2014tb}: as their angular dependence differs from the one discussed here, it can constitute yet another tool in order to realize complicated, fine-tuned interaction patterns. Secondly, the present proposal, which generates pure gauge theories, can be combined in a modular way with previous ones~\cite{Banerjee:2012cl} in such a way that either fermionic or bosonic matter can be included into the dynamics. Microscopically, this would require an additional fermionic (bosonic) species to be trapped onto a lattice whose minima sits at the centre of each vertex. The combination of gauge fields and dynamical matter fields will then allow to investigate scenarios such as 2D quantum electrodynamics in a quantum link formulation in the fermionic case, or the Fradkin-Shenker scenario of Higgs physics in the bosonic one. Finally, cold atom realizations can also provide a suitable platform for the investigation of dynamical effects in quantum dimer models and gauge theories in general; it'd be interesting to see whether simple observables and experimental procedures can be implemented, to described complex many-body phenomena such as string dynamics~\cite{R17} in the presence of static charges~\cite{Banerjee:2013fl}, or the dynamical properties of thermally activated monopoles on top of a vacuum state.

\begin{acknowledgments}
We thank T. Pfau for stimulating discussions in the initial state of this work. We also thank I. Bloch, Ch. Gro{\ss}, M. Hennrich, A. L\"auchli, E. Rico and F. Schreck for helpful discussions. Furthermore, discussions with all members of the R-ION and UQUAM consortium are kindly acknowledged. This project was supported in parts by the ERC Synergy Grant UQUAM, SIQS, the SFB FoQuS (FWF Project No. F4006-N16), and the ERA-NET CHIST-ERA (R-ION consortium). R.M. acknowledges the Helmholtz Virtual Institute "New States of Matter and Their Excitations''.
\end{acknowledgments}

\appendix

\section{Effect of the AC-Stark lasers on the ground state}\label{app:aclattice}
\begin{figure}[tb]  
\centering 
\includegraphics[width= .9\columnwidth]{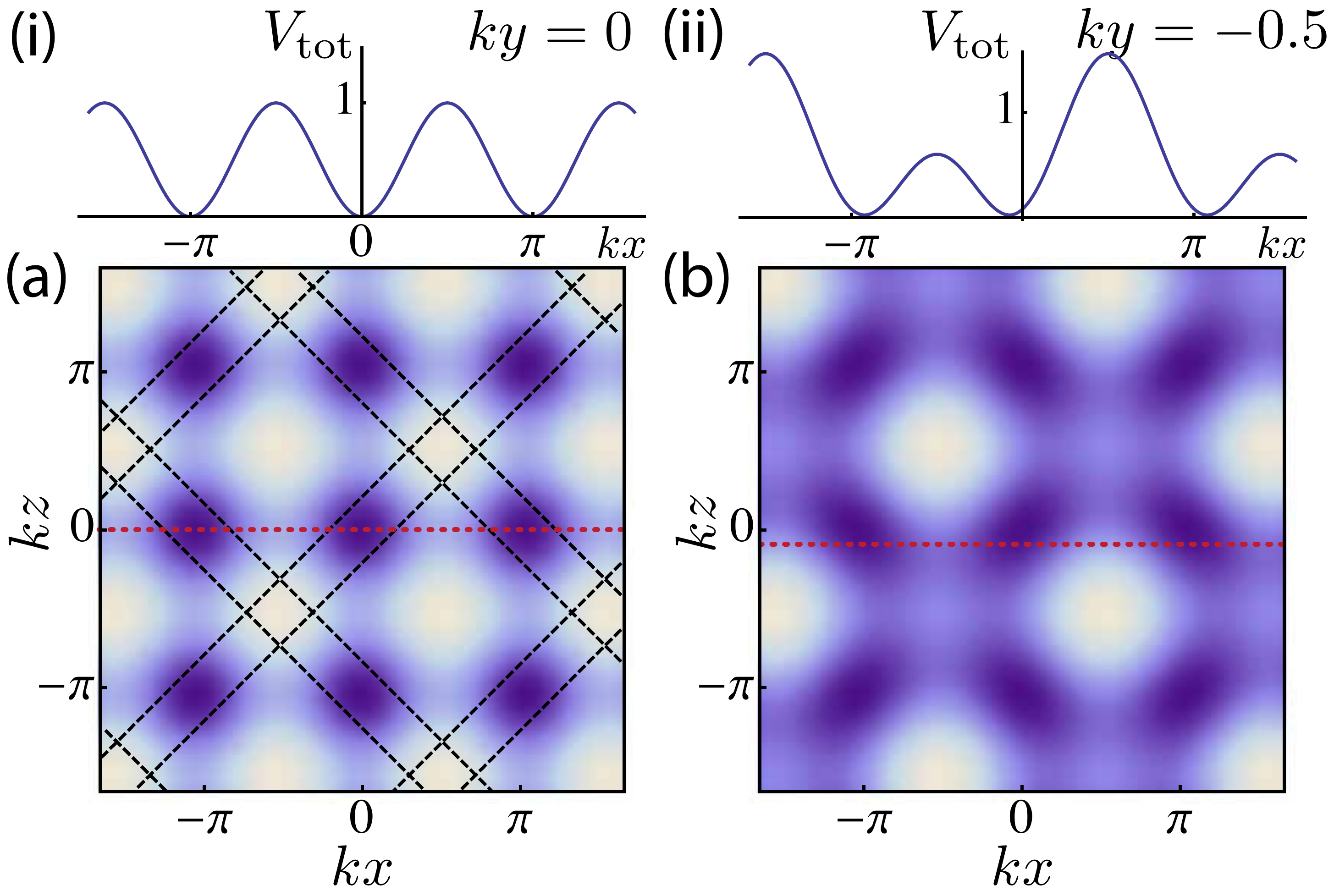}
\caption{\small{Contour plots of the total trapping potential, $V_{\rm tot}(x,z)$. (a)~without the AC Stark potential ($\alpha=0$) and (b)~with the AC Stark potential ($\alpha=1$). Black dashed lines in (a) show the 0.9 level lines of the AC Stark potential, $V_{AC}$. The insets (i) and (ii) show the 1D potential along the red dotted lines for (i) $ky=0$ and  (ii) $ky=-0.5$. }}
\label{fig:aclattice} 
\end{figure}

The AC Stark lasers introduced in Sec.~\ref{sec:single} will create an additional trapping potential, $V_{AC}(\mathbf{r}_i)|g\rangle\langle g|_i$, for ground state atoms with minima not commensurate with the initial trapping lattice. In order to not distort the desired lattice structure this additional potential must not be larger than the initial trapping potential. The dominant effect comes from a second order Stark effect by off-resonantelly coupling the $5S$ state to the first excited state $5P$ and is given by \mbox{$V_{AC}=\Omega_{5s5p}^2/(2\Delta_{5s5p})$} with Rabi frequency $\Omega_{5s5p}=2 d_{5s5p}\mathcal{E}/\hbar$ and detuning $\Delta_{5s5p}=2\pi c(\lambda_{5s5p}^{-1}-\lambda_{AC}^{-1})$. Here, $d_{5s5p}=\langle 5S|d|5P\rangle$ is the transition dipole matrix element and $\lambda_{5s5p}$ the transition wavelength. Fig.~\ref{fig:aclattice}(a) shows the desired trapping lattice created by two counter propagating laser beams which form a ground state potential \mbox{$V_{\rm trap}(z,x)=\cos^2kz+\cos^2kx$}. The dashed black lines indicate the 0.9 level lines of the AC-Stark potential \mbox{$V_{AC}(z,x)=\cos^2[k_{AC}(x-y)/\sqrt{2}]+\sin^2[k_{AC}(x+y)/\sqrt{2}]$} with $k_{AC}=k/\sqrt{2}$. The maxima are localized at the $\medbullet$ and $\blacksquare$ lattice sites, respectively, as required in Sec.~\ref{sec:single}. Fig.~\ref{fig:aclattice}(b) shows the total potential, \mbox{$V_{\rm tot}=V_{\rm trap}+\alpha V_{AC}$}, in the case of equal strength, i.e. $\alpha=1$. The insets on top show the 1D potential along the (i) $ky=0$ and (ii) $ky=-0.5$ lines [red dotted lines in Fig.~\ref{fig:aclattice}(a) and (b)]. In the case of equal strength of the trapping lattice and additional lattice created by the AC-Stark lasers the potential minima are still located at the same position, but slightly elongated. Note that the potential barrier between neighboring lattice sites is about 1/2 smaller than without the additional AC Stark laser. This will lead to higher tunneling rates compared to the case without the AC Stark laser.

\section{Global Rydberg laser excitation}\label{app:laser}
In the following we show that it is possible to weakly admix the locally polarized Rydberg states of Sect.~\ref{sec:single} to the electronic ground state $|g\rangle$ using a {\it single} laser with a wave vector $\mathbf{k}\sim \mathbf{y}$ and polarization $\sigma_+$ (see Fig.~\ref{fig:tensor})
\begin{equation}
\begin{split}
H_L=\frac{\Omega_R}{2}\left[|g\rangle_y\;_y\langle n{}^2P_{3/2},3/2|+{\rm h.c.}\right].
\end{split}
\end{equation}
\begin{widetext}
In the local $x$- and $z$- basis this laser will couple to all four $m_j$-levels with different weights, i.e.
\begin{align}
&|{\textstyle\frac{3}{2}}\rangle_y=\frac{1}{2\sqrt{2}} \left[|{\textstyle\frac{3}{2}}\rangle_z+i\sqrt{3}|{\textstyle\frac{1}{2}}\rangle_z-\sqrt{3}|{\textstyle-\frac{1}{2}}\rangle_z-i|{\textstyle-\frac{3}{2}}\rangle_z\right],\\
&|{\textstyle\frac{3}{2}}\rangle_y=\frac{1}{2\sqrt{2}}\left[
e^{-3i\pi/4}|{\textstyle\frac{3}{2}}\rangle_x+\sqrt{3} e^{-i\pi/4}|{\textstyle\frac{1}{2}}\rangle_x+\sqrt{3} e^{i\pi/4}|{\textstyle-\frac{1}{2}}\rangle_x+e^{3i\pi/4}|{\textstyle-\frac{3}{2}}\rangle_x\right].
\end{align}
where we used the irreducible representation of a rotation in the $j=3/2$ subspace, $D[\mathcal{R}(\alpha,\beta,\gamma)]=e^{-i\alpha J_z}e^{-i\beta J_y}e^{-i\gamma J_z}$, with matrix elements
\begin{equation*}
d_{m,m'}^{(3/2)}(\beta)=\left(
\begin{array}{cccc}
\cos\left(\frac{\beta }{2}\right)^3 &
-\sqrt{3} \cos\left(\frac{\beta }{2}\right)^2 \sin\left(\frac{\beta }{2}\right) &
\sqrt{3} \cos\left(\frac{\beta }{2}\right) \sin\left(\frac{\beta }{2}\right)^2 &
-\sin\left(\frac{\beta }{2}\right)^3\\
\sqrt{3} \cos\left(\frac{\beta }{2}\right)^2 \sin\left(\frac{\beta }{2}\right) &
\frac{1}{2} \cos\left(\frac{\beta }{2}\right) (3 \cos\left(\beta\right)-1) &
-\frac{1}{2} (1+3 \cos\left(\beta\right)) \sin\left(\frac{\beta }{2}\right) &
\sqrt{3} \cos\left(\frac{\beta }{2}\right) \sin\left(\frac{\beta }{2}\right)^2\\
\sqrt{3} \cos\left(\frac{\beta }{2}\right) \sin\left(\frac{\beta }{2}\right)^2 &
\frac{1}{2} (1+3 \cos\left(\beta\right)) \sin\left(\frac{\beta }{2}\right) &
\frac{1}{2} \cos\left(\frac{\beta }{2}\right) (3 \cos\left(\beta\right)-1) &
-\sqrt{3} \cos\left(\frac{\beta }{2}\right)^2 \sin\left(\frac{\beta }{2}\right)\\
\sin\left(\frac{\beta }{2}\right)^3 &
\sqrt{3} \cos\left(\frac{\beta }{2}\right) \sin\left(\frac{\beta }{2}\right)^2 &
\sqrt{3} \cos\left(\frac{\beta }{2}\right)^2 \sin\left(\frac{\beta }{2}\right) &
\cos\left(\frac{\beta }{2}\right)^3
\end{array}
\right).
\end{equation*}
\end{widetext}
Since the states $|m\neq 3/2\rangle_{z,x}$ are energetically separated by at least $E_{\rm AC}$ from the $|m=3/2\rangle$ state a laser with detuning $\Delta_R\ll E_{\rm AC}$ and wave vector $\mathbf{k}\sim \mathbf{y}$ will selectively admix the states $|3/2\rangle_z$ and $|3/2\rangle_z$ at lattice sites $\medbullet$ and $\blacksquare$, respectively, to the ground state $|g\rangle$ with an effective Rabi frequency $\Omega_R/(2\sqrt{2})$.

\section{Van der Waals interactions}\label{app:vdw}
In this appendix we briefly summarize the technical details in order to calculate the angular dependent van der Waals interactions of Sec.~\ref{sec:pstates}. Due to the odd parity of the electric dipole operators $d^{(i)}_{\mu}$ and $d^{(j)}_{\nu}$, the dipole-dipole interaction, $V_{\rm dd}$, of Eq.~\eqref{eq:dipdip} can only couple  states with initial angular (total) momentum $\ell$ ($j$) to states with new angular (total) momentum $\ell\pm1$ ($j$ or $j\pm 1$). Therefore, the number of possible ``channels'' $n\ell j m_1 +n\ell j m_2\longrightarrow n'\ell' j' m' +n''\ell'' j'' m''$ for which the matrix element $\langle n\ell j m_1;n\ell j m_2|V_{\rm dd}^{(ij)}|n'\ell' j' m' ;n''\ell'' j'' m''\rangle$ is non-zero are limited.
While there is no selection rule for possible final principal quantum numbers $n'$ and $n''$ which solely determine the overall strength of the matrix element, the dipole-dipole matrix element is only non-zero if the magnetic quantum numbers and the spherical component of the dipole operator fulfill $m_1+\mu=m'$ and $m_2+\nu=m''$. If the energy difference 
$
\delta_{\alpha\beta}=E(\alpha)+E(\beta)-2E(n \ell j),
$
between the initial states $n\ell j$ and the intermediate states $\alpha\equiv n_\alpha \ell_\alpha j_\alpha m_\alpha$ and $\beta\equiv n_\beta \ell_\beta j_\beta m_\beta$ of the atoms is larger than the dipole-dipole matrix element connecting those states the dominant interaction is of van der Waals type which arises from $V_{\rm dd}$ in second order perturbation
\begin{equation}
\begin{split}
\hat V_{\rm vdW}=\hat P_{12}\sum_{\alpha\beta}\frac{\hat V_{\rm dd}\hat Q_{\alpha,\beta}\hat V_{\rm dd}}{\delta_{\alpha\beta}}\hat P_{34}.
\label{eq:vdwop}
\end{split}
\end{equation}
Here, $\hat V_{\rm vdW}$ is an operator acting in the degenerate manifold of magnetic sublevels with $\hat P_{ij}=| n\ell j m_i,n\ell j m_j\rangle\langle n\ell j m_i,n\ell j m_j|$ a projector into the $n\ell j$-manifold and $\hat Q_{\alpha,\beta}= |\alpha,\beta\rangle\langle \alpha,\beta|$ a projector on a specific state in the complementary space. The sum is over all two-atom energy levels, where the indices $\alpha\equiv n_\alpha \ell_\alpha j_\alpha m_\alpha$ and $\beta\equiv n_\beta \ell_\beta j_\beta m_\beta$ denote a full set of quantum numbers that specify the states. 
Due to the electric dipole selection rules discussed above this sum can be split up into channels denoted by $\nu=(\ell_\alpha,j_\alpha;\ell_\beta,j_\beta)$. Eq.~\eqref{eq:vdwop} can be written as $\hat V_{\rm vdW}=\sum_{\nu} C_6^{(\nu)} \mathcal{D}_\nu(\vartheta,\varphi)/r^6$, where $C_6^{(\nu)}$ contains the radial part of the matrix elements 
\begin{equation}
C_6^{(\nu)}=\sum_{n_\alpha,n_\beta}\frac{\mathcal{R}_1^\alpha\mathcal{R}_2^\beta\mathcal{R}_3^\alpha\mathcal{R}_4^\beta}{\delta_{\alpha\beta}}
\end{equation}
which accounts for the overall strength of the interaction and is independent of the magnetic quantum numbers. Here, 
$\mathcal{R}_{i}^{j}=\int dr r^2 \psi_{n_i,\ell_i,j_i}(r)^* r\,\psi_{n_j,\ell_j,j_j}(r)$ is the radial integral calculated with radial wave functions $\psi_{n_j,\ell_j,j_j}(r)$ obtained using the model potential from~\cite{Marinescu:1994cw}.
The matrix
\begin{equation}
\mathcal{D}_\nu(\vartheta,\varphi)=\hat P_{12}\sum_{m_\alpha,m_\beta}\mathcal{M}_\nu \hat Q_{\alpha,\beta}\mathcal{M}_\nu\, \hat P_{34}
\end{equation}
\begin{widetext}
on the other hand is a matrix in the subspace of magnetic quantum numbers which contains the relative angles between the two atoms ($s=1/2$)
\begin{equation}
\begin{split}
\langle m_1,m_2|\mathcal{M}_\nu|m_\alpha,m_\beta\rangle=&
(-)^{s-m_1}\sqrt{\prod_{i={1,\alpha}}(2\ell_i+1)(2j_i+1)}\ws{\ell_1,\ell_\alpha, 1,j_\alpha,j_1,s}\wt{\ell_\alpha,1,\ell_1,0,0,0}\\
\times&(-)^{s-m_2}\sqrt{{\prod_{i=2,\beta}}(2\ell_i+1)(2j_i+1)}\ws{\ell_2,\ell_\beta, 1,j_\beta,j_2,s}\wt{\ell_\beta,1,\ell_2,0,0,0}\\
\times&\left(-\sqrt{\frac{24 \pi}{5}} \sum_{\mu,\nu}C_{\mu,\nu;\mu+\nu}^{1,1;2} \wt{j_\alpha,1,j_1,m_\alpha,\mu,-m_1}\wt{j_\beta,1,j_2,m_\beta,\nu,-m_2} 
 Y_{2}^{\mu+\nu}(\vartheta,\varphi)^*\right).
\end{split}
\end{equation}
As an example we show the $\mathcal{D}_1$ matrix for the first channel $P_{3/2}+P_{3/2}\longrightarrow S_{1/2}+S_{1/2}$
\begin{equation}
\mathcal{D}_1=\left(
\begin{array}{cccc}
 \frac{1}{4} \sin ^4\vartheta & -\frac{1}{2 \sqrt{3}} \cos \vartheta  \sin ^3\vartheta & \frac{\sin ^2\vartheta}{24 \sqrt{3}} (3 \cos 2 \vartheta +1) & 0 \\
 -\frac{1}{2 \sqrt{3}}\cos \vartheta  \sin ^3\vartheta  & \frac{1}{12} \left(\sin ^4\vartheta +\sin ^22 \vartheta \right) & -\frac{1}{9} \cos\vartheta  \sin \vartheta  & \frac{\sin ^2\vartheta }{24 \sqrt{3}} (3 \cos 2 \vartheta +1) \\
 \frac{ \sin ^2\vartheta}{24 \sqrt{3}} (3 \cos 2 \vartheta +1)& -\frac{1}{9} \cos \vartheta  \sin \vartheta  & \frac{1}{864} (12 \cos 2 \vartheta -27 \cos 4 \vartheta +47) & -\frac{\sin 2 \vartheta }{24 \sqrt{3}} (3 \cos 2 \vartheta +1) \\
 0 & \frac{\sin ^2\vartheta}{24 \sqrt{3}} (3 \cos 2 \vartheta +1)  & -\frac{\sin 2 \vartheta }{24 \sqrt{3}} (3 \cos 2 \vartheta +1) & \frac{1}{144} (3 \cos 2 \vartheta +1)^2 \\
\end{array}
\right)
\end{equation}
\end{widetext}
in the subspace of states $|\frac32,\frac32\rangle$, $|\frac32,\frac12\rangle$, $|\frac32,-\frac12\rangle$ and $|\frac32,-\frac32\rangle$ where the first atom is fixed in the $m=3/2$ state.
In general one has to diagonalize the operator $\hat V_{\rm vdW}$ in the degenerate Zeeman subspace in order to obtain the new eigenenergies and eigenstates in the presence of interactions. If an external electric or magnetic field separates an initial two atom state $|m_1,m_2\rangle$ from all other Zeeman sublevel such that the energy difference is larger than the vdW coupling matrix elements then it is possible to simply take expectation values of $V_{m_1,m_2}^{(n)}(\mathbf{r})=\langle m_1,m_2|\hat V_{\rm vdW}|m_1,m_2\rangle$ in order to obtain the interaction potential of two atoms initially in the $|m_1,m_2\rangle$ state.

\section{Mixed interactions}\label{app:mixed}
\begin{figure}[tb]  
\centering 
\includegraphics[width= .9\columnwidth]{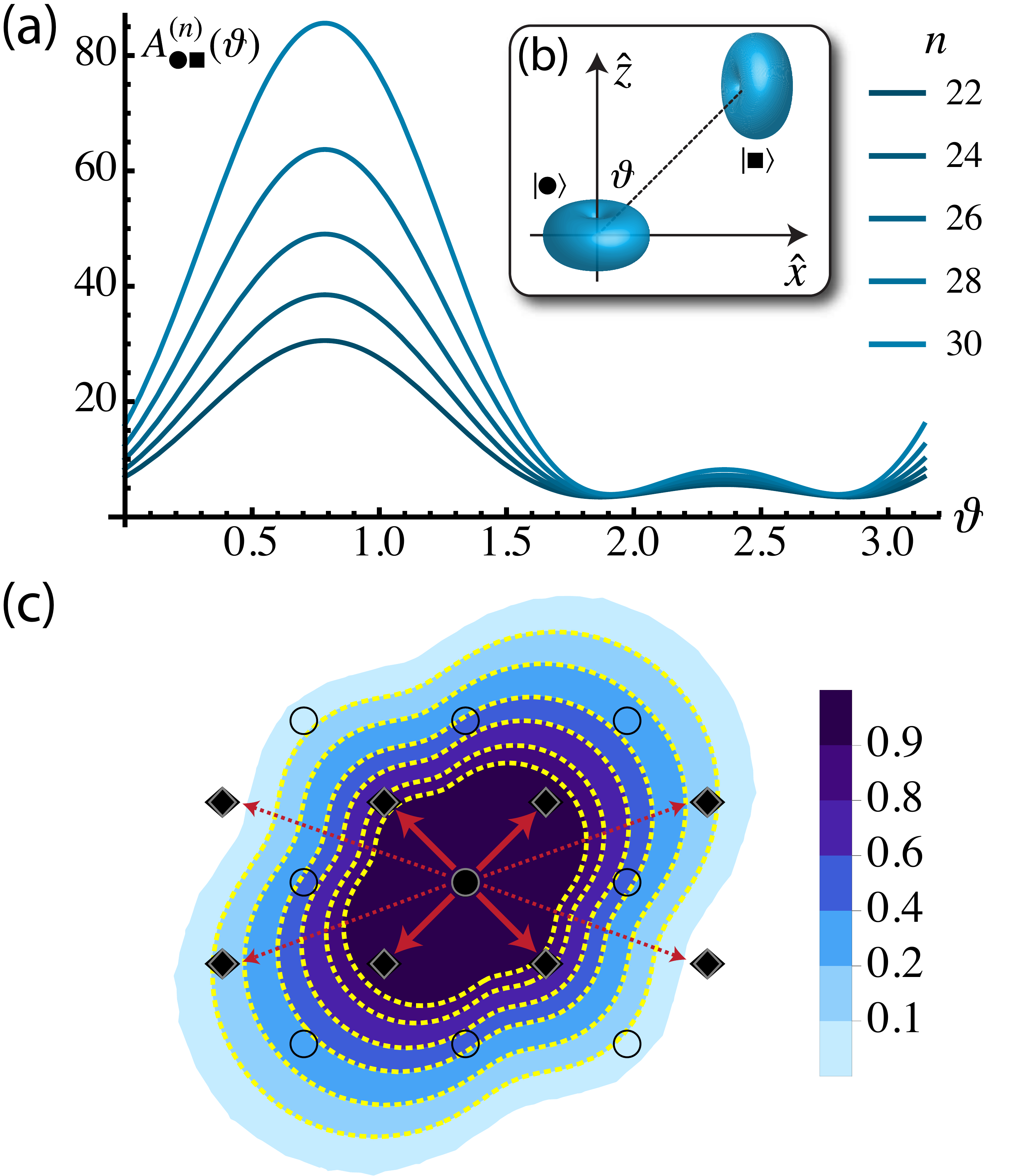}
\caption{\small{(a) Angular part, $A^{(n)}_{\medbullet\blacksquare}(\vartheta)$, of the van der Waals interaction, $V^{(n)}_{\medbullet\blacksquare}(r,\vartheta)=(n-\delta_{n\ell j})^{11} A^{(n)}_{\medbullet\blacksquare}(\vartheta)/r^6$, between a pair of $^{87}$Rb atoms in the  $|r_\medbullet\rangle=|n{}^2P_{3/2},3/2\rangle_z$ and $|r_\blacksquare\rangle=|n{}^2P_{3/2},3/2\rangle_x$ Rydberg states of $^{87}$Rb (solid lines). We plot the rescaled interaction energy,  $A^{(n)}_{\medbullet\blacksquare}(\vartheta)$ as a function of the angle $\vartheta$ for various values of the principal quantum number $n$, with $\delta$ the quantum defect. (b) Cartoon of the states and definition of the angle $\vartheta$. (c) Contour plot of the effective interaction $\tilde V_{\medbullet\blacksquare}^{(n)}(r,\vartheta)/\tilde V_0$ between the dressed ground state atom $|\medbullet\rangle$ in the middle and the NN $|\blacksquare\rangle$ atoms (red arrows) the NNN $|\blacksquare\rangle$ atoms (red dotted arrows). }}
\label{fig:int45} 
\end{figure}
In the following we show how to calculate the mixed interactions, $V_{\medbullet\blacksquare}(r,\vartheta)$, introduced in Sec.~\ref{sec:pstates} between the {\it locally} polarized Rydberg states  \mbox{$|\medbullet\rangle\equiv  |n{}^2P_{3/2}, 3/2\rangle_\mathbf{z}$} and \mbox{$|\blacksquare\rangle\equiv|n{}^2P_{3/2}, 3/2\rangle_\mathbf{x}$}. Here, the indices $z$ and $x$ denote the local quantization axis of the state. In the following we work in the $z$-basis. Rotating the latter state into the $z$ basis using the irreducible representation $D^{(3/2)}[\mathcal{R}(\mathbf{\hat y},\pi/2)]^{-1}$ of a rotation around $\mathbf{y}$ by an angle of $\pi/2$  in the $j=\frac{3}{2}$ space yields
\begin{equation}
|\blacksquare\rangle=\frac{1}{2\sqrt{2}}\left[|{\textstyle\frac{3}{2}}\rangle_{\mathbf{z}}-\sqrt{3}|{\textstyle\frac{1}{2}}\rangle_{\mathbf{z}}+\sqrt{3}|{\textstyle-\frac{1}{2}}\rangle_{\mathbf{ z}}-|{\textstyle-\frac{3}{2}}\rangle_{\mathbf{ z}}\right],
\label{eq:cm}
\end{equation}
where $|m\rangle_\mathbf{z}\equiv |n{}^2P_{3/2},m\rangle_\mathbf{z}$. The state $|\blacksquare\rangle=|n{}^2P_{3/2}, 3/2\rangle_\mathbf{x}=\sum_m c_m |n{}^2P_{3/2},m\rangle_\mathbf{z}$ is thus a superposition of different $m_j$-states in the $z$ basis. Interactions between two atoms in a $|\blacksquare\medbullet\rangle$ or $|\blacksquare\blacksquare\rangle$ state can be calculated by evaluating the corresponding matrix elements of Eq.~\eqref{eq:vdwop} which requires to compute van der Waals interactions between atoms in different $m_j$ states, e.g.
\begin{equation}
\begin{split}
\langle \medbullet \blacksquare| V_{\rm vdW} |\medbullet \blacksquare\rangle=\sum_{m,m'} c_{m'} c^*_m \langle{\textstyle \frac32},m | V_{\rm vdW} |{\textstyle \frac32},m'\rangle.
\end{split}
\end{equation}
The angular dependence of the van der Waals interaction between two Rydberg atoms in a $|\medbullet\medbullet\rangle$ or in a $|\blacksquare\blacksquare\rangle$ state, $V_{\medbullet\medbullet}(r,\vartheta)=V_{\blacksquare\blacksquare}(r,\vartheta-\pi/2)\sim \sin^4\vartheta/r^6$ are the same up to a rotation by 90 degrees and show the typical anisotropic behavior discussed in Sec.~\ref{sec:pstates} [see solid lines in Fig.~\ref{fig:pstateint}(a)] On the other hand, the angular dependence of the mixed interactions between two Rydberg atoms in a $|\blacksquare\medbullet\rangle$ state, shown in Fig.~\ref{fig:int45}(a), exhibits two asymmetric maxima at $\vartheta=\pm\pi/4$. The asymmetry arises from off-diagonal matrix elements, e.g. $\langle{\textstyle \frac{3}{2},\frac{1}{2}}|\hat V_{\rm vdW}|{\textstyle \frac{3}{2},-\frac{1}{2}}\rangle\sim \sin 2\vartheta$. Note that the actual strength of the interaction only affects the Condon radius, $r_c$ [see Eq.~\eqref{eq:rc}], but not the energy  shift $\tilde V_0$ [see Eq.~\eqref{eq:V0}] for $r\rightarrow 0$. Panel (c) of Fig.~\ref{fig:int45} shows a contour plot of the mixed interaction,$\tilde V_{\medbullet\blacksquare}/\tilde V_0$ of Eq.~\eqref{eq:dressedpot1}, between the dressed ground state atoms $|\medbullet\rangle$ in the middle and the surrounding $|\blacksquare\rangle$ atoms. Interactions with the neighboring $|\blacksquare\rangle$ atoms (red solid arrows) are strong, $\sim \tilde V_0$, while interactions with next-nearest-neighbor $|\blacksquare\rangle$ atoms (red dotted arrows) are strongly suppressed due to the plateau structure of the potential.

\section{Fourth order Brillouin-Wigner perturbation theory for N atoms}
\label{app:softcore}
In this appendix we review Brillouin-Wigner perturbation theory for N atoms in order to obtain the effective ground state potentials of Sec.~\ref{sec:softcore}. We are interested in finding the new ground state $|\tilde G\rangle$ of the Hamiltonian~\eqref{eq:Htot} which is adiabatically connected to $|G\rangle=\bigotimes_{i=1}^N|g\rangle_i$ for $\Omega\rightarrow 0$. \\
\\
For the following analysis it is convenient to split the dynamics in subspaces containing 0, 1, 2 etc. Rydberg excitations
\begin{equation}
\begin{split}
&\mathcal{H}_0=\{|g_1,\ldots,g_N\rangle\},\\
&\mathcal{H}_1=\{|g_1,\ldots,g_{i-1},r_i,g_{i+1},\ldots,g_N\rangle\},\\
&\mathcal{H}_2=\{|g_1,\ldots,g_{i-1},r_i,g_{i+1},\ldots ,g_{j-1},r_j,g_{j+1},\ldots,g_N\rangle\},\\
&\ldots\\
&\mathcal{H}_N=\{|r_1,\ldots,r_N\rangle\},
\end{split}
\end{equation}
with dimension $\text{dim } \mathcal{H}_n=\binom{N}{n}$. Only adjacent subspaces $\mathcal{H}_n$ and $\mathcal{H}_{n+1}$ are coupled via the single atom laser Hamiltonian proportional to $\Omega$. We can group the Hamiltonian into sectors containing only a specific number of Rydberg excitations
\begin{equation}
\begin{split}
H&=\left(\begin{array}{ccccc}
\mathbf{H}_0 & 0 & 0 & 0 &\\
0 & \mathbf{H}_1 & 0 & 0 & \\
0 & 0 & \mathbf{H}_2 & 0 & \\
0 & 0 & 0 & \mathbf{H}_3 & \\
 &  & &  & \ddots
\end{array}\right)+
\left(\begin{array}{ccccc}
0 & \mathbf{\Omega}_1 & 0 & 0  &\\
\mathbf{\Omega}_1^\dag & 0 & \mathbf{\Omega}_2 & 0 &\\
0 & \mathbf{\Omega}_2^\dag & 0 & \mathbf{\Omega}_3 & \\
0 & 0 & \mathbf{\Omega}_3^\dag &0 & \\
 &  & &  & \ddots
\end{array}\right)\\
&\equiv H_0+H_1.
\label{eq:block}
\end{split}
\end{equation}
We denote the matrices of size $(\text{dim } \mathcal{H}_n)^2$ on the diagonal with $\mathbf{H}_n$. They describe the dynamics in the subspace $\mathcal{H}_n$ with a fixed number of Rydberg states $n$, while the $\mathbf{\Omega}_n$ matrices of size $(\text{dim } \mathcal{H}_{n-1} \times \text{dim } \mathcal{H}_n)$ describe the coupling between adjacent sectors $n$ and $n-1$ due to the laser. Note, that only subspaces $\mathbf{H}_{n\geq 2}$ contain the interaction potentials $V_{ij}$ since we assume that ground and Rydberg states do not significantly interact. E.g. we find $\mathbf{H}_0=0$ and $\mathbf{H}_1=-\Delta 1_N$, where $1_N$ is the $N\times N$ identity matrix. Finally, we split the Hamiltonian into two parts, $H_0$ and $H_1$, accounting for the diagonal terms and the laser coupling, respectively.\\
\\
In the following we use Brillouin-Wigner perturbation theory to find the energy of the ground state of $H$ up to fourth order in $H_1/H_0$. Therefore, we define the projector $\mathbb{P}=|G\rangle\langle G|$ and its complement $\mathbb{Q}=1-\mathbb{P}$ for which $\mathbb{P} H_0 \mathbb{Q}=0$. Splitting the Hamiltonian in the corresponding subspaces  yields
\begin{equation}
\begin{split}
H_0
=\left(\begin{array}{c|cccc}
\mathbf{H}_0 & 0 & 0 & 0 & \\
\hline
0 & \mathbf{H}_1 & 0 & 0 & \\
0 & 0 & \mathbf{H}_2 & 0 & \\
0 & 0 & 0 & \mathbf{H}_3 & \\
&  & &  & \ddots
\end{array}\right)=
\left(\begin{array}{c|cccc}
\mathbb{P}H_0\mathbb{P} & &\mathbb{P}H_0\mathbb{Q} & \\
\hline
& &  &\\
\mathbb{Q}H_0\mathbb{P} & & \mathbb{Q}H_0\mathbb{Q} &\\
& & &
\end{array}\right)
\end{split}
\end{equation}
and 
\begin{equation}
\begin{split}
H_1
=\left(\begin{array}{c|cccc}
0 & \mathbf{\Omega}_1 & 0 & 0 &\\
\hline
\mathbf{\Omega}_1^\dag & 0 & \mathbf{\Omega}_2 & 0  &\\
0 & \mathbf{\Omega}_2^\dag & 0 & \mathbf{\Omega}_3 &\\
0 & 0 & \mathbf{\Omega}_3^\dag &0 & \\
&  & &  & \ddots
\end{array}\right)=
\left(\begin{array}{c|ccc}
\mathbb{P}H_1\mathbb{P} & &\mathbb{P}H_1\mathbb{Q} & \\
\hline
& &  &\\
\mathbb{Q}H_1\mathbb{P} & & \mathbb{Q}H_1\mathbb{Q} &\\
& & &
\end{array}\right)
\end{split}
\end{equation}
For the zero order term we find
$
E_{G}^{(0)}=\mathbb{P}H_0\mathbb{P}=\mathbf{H}_0=0.
$
The first order contribution is also zero, since
$
E_{G}^{(1)}=\mathbb{P}H_1\mathbb{P}=0.
$
In order to calculate higher order terms we first need to calculate the resolvent operator in the $\mathbb{Q}$ space
\begin{equation}
\begin{split}
R=\frac{1}{E_{G}^{(0)}-\mathbb{Q}H_0\mathbb{Q}}=-
\left(\begin{array}{ccccc}
\mathbf{H}_1^{-1} &0 &0 &\\
0& \mathbf{H}_2^{-1} &0 &\\
0&0&\mathbf{H}_3^{-1} & \\
 & & & \ddots\\
\end{array}\right)
\end{split}
\end{equation}
The second order contribution to the ground state energy is 
$
E_{G}^{(2)}=\mathbb{P}H_1\mathbb{Q}R\mathbb{Q}H_1\mathbb{P}.
$
Using 
$
\mathbb{P}H_1\mathbb{Q}=\left(\mathbf{\Omega}_1,0,0\right),
$
and $\mathbf{\Omega}_1=\Omega\mathbf{1}^T$, where $\mathbf{1}$ is a $n\times 1$ vector containing only 1's, we get
$
E_{G}^{(2)}=\mathbf{\Omega}_1(-\mathbf{H}_1^{-1})\mathbf{\Omega}_1^\dag=N\frac{\Omega^2}{\Delta}.
$
The second order contribution accounts for the single-particle light shift due to the laser, $E_{G}^{(2)}/N$. 
The third order contribution to the ground state energy vanishes
$
E_{G}^{(3)}=\mathbb{P}H_1\mathbb{Q}R\left(\mathbb{Q}H_1\mathbb{Q}-E_{G}^{(1)}\right)R\mathbb{Q}H_1\mathbb{P}=0.
$
For the fourth order contribution we obtain
\begin{equation}
\begin{split}
E_{G}^{(4)}&=\mathbb{P}H_1\mathbb{Q}R\left(\mathbb{Q}H_1\mathbb{Q}-E_{G}^{(1)}\right)R\left(\mathbb{Q}H_1\mathbb{Q}-E_{G}^{(1)}\right)R\mathbb{Q}H_1\mathbb{P}\\
&+\mathbb{P}H_1\mathbb{Q}R\left(-E_{G}^{(2)}\right)RQH_1\mathbb{P}.
\end{split}
\end{equation}
Using the matrix representation in the subspaces yields
\begin{equation}
\begin{split}
E_{G}^{(4)}=&-\mathbf{\Omega}_1\mathbf{H}_1^{-1}\mathbf{\Omega}_2\mathbf{H}_2^{-1}\mathbf{\Omega}_2^\dag\mathbf{H}_1^{-1}\mathbf{\Omega}_1^\dag\\
&+\mathbf{\Omega}_1\mathbf{H}_1^{-1}\mathbf{\Omega}_1\mathbf{H}_1^{-1}\mathbf{\Omega}_1^\dag\mathbf{H}_1^{-1}\mathbf{\Omega}_1^\dag.
\label{eq:fourth}
\end{split}
\end{equation}
The first line of the latter equation contains (for the first time) the interaction potential $V_{ij}$ which is present in all terms $\mathbf{H}_{n\geq 2}$. The last line can be simplified to $-N^2\frac{\Omega^4}{\Delta^3}$, while the explicit form of the first one depends on the number of particles.

In the following we will illustrate the explicit form of the ground state energy for $N=3$ particles:
\begin{widetext}
\begin{equation}
\begin{split}
H=\left(\begin{array}{c|ccc|ccc|c}
0& \Omega & \Omega & \Omega & 0& 0& 0& 0\\
\hline
\Omega & -\Delta & 0& 0& \Omega & \Omega & 0& 0\\
\Omega & 0& -\Delta & 0& \Omega & 0& \Omega & 0\\
\Omega & 0& 0& -\Delta & 0& \Omega & \Omega & 0\\
\hline
0& \Omega & \Omega & 0& V_{12}-2 \Delta & 0& 0& \Omega \\
0& \Omega & 0& \Omega & 0& V_{13}-2 \Delta & 0& \Omega \\
0& 0& \Omega & \Omega & 0& 0& V_{23}-2 \Delta & \Omega \\
\hline
0& 0& 0& 0& \Omega & \Omega & \Omega & V_{12}+V_{13}+V_{23}-3 \Delta
\end{array}\right),
\end{split}
\end{equation}
with lines indicating the blocks of Eq.~\eqref{eq:block}. Using Eq.~\eqref{eq:fourth} we obtain in fourth order a sum of binary interactions
\begin{equation}
\begin{split}
E_{ggg}^{(4)}=&-4\frac{\Omega^4}{\Delta^2}\left[\frac{1}{V_{12}-2\Delta}+\frac{1}{V_{13}-2\Delta}+\frac{1}{V_{23}-2\Delta}\right]-9\frac{\Omega^4}{\Delta^3}.
\end{split}
\end{equation}

\section{Finite-size clusters}\label{app:Clusters}
In our Exact Diagonalizations we have considered the following checkerboard clusters with periodic boundary conditions:
\be
\begin{array}{cccccccc}
N & \vec{T}_1& \vec{T}_2 &  \vec{G}_1 & \vec{G}_2 & \mc{D}_{\text{full}}^{S_z=0} & \mc{D}_{\text{spin-ice}} \\
\hline
16 & (2,2) & (-2,2) & (\frac{\pi}{2},\frac{\pi}{2}) & (-\frac{\pi}{2},\frac{\pi}{2}) & 12870 & 90 \\
32 & (4,0) & (0,4) & (\frac{\pi}{2},0) & (0,\frac{\pi}{2}) & 601080390 & 2970  \\
36 & (3,3) & (-3,3) & (\frac{\pi}{3},\frac{\pi}{3})  & (-\frac{\pi}{3},\frac{\pi}{3}) & 9075135300& 6840 \\
64 & (4,4) & (-4,4) & (\frac{\pi}{4},\frac{\pi}{4}) & (-\frac{\pi}{4},\frac{\pi}{4}) & 1832624140942590534 & 2891562 \\
72 & (6,0) & (0,6) & (\frac{\pi}{3},0) & (0,\frac{\pi}{3}) & 442512540276836779204 & 16448400 
\end{array}
\ee
\end{widetext}
where $N$ is the number of lattice sites, $\vec{T}_{1,2}$ are the spanning vectors of the cluster, $\vec{G}_{1,2}$ are the reciprocal vectors, $\mc{D}_{\text{full}}^{S_z\!=\!0} $ is the size of the full Hilbert space in the total magnetization $S_z=0$ sector, and $\mc{D}_{\text{spin-ice}}$ is the dimensionality of the spin ice manifold. 
Note that, in order to evaluate $J_{ij}$ across the periodic boundaries in a consistent manner~\footnote{It is not enough to choose the minimum distance as a criterion, since it may happen that the distance between $i$ and $j$ and e.g. that between $i$ and $j+\vec{T}_{1}$ is the same but the corresponding amplitudes are not the same.}, we keep the maximum amplitude among the set $\{J_{i,j+\epsilon_1 \vec{T}_1+\epsilon_2\vec{T}_2}, \epsilon_{1,2}\!=\!-1,0,1\}$, where $\vec{T}_1$ and $\vec{T}_2$ are the spanning vectors of the cluster.

\section{Classical Minimization}\label{app:ClassMin}
Table~\ref{tab:ClassMin} summarizes the main results from the classical minimization procedure of Sec.~\ref{Sec:ClassMin} for the finite clusters considered in our ED study but also for the thermodynamic limit (last line). The main findings are as follows: \\

\noindent
{\it (i) $J_c=0.3$, all clusters} --- 
Here the minimum sits at $\vec{Q}=(\pi,\pi)$, with $\vec{v}_1(\vec{Q})=\frac{1}{\sqrt{2}}(1,1)$ and $\lambda_1(\vec{Q}) < \lambda_2(\vec{Q})$. The minimum energy is achieved by
\bea
\sigma_{\vec{k},\alpha}=\sqrt{2N_{uc}}~ v_{1\alpha}(\vec{Q})\delta_{\vec{k},\vec{Q}} 
\Rightarrow
\sigma_{\vec{R},\alpha}=\sqrt{2}~ v_{1\alpha}(\vec{Q}) e^{i\vec{Q}\cdot\vec{R}}
\eea
where $N_{uc}\!=\!N/2$ stands for the number of unit cells, and the constants have been chosen to satisfy the spin length constraint. The energy is given by $E'/N_{uc}\!=\!\lambda_1(\vec{Q})$.\\

\noindent
{\it (ii) $J_c=0.1$, all clusters except N=36} --- 
Here we have two optimal wavevectors $\vec{Q}_1\!=\!(0,\pi)$ and $\vec{Q}_2\!=\!(\pi,0)$ with $\lambda_1(\vec{Q}_1)=\lambda_1(\vec{Q}_2)$ and $\vec{v}_1(\vec{Q}_1)\!=\!(1,0)$, $\vec{v}_1(\vec{Q}_2)\!=\!(0,1)$, and  $\lambda_1(\vec{Q}_j) \!<\! \lambda_2(\vec{Q}_j)$. Then the solutions that satisfy the spin length constraint are 
\be
\sigma_{\vec{R},1}=\pm e^{i \vec{Q}_1\cdot\vec{R}}, ~~\sigma_{\vec{R},2}=\pm e^{i \vec{Q}_2\cdot\vec{R}}
\ee
i.e., we have four ground states, with energy $E'/N_{uc}\!=\!\lambda_1(\vec{Q}_1)$.\\

\noindent
{\it (iii) $J_c\leq\!0.01$, all clusters except N=36 and 72} --- Here the minima sit at $\pm\vec{Q}\!=\!\pm (-\frac{\pi}{2},\frac{\pi}{2})$ with eigenvectors $\vec{v}_1(\vec{Q})\!=\!(1,\mp i)/\sqrt{2}$, and again $\lambda_1(\vec{Q})\!<\!\lambda_2(\vec{Q})$. Let us try the ansatz:
\bea
\sigma_{\vec{k},\alpha}&=&\frac{\sqrt{N_{uc}}}{\sqrt{2}} \Big( v_{1\alpha}(\vec{Q})\delta_{\vec{k},\vec{Q}} + v_{1\alpha}^\ast(\vec{Q})\delta_{\vec{k},-\vec{Q}} \Big) \\
\Rightarrow
\sigma_{\vec{R},\alpha}&=&\frac{1}{\sqrt{2}}\Big[ v_{1\alpha}(\vec{Q}) e^{i\vec{Q}\cdot\vec{R}}+v_{1\alpha}(-\vec{Q}) e^{-i\vec{Q}\cdot\vec{R}}\Big] \\
&=& \sqrt{2}~ \text{Re} [ v_{1\alpha}(\vec{Q}) e^{i\vec{Q}\cdot\vec{R}} ] 
= \Big( \cos(m-n)\frac{\pi}{2}, ~\sin(m-n)\frac{\pi}{2} \Big)
\eea
where we labeled $\vec{R}\!=\!n \vec{e}_x+ m\vec{e}_y$, and $n$, $m$ are integers. This ansatz does not satisfy the spin length constraint at all sites. Another ansatz is $\sigma_{\vec{R},\alpha}= \Big( -\sin(m-n)\frac{\pi}{2}, ~\cos(m-n)\frac{\pi}{2} \Big)$, which results from the first ansatz by replacing $\vec{v}_{1}(\vec{Q})\!\to\! i \vec{v}_1(\vec{Q})$. To get a solution that satisfies the spin constraint we combine the two: 
\begin{widetext}
\bea
\sigma_{\vec{k},\alpha}&=&\frac{\sqrt{N_{uc}}}{\sqrt{2}} \Big( (\epsilon_1 + i\epsilon_2) v_{1\alpha}(\vec{Q})\delta_{\vec{k},\vec{Q}} + 
(\epsilon_1 - i\epsilon_2) v_{1\alpha}^\ast(\vec{Q})\delta_{\vec{k},-\vec{Q}} \Big) \\
\Rightarrow
\sigma_{\vec{R},\alpha}&=&\frac{1}{\sqrt{2}}\Big[ (\epsilon_1 + i\epsilon_2) v_{1\alpha}(\vec{Q}) e^{i\vec{Q}\cdot\vec{R}}
+(\epsilon_1 - i\epsilon_2) v_{1\alpha}(-\vec{Q}) e^{-i\vec{Q}\cdot\vec{R}}\Big] 
= \sqrt{2}~ \text{Re} [ (\epsilon_1 + i\epsilon_2) v_{1\alpha}(\vec{Q}) e^{i\vec{Q}\cdot\vec{R}} ] \\
&=& \Big( \epsilon_1\cos(m-n)\frac{\pi}{2}-\epsilon_2\sin(m-n)\frac{\pi}{2}, ~\epsilon_1\sin(m-n)\frac{\pi}{2} + \epsilon_2\cos(m-n)\frac{\pi}{2}\Big)
\eea
where $\epsilon_{1,2}\!=\!\pm 1$, i.e. we have four possible solutions, all with energy $E'/N_{uc}\!=\!\lambda_1(\vec{Q})$. \\
\end{widetext}

\noindent
{\it All remaining cases}--- Here we cannot satisfy the spin length constraint. This is what happens e.g. for N=36 and $J_c\!\leq\!0.1$, for N=72 and $J_c\!\leq\!0.01$, and for $N=\infty$ and $J_c\!\leq\! 0.01$. In these cases, $\lambda_1(\vec{Q})$ serves only as a lower bound of the energy (see comparison with ED data in Table~\ref{tab:ClassMin}). From these results we can infer that the best choice for the cutoff is 0.001, and the best finite size cluster for the investigation of the $t\!=\!0$ ground states is the N=64 cluster, whose ground state energy per unit cell, $E'/N_{uc}\!=\!-2.108984$, is very close to the one for $N\!=\!\infty$ ($E'/N_{uc}\!=\!-2.11938$). The next good cluster (again in terms of energy) is N=32 which has $E'/N_{uc}\!=\!-2.056628$.

\begin{widetext}

\begin{table}[!tb] \ra{1.3}
\caption{Results for the classical ground state at $t=0$ from the classical minimization method of Sec.~\ref{Sec:ClassMin}, and Exact Diagonalizations (ED). Here $N_{\text{ints}}$ is the total number of interaction terms (of the type $S_i^zS_j^z$) in the Hamiltonian, $\vec{Q}$ is the minimum of $\lambda_1(\vec{k})$ over the BZ, $\vec{v}_1(\vec{Q})$ and $\lambda_1(\vec{Q})$ are the corresponding eigenvector (dashes indicate when the eigenvectors $\vec{v}_1(\vec{Q})$ cannot satisfy the spin length constraint) and eigenvalue, respectively. The last line for each given N gives the corresponding ground state energies per unit cell (multiplied by a factor of 4 to account for the unit spin length) as found by ED. Bold numbers indicate the cases with $N_{\text{ints}}(N)\!<\!N_{\text{ints}}(\infty)$ (for the given cutoff $J_c$) due to the finite size, showing that it is not safe to decrease the cutoff further.}\label{tab:ClassMin}
\begin{ruledtabular}
\begin{tabular}{cc|lccc} 
N &&\text{cutoff}=0.3 & 0.1 & 0.01  & 0.001 \\ 
\hline
16  & $N_{\text{ints}}/N$ & 3  & {\bf 5} & {\bf 6.5} & {\bf 7.5} \\ 
      & $\vec{Q}$ & $(\pi,\pi)$  & $(0,\pi), (\pi,0)$ & $\pm(-\pi/2,\pi/2)$ & $\pm(-\pi/2,\pi/2)$  \\ 
       &$\vec{v}_1(\vec{Q})$ & $(1,1)/\sqrt{2}$  & $(1,0)$, $(0,1)$ & $(1,\mp i)/\sqrt{2}$ & $(1,\mp i)/\sqrt{2}$ \\ 
       &$\lambda_1(\vec{Q})$ & -2.211504  & -1.97784 & -2.011411 & -1.995727 \\ 
       &ED & -2.21150 &-1.97784&-2.011411& -1.995727 \\ 
\hline
32   & $N_{\text{ints}}/N$& 3  & 7 & {\bf 10.5} & {\bf 15.5} \\ 
       &$\vec{Q}$ & $(\pi,\pi)$  & $(0,\pi), (\pi,0)$ & $\pm(-\pi/2,\pi/2)$ & $\pm(-\pi/2,\pi/2)$ \\ 
       &$\vec{v}_1(\vec{Q})$& $(1,1)/\sqrt{2}$  & $(1,0)$, $(0,1)$ & $(1,\mp i)/\sqrt{2}$ & $(1,\mp i)/\sqrt{2}$  \\
       &$\lambda_1(\vec{Q})$ & -2.211504 & -2.271598 & -2.089309 & -2.056628 \\
       &ED & -2.21150 & -2.271598 & -2.089309 & -2.056628 \\
\hline
36   & $N_{\text{ints}}$& 3  & 7 & {\bf 11.5} & {\bf 16.5} \\
       &$\vec{Q}$ & $(\pi,\pi)$  & $\pm(-\pi,\pi/3)$ & $\pm(\pi/3,-\pi/3)$ & $\pm(\pi/3,-\pi/3)$ \\
       &$\vec{v}_1(\vec{Q})$& $(1,1)/\sqrt{2}$  & --- & --- & --- \\
       &$\lambda_1(\vec{Q})$ & -2.211504  & -2.03528 & -1.95403 &  -1.94119 \\
       &ED & -2.21150 & -1.8845151 & -1.8803936 & -1.8731703 \\
\hline
64   & $N_{\text{ints}}/N$& 3  & 7 & 12 & {\bf 25.5} \\
       &$\vec{Q}$ & $(\pi,\pi)$  & $(0,\pi), (\pi,0)$ & $\pm(-\pi/2,\pi/2)$ & $\pm(-\pi/2,\pi/2)$ \\
       &$\vec{v}_1(\vec{Q})$& $(1,1)/\sqrt{2}$  & $(1,0)$, $(0,1)$ & $(1,\mp i)/\sqrt{2}$ & $(1,\mp i)/\sqrt{2}$ \\
       &$\lambda_1(\vec{Q})$ & -2.211504  & -2.271598 & -2.166198 & -2.108984  \\
       &ED & -2.21150 & -2.271598 & 2.166198& -2.108984 \\
\hline
72   & $N_{\text{ints}}/N$& 3  & 7 & 12 & {\bf 27.5} \\
       &$\vec{Q}$ & $(\pi,\pi)$  & $(0,\pi), (\pi,0)$ & $\pm(2\pi/3,-\pi/3)$ & $\pm(2\pi/3,-\pi/3)$ \\
       &$\vec{v}_1(\vec{Q})$& $(1,1)/\sqrt{2}$  & $(1,0)$, $(0,1)$ & --- & --- \\
       &$\lambda_1(\vec{Q})$ & -2.211504  & -2.271598 & -2.075198 & -2.03849 \\
       &ED & -2.21150 & -2.271598 & -2.04433866 & -2.0137376 \\
\hline
$\infty$ & $N_{\text{ints}}/N$& 3   & 7 & 12 &  31 \\
       &$\vec{Q}$ & $(\pi,\pi)$  & $(0,\pi)$, $(\pi,0)$ & $\pm 0.473296 (-\pi,\pi)$ & $\pm 0.4573374 (-\pi,\pi)$ \\
       &$\vec{v}_1(\vec{Q})$&$(1,1)/\sqrt{2}$&$(1,0)$, $(0,1)$&---&--- \\
       &$\lambda_1(\vec{Q})$& -2.21150  & -2.27159 & -2.17273 & -2.11938 \\
       &$\lambda_1(\frac{-\pi}{2},\frac{\pi}{2})$& -1.93452 & -1.93452 & -2.16620 & -2.10609 \\
\end{tabular}
\end{ruledtabular}
\end{table}
\end{widetext}

\section{Technical details on classical Monte Carlo simulations}\label{app:CMC}
Low energy configurations are generated by thermal annealing consisting of a million sweeps per site, using a combination of single-spin flip and long-loop updates. The former can annihilate all types of defects (3up-1down, 3down-1up, 4up, and 4down) but suffers from very low acceptance ratios at low temperatures, while the loop updates have much higher acceptance ratios ($\sim$16\%) and can annihilate all defects except the 3up-1down (or 3down-1up). So combining both types of updates gives sufficiently large acceptance ratios and can annihilate all defects.

\bibliographystyle{apsrev4-1}
\bibliography{libraryAll.bib}

\end{document}